\documentclass[reprint,superscriptaddress,aps,prl,longbibliography,nofootinbib]{revtex4-2}

\usepackage{amsmath}
\usepackage{amssymb}
\usepackage{amsthm}
\usepackage{mathtools}
\usepackage{mathdots}
\usepackage{amsfonts}
\usepackage{bbm}
\usepackage{xr}
\usepackage{xcolor}
\usepackage{bbold}
\usepackage{epsfig,color,dsfont,upgreek,physics}
\usepackage{mathrsfs}
\usepackage{multirow}
\usepackage{graphicx}
\usepackage{dcolumn} 
\usepackage{bm} 
\usepackage{float}
\usepackage[driverfallback=dvipdfm]{hyperref} 
\usepackage{longtable}
\usepackage{ulem}  
\usepackage{enumerate}
\usepackage{url}
\allowdisplaybreaks

\begin{document}

\title{Unified Theory of Spin and Orbital Inverse Faraday Effects in Rashba Metals}

\author{Jaglul Hasan}
\affiliation{Department of Physics and Astronomy, Iowa State University, Ames, Iowa 50011, USA}
\affiliation{Ames National Laboratory, U.S. Department of Energy, Ames, Iowa 50011, USA}

\author{Maxim Dzero}
\affiliation{Department of Physics, Kent State University, Kent, Ohio 44242, USA}

\author{Chandan Setty}
\affiliation{Department of Physics and Astronomy, Iowa State University, Ames, Iowa 50011, USA}
\affiliation{Ames National Laboratory, U.S. Department of Energy, Ames, Iowa 50011, USA}
\begin{abstract}
The inverse Faraday effect (IFE) offers a promising route toward optical control of magnetic moments in quantum materials. Yet experiments on spin--orbit--coupled metals indicate that no single microscopic mechanism suffices, implicating orbital currents alongside spin polarization. We derive both responses in a disordered Rashba metal using quantum kinetics and diagrammatics with impurity--renormalized optical vertices, providing the first unified diagrammatic derivation of spin and orbital IFE. Treating impurity scattering consistently, our number-conserving kinetic formulation restores full phase-space derivatives and the optical magnetic field neglected in earlier treatments. We connect nonlinear orbital magnetization to absorptive linear optical conductivity, revealing that the magnetic Lorentz force supplies half the orbital response. Our unified treatment resolves both channels across the Rashba resonance: spin IFE changes sign as frequency increases, while the orbital IFE dominates, keeping the direction of the total magnetization unchanged.
\end{abstract}
\date{November 4, 2026}
\maketitle


\textit{Introduction.---}The inverse Faraday effect (IFE) enables optical control of magnetism by coupling the angular momentum of circularly polarized light to electronic spins and orbital motion~\cite{Pitaevskii:1960,Pershan:1963,Malmstrom:1965,Malmstrom:1966,Pisarev:Nature2005,Kirilyuk2006,Kruglyak2007,Satoh2010,AHMReid2010,RevModPhys:2010,Mikhaylovskiy2012,Hinzke2015,UlrichNowak2021,Kohlmann2025,Huber2026,GiovanniDeNinnoPRL2022,JinZuanming2011,Pavlov2018,Guo2016,Ortiz2023,Nikitin1990,Perroni2006,PerroniLiebsch2006,Pisarev2008,Popova2012,Takayoshi2014,BattiatoSweden2014}. Yet, despite much work, the microscopic origin of the IFE even in nominally simple spin--orbit--coupled (SOC) metals remains unsettled: femtosecond measurements in terbium gallium garnet (TGG) point to orbital magnetization~\cite{Mikhaylovskiy2012}, while transition-metal Kerr measurements reveal material trends captured by neither free-electron orbital estimates nor interband spin polarization calculations~\cite{Ortiz2023}. This raises a basic theoretical question: how do spin and orbital IFE channels coexist in a SOC metal, and what determines their relative weight in the total light-induced magnetization? In this Letter, we address this question by developing a unified microscopic theory of spin and orbital IFEs in a disordered Rashba two-dimensional electron gas (2DEG).

The difficulty is that existing microscopic theories have treated these channels largely separately. Edelstein's high-frequency spin-IFE theory, closely related to current-induced magnetoelectric coupling~\cite{Edelstein:1990,Edelstein:1995,Edelstein:2005}, requires dissipation on time-reversal-symmetry grounds~\cite{Edelstein:1998,Edelstein:2000}. Other spin IFE theories address low-frequency orbital-to-spin angular-momentum transfer through perturbative SOC~\cite{Taguchi:2011,Taguchi:2012} and resonant or lattice-model responses~\cite{Titov2016,Tanaka2020,Tanaka2024}, but neither evaluate orbital magnetization nor establish a common microscopic impurity-scattering description of the joint frequency and Rashba-coupling dependence across these regimes.

Orbital IFE persists without SOC~\cite{Hertel:2006,Zhang2009,Hertel2015,Yoshino2011,Balatsky2024}, with related phenomena studied across diverse physical systems~\cite{Petrila2015,Mokrousov2023,Yokoyama:2018,Mironov2021,Majedi:2021,Buzdin:2022,BuzdinAVortices:2022,Tokman2020,Banerjee2022,ShabalaGeilhufe2024,Belotelov2010,Nadarajah2017,Oppeneer:2018,Gonzalez2024,Parchenko2025,Hareau2025,Hamid2015,SongJinIm2017,CholSong2019,Krichevsky2024:1,Krichevsky2024:2,Mou2023,Yang2023,XingyuYang2024,YoshidaYokoyama2026,Maxim:2024,WongBalatsky2025}. A Rashba kinetic theory~\cite{Maxim:2025} nevertheless evaluates only the orbital IFE within a nonconserving approximation that explicitly neglects collision-integral feedback and part of the full phase-space derivative structure. Its response vanishes without SOC, can reverse with frequency, and has an $\omega^{-5}$ high-frequency tail, whereas the ordinary-metal orbital response retains its sign and decays as $\omega^{-3}$~\cite{Hertel:2006,Balatsky2024}. First-principles calculations predict material-specific spin and orbital IFEs~\cite{Berritta2016,Adamantopoulos2025}, but do not resolve their joint dependence on spin precession and microscopic impurity relaxation. The optical magnetic-field contribution is absent from their electric-dipole formulations and neglected in previous orbital treatments because the magnetic force is considered negligible compared with the electric force~\cite{Hertel:2006,Balatsky2024,Maxim:2025}. A unified microscopic theory must therefore explain how SOC modifies spin and orbital IFEs and how their interplay determines the total magnetization under both optical fields.

Our number-conserving kinetic formulation and diagrammatic calculations treat spin and orbital IFEs within a common microscopic impurity-scattering framework, providing the first diagrammatic derivation of orbital IFE.  We identify an overlooked physical mechanism: The optical magnetic field deflects the electrically driven current, producing as much orbital magnetization as the electric force. We relate this orbital moment to absorptive linear optical conductivity, explaining how SOC reshapes its frequency dependence while preserving its sign. Rashba spin precession and impurity relaxation instead produce a dispersive spin IFE that reverses as frequency crosses the Rashba resonance. Near resonance, the enhanced orbital contribution dominates and keeps the total magnetization's direction unchanged. The full orbital response connects continuously to the ordinary-metal limit as SOC vanishes and retains its $\omega^{-3}$ high-frequency tail. These results provide a microscopic basis for interpreting light-induced magnetization and exploring nonlinear optical control in low-dimensional electronic systems~\cite{AdvQuantTech:2022,Sc.Reps:2017,FerromagMoire:2022,SCCuprate:2011,PhysRevMaterials.1.014401,CDWNatPhys:2020,Metals:2019,RoadMap:2022}.\newline

\textit{Model and observables.---}We consider a disordered Rashba 2DEG with both helical branches occupied,
\begin{equation}
\begin{aligned}
 \hat H={}&\frac{\boldsymbol\pi^2}{2m}-\varepsilon_F
 +\alpha_{\mathrm{so}}(\boldsymbol\pi\times\mathbf{c})\cdot\boldsymbol{\sigma}
 +e\varphi+U(\mathbf r),\\
 \boldsymbol\pi={}&-i\boldsymbol\nabla-\frac e{c_0}\mathbf A .
\end{aligned}
\label{Eq:H}
\end{equation}
Here $e<0$, $\mathbf{c}=\mathbf e_z$, $c_0$ is the speed of light, and Gaussian units with $\hbar=k_B=1$ are used. Broken inversion symmetry permits the Rashba splitting $E_\lambda(p)=p^2/(2m)-\varepsilon_F+\lambda\alpha_{\mathrm{so}}p$, $\lambda=\pm1$. Short-range scalar disorder gives a momentum-relaxation rate $\Gamma=\tau^{-1}$ and single-particle width $\Gamma/2$ in the self-consistent Born approximation (SCBA). The scales are $p_F=\sqrt{2m\varepsilon_F}$, $v_F=p_F/m$, $\Delta_R=2\alpha_{\mathrm{so}}p_F$, $\delta=\alpha_{\mathrm{so}}/v_F$, and $\nu_0=m/(2\pi)$ per spin. At $T=0$ and fixed chemical potential, the analytical response retains arbitrary $\omega\tau$ and $\Delta_R\tau$ within $\Gamma,\omega,\Delta_R\ll\varepsilon_F$ and $\delta\ll1$, using broadened equilibrium propagators and impurity ladders to leading metallic order (SM Sec.~SI and Appendix~A~\cite{SM}).

The physical field is $\mathbf{E}_{\mathrm{phys}}(\mathbf r,t)=\mathbf{E}(\mathbf r)e^{-i\omega t}+\mathrm{c.c.}$, where $\mathbf{E}(\mathbf r)=E_0(\mathbf r)(\mathbf e_x+i\ell\mathbf e_y)e^{i\mathbf q\cdot\mathbf r}/\sqrt2$, $\ell=\pm1$, and $E_0(\mathbf r)$ varies slowly in plane. Its helicity is $\mathcal{H}=i\mathbf{c}\cdot(\mathbf{E}\times\mathbf{E}^*)=\ell|E_0|^2$. With $\hat w=-iG^</(2\pi)$ and $\int_{\mathbf p,\epsilon}=\int d\epsilon\,d^2p/(2\pi)^2$,
\begin{equation}
\begin{aligned}
 S_z&=\frac12\int_{\mathbf p,\epsilon}\operatorname{Tr}(\sigma_z\hat w),
 \qquad M_{{\mathrm{spin}},z}=-g\mu_B S_z,\\
 \mathbf j&=e\int_{\mathbf p,\epsilon}\operatorname{Tr}(\hat{\mathbf v}\hat w),\\
 \hat{\mathbf v}&=\frac{\mathbf{p}}{m}+\alpha_{\mathrm{so}}(\mathbf{c}\times\boldsymbol{\sigma}),
\end{aligned}
\label{eq:phys-observables}
\end{equation}
where $\mu_B=|e|/(2m_ec_0)>0$. Free-carrier circulation defines the orbital sheet moment,
\begin{equation}
 \mathbf j^M=c_0\boldsymbol\nabla\times(M_{{\mathrm{orb}},z}\mathbf{c}),
 \qquad M_{{\mathrm{tot}},z}=M_{{\mathrm{spin}},z}+M_{{\mathrm{orb}},z}.
\label{Eq:jIFEtoM}
\end{equation}
This complements crystalline orbital-magnetization theory~\cite{Thonhauser2005,Shi2007}.

\textit{Conserving quantum kinetics.---}The Keldysh parametrization $G^K=G^R\circ\hat h-\hat h\circ G^A$, with Wigner convolution $\circ$, defines the distribution $\hat h$. Its local-force structure at the retained gradient order is~\cite{Kita:2001,MSH:2004,Andrey2006,DzeroLevchenko2025,Maxim:2025}
\begin{equation}
\begin{aligned}
 \partial_t\hat h+\tfrac12\{\hat v_i,\partial_{R_i}\hat h\}
 +i[\mathbf b_p\cdot\boldsymbol{\sigma},\hat h]
 &+eE_i\mathscr D_i\hat h\\
 +\frac e{2c_0}\{(\hat{\mathbf v}\times\mathbf{B})_i,\partial_{p_i}\hat h\}
 &=\hat I_{\mathrm{B}}[\hat h],
\end{aligned}
\label{eq:kineticeqn}
\end{equation}
with $\mathbf b_p=\alpha_{\mathrm{so}}\mathbf{p}\times\mathbf{c}$ and
\begin{equation}
 \mathscr D_i\hat h=
 \partial_{p_i}\hat h+\tfrac12\{\hat v_i,\partial_\epsilon\hat h\}.
\label{eq:phase-space}
\end{equation}
Unlike the field-operator reduction in Ref.~\cite{Maxim:2025}, $\mathscr D_i$ retains the full momentum and energy derivatives in the linear and quadratic sources. At quadratic order they also differentiate velocities, precession denominators, and the driven distribution before energy integration.

The magnetic term in Eq.~\eqref{eq:kineticeqn} retains light's Lorentz force, neglected in electric-field-only orbital treatments~\cite{Hertel:2006,Balatsky2024,Maxim:2025}. Acting on the first-order electrically driven distribution, it generates a quadratic dc circulating current at the same envelope-gradient order as the electric density force.

Here $\hat I_{\mathrm B}$ is the SCBA impurity collision integral,
\begin{equation}
\begin{aligned}
 \hat I_{\mathrm B}[\hat h]
 &=i\hat\Sigma^K-i(\hat\Sigma^R\circ\hat h-\hat h\circ\hat\Sigma^A),\\
 \hat\Sigma^a
 &=\frac{\Gamma}{2\pi\nu_0}\int\frac{d^2p}{(2\pi)^2}\hat G^a,
 \qquad a=R,A,K.
\end{aligned}
\label{eq:born-collision}
\end{equation}
Both scattering-in $i\hat\Sigma^K$ and scattering-out are retained in the driven response. Their feedback enforces charge continuity and generates impurity vertex corrections. Reference~\cite{Maxim:2025} retains lifetime broadening but explicitly neglects this driven feedback. The collision kernel retains the finite-frequency shifts
$\epsilon\pm\omega/2$; the local-force notation does not
impose $\omega\tau\ll1$ (SM Sec.~SIII).

In the SOC-free scalar limit, $I[f]=\Gamma(\langle f\rangle_\phi-f)$ cancels density relaxation while the current relaxes at $\Gamma$, yielding
\begin{equation}
 -i\omega\,\delta n^{(1)}
 +\frac{i}{e}\mathbf q\cdot\mathbf j^{(1)}=0 .
\label{eq:phys-charge-gain}
\end{equation}
Loss alone spuriously replaces $-i\omega$ by $\Gamma-i\omega$. The conserved density harmonic feeds the circulating current through the quadratic electric force $e\delta n^{(1)}\mathbf{E}$ despite having no direct current moment.

\begin{figure*}[t]
\centering
\includegraphics[width=\textwidth]{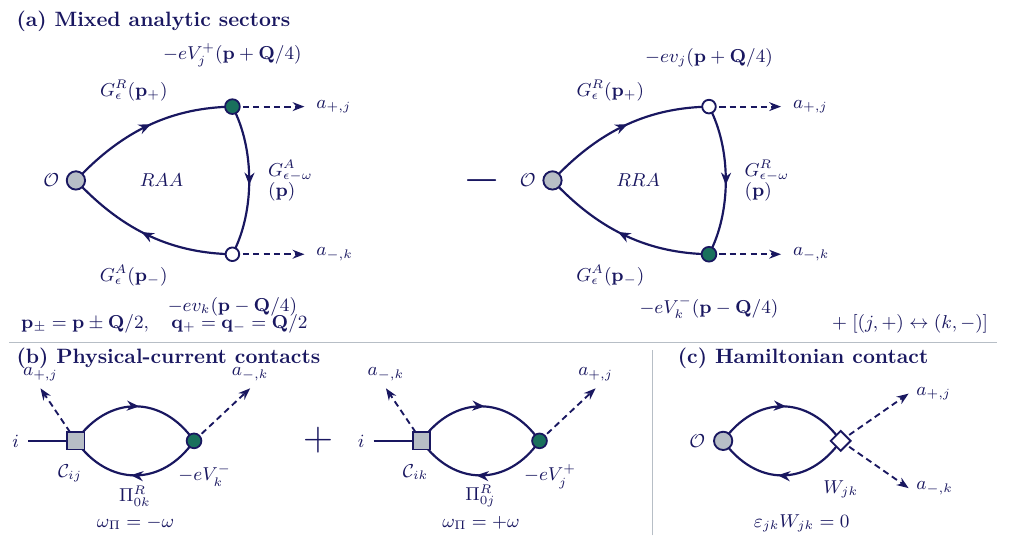}
 \caption{Spin and orbital IFE diagrams, with $a_\pm=A_\pm/c_0$. (a) $RAA-RRA$ sectors and exchanged optical ordering. The measurement vertex is dressed $\widetilde\sigma_z/2$ or dressed $eJ_i$. Green circles denote impurity dressing on an $RA$ pair; the other optical vertex is bare. (b) Two physical-current contacts, $\mathcal C_{ij}=-e^2\delta_{ij}\mathbf1/m$, each attach one optical leg to the measurement vertex; bubbles denote the full retarded density--current response. (c) The quadratic Hamiltonian contact $W_{jk}=e^2\delta_{jk}\mathbf1/m$ vanishes in the helicity-odd response.}
 \label{fig:IFEdiagrams}
\end{figure*}

\textit{Diagrammatic response.---}Independently, expansion to second order in the optical field gives the spin-density and physical-current diagrams in Fig.~\ref{fig:IFEdiagrams}.
The spin insertion measures spin polarization; the orbital response follows from the current linear in output momentum $\mathbf Q$ before setting $\mathbf Q\to0$. This envelope momentum differs from the optical carrier momentum $\mathbf q$. Both optical orderings and the field-dependent physical-current contacts in Fig.~\ref{fig:IFEdiagrams} enter the response. These contacts follow from differentiating the current with respect to $\mathbf A$ and complete the electromagnetic response~\cite{AversaSipe1995,SipeShkrebtii2000} (SM Secs.~SVII--SVIII).

Impurity scattering renormalizes the velocity operator in the optical vertex according to
\begin{equation}
\begin{aligned}
 \hat V_i&=\hat v_i+
 \frac{\Gamma}{2\pi\nu_0}\int\frac{d^2p}{(2\pi)^2}
 G_+^R\hat V_iG_-^A\\
 &=\frac{p_i}{m}+\alpha_{\mathrm{so}}s(\mathbf{c}\times\boldsymbol{\sigma})_i .
\end{aligned}
\label{eq:phys-optical-vertex}
\end{equation}
Here $G_\pm^{R,A}=G^{R,A}(\mathbf{p},\epsilon\pm\omega/2)$, $z=\Gamma-i\omega$, and
\begin{equation}
\begin{aligned}
 w&=\frac{\Gamma}{z},&
 u&=\frac{\Gamma z}{z^2+\Delta_R^2},\\
 v&=\frac{w+u}{2},&
 s&=\frac{1-w}{1-v},
\end{aligned}
\label{eq:ladders}
\end{equation}
where the amplitudes $w,u,v$ describe charge, out-of-plane-spin, and in-plane-spin propagation respectively. Edelstein's high-frequency treatment dresses the spin-output vertex but neglects optical dressing~\cite{Edelstein:1998}. Here $s(\omega)$ is retained in both the spin coefficient and optical conductivity [Eqs.~\eqref{eq:spin-result} and~\eqref{eq:sigma}], linking both IFE channels to impurity-dressed Rashba precession. The kinetic scattering-in feedback yields the same factor (SM Sec.~SIV).  The static current measurement vertex is also impurity dressed. At $\mathbf Q=0$, the ladder cancels its anomalous Rashba
component, giving $\hat J_i=(p_i/m)\mathbf 1$, while the
finite-frequency optical vertex retains $s(\omega)$.
The finite-$\mathbf Q$ dependence is retained until the curl derivative is taken (SM Sec.~SVIII).

\begin{figure*}[t]
 \centering
 \includegraphics[width=\textwidth]{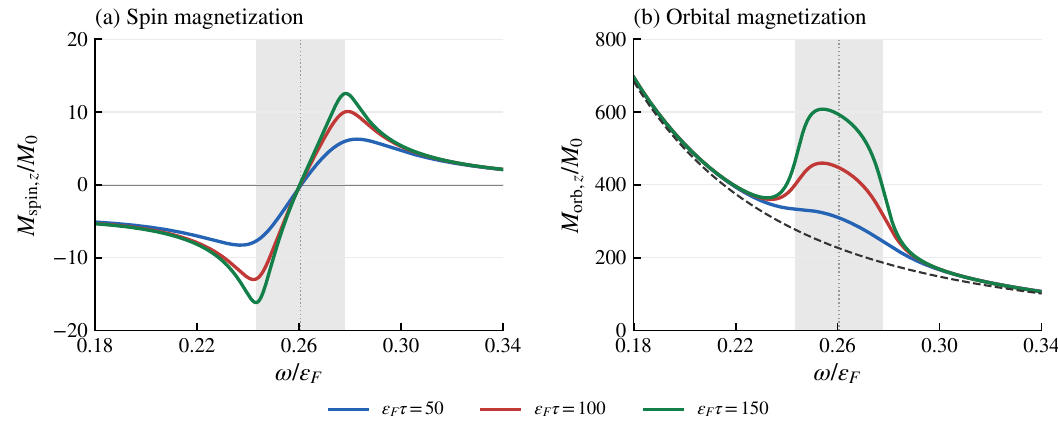}
 \caption{Full-momentum numerical IFE. (a) Spin magnetization reverses across the Rashba transition window (gray). (b) Orbital magnetization remains positive and dominates; the dashed curve is the SOC-free Drude result at $\varepsilon_F\tau=100$. Colors denote $\varepsilon_F\tau=50,100,150$; $\alpha_{\mathrm{so}}p_F/\varepsilon_F=0.13$, $\ell=-1$, $g=2$, and $m=m_e$, with common normalization $M_0$. The reservoir rate is $\eta/\Gamma=0.0025$ and the kinetic-energy cutoff is $4\varepsilon_F$. Curves interpolate 81 calculated frequencies per disorder value; convergence and analytical comparisons appear in SM Appendix~D.}
 \label{fig:channels}
\end{figure*}

\textit{Orbital IFE benchmark.---}Without SOC, spin generation vanishes but charge circulation survives. The Drude amplitudes $\mathbf v_\omega=e\mathbf{E}/[m(\Gamma-i\omega)]$ and $\mathbf r_\omega=\mathbf v_\omega/(-i\omega)$ give, with $n_0=p_F^2/(2\pi)$,
\begin{equation}
\begin{aligned}
 L_z^{(0)}&=n_0m\,\overline{(\mathbf r\times\mathbf v)_z}
 =\frac{2n_0e^2}{m\omega(\Gamma^2+\omega^2)}\mathcal{H},\\
 M_{{\mathrm{orb}},z}^{(0)}&=\frac{e}{2mc_0}L_z^{(0)}
 =\frac{n_0e^3}{m^2c_0\omega(\Gamma^2+\omega^2)}\mathcal{H} .
\end{aligned}
\label{eq:orbIFEnormal}
\end{equation}
Both microscopic calculations recover this circulation~\cite{Hertel:2006,Balatsky2024} (SM Sec.~SII), scaling as $\omega^{-1}$ for $\omega\ll\Gamma$ and $\omega^{-3}$ for $\Gamma\ll\omega$. The low-frequency periodic-drive limit requires displacement small compared with the envelope scale.

The normalization is $M_0=n_0\mu_B[e|E_0|/(p_F\varepsilon_F)]^2$. Representative InAs values $m=0.048m_e$ and $n_0=2.5\times10^{12}\,\mathrm{cm}^{-2}$ give $\varepsilon_F\simeq125\,\mathrm{meV}$~\cite{Matsuda2020,Bennett2019} and $M_0/(n_0\mu_B)\simeq4\times10^{-8}$ for $|E_0|=100\,\mathrm{V/cm}$. Relative channel weights depend on $g$ and $m/m_e$.

\textit{Spin IFE in a Rashba 2DEG.---}Electric driving displaces the Rashba spin texture, inducing in-plane Edelstein polarization~\cite{Edelstein:1990}. Circular driving rotates the effective SOC field, producing interbranch coherence and a perpendicular dc spin component. Both the kinetic equation and the spin-insertion diagrams give
\begin{equation}
\begin{aligned}
 M_{{\mathrm{spin}},z}&=g\mu_B e^2\nu_0\mathcal C(\omega)\mathcal{H},\\
 \mathcal C(\omega)&=\operatorname{Im}\left[
 s\frac{2\alpha_{\mathrm{so}}^2/z-\Gamma/p_F^2}{z^2+\Delta_R^2}
 \right].
\end{aligned}
\label{eq:spin-result}
\end{equation}
Here $s$ is the optical-vertex factor in Eq.~\eqref{eq:ladders}; $-\Gamma/p_F^2$ arises from radial phase-space and output-vertex terms (SM Secs.~SV and SVII).

For resolved Rashba splitting away from resonance,
\begin{equation}
 \mathcal C(\omega)\simeq
 \frac{\Delta_R^2}{2p_F^2\omega(\Delta_R^2-\omega^2)},
 \qquad \Gamma\ll\omega,\ |\omega-\Delta_R|.
\label{eq:spin-clean}
\end{equation}
Virtual interbranch mixing changes sign across the splitting. Near resonance, the dressed precession pole produces the dispersive term $-(\omega-\Delta_R)/[(\omega-\Delta_R)^2+(3\Gamma/4)^2]$, with a background-dependent zero.

At $\max(\Gamma,\Delta_R)\ll\omega\ll\varepsilon_F$,
\begin{equation}
 \mathcal C(\omega)\simeq
 -\frac{\Delta_R^2+4\Gamma^2}{2p_F^2\omega^3}.
\label{eq:phys-spin-high}
\end{equation}
The coefficient crosses from SOC-controlled precession to frequent-scattering scaling for $\Delta_R\ll\Gamma$, as in Edelstein's optical limit~\cite{Edelstein:1998,Edelstein:2000}; matched spin-relaxation and field conventions determine the prefactors (SM Sec.~SIX). A low-frequency SOC expansion cannot resolve this resonance~\cite{Taguchi:2012}. Exactly zero SOC has no spin source, but the stationary $\alpha_{\mathrm{so}}\to0$ limit is nonuniform because spin relaxation also vanishes. We plot the spin magnetization in Fig.~\ref{fig:channels}(a).

\textit{Orbital IFE in a Rashba 2DEG.---}The momentum moment of Eq.~\eqref{eq:kineticeqn} combines gain and loss into $F_{{\mathrm{imp}},i}=-m\Gamma j_i/e$, including the anomalous Rashba current missed by loss alone. The stationary Rashba spin equation cancels the uniform helicity-odd momentum flux, leaving the current balance
\begin{equation}
\begin{aligned}
 \Gamma\mathbf j^{(2),H}
 =\Bigg[&\frac{e^2}{m}\sum_a n_{-a}^{(1)}\mathbf{E}_a+\frac{e}{mc_0}\sum_a\mathbf j_{-a}^{(1)}\times\mathbf{B}_a
 \Bigg]^H .
\end{aligned}
\label{eq:force-balance}
\end{equation}
where $\omega_a=a\omega$, $a=\pm$, and $H$ denotes the helicity-odd part. Charge conservation and Faraday's law give
$e\omega_a n_a^{(1)}=\mathbf q_a\cdot\mathbf j_a^{(1)}$ and
$\mathbf{B}_a=c_0\mathbf q_a\times\mathbf{E}_a/\omega_a$, with $\mathbf j_a^{(1)}=\sigma(a\omega)\mathbf{E}_a$. Both forces occur at the same gradient order in the in-plane envelope.

The optical momenta are
$\mathbf q_\pm=\pm\mathbf q+\mathbf Q/2$.
For zero in-plane carrier momentum, the symmetric envelope
extraction uses $\mathbf q_\pm=\mathbf Q/2$;
the $\mathbf Q$-linear current gives
\begin{equation}
 \boxed{\begin{gathered}
 M_{{\mathrm{orb}},z}
 =\frac{e}{mc_0\Gamma\omega}\operatorname{Re}\sigma(\omega)\mathcal{H},\\
 M_{E,z}=M_{B,z}=\tfrac12M_{{\mathrm{orb}},z}.
 \end{gathered}}
\label{eq:orbital-conductivity}
\end{equation}
The electric field acts on the induced density; the magnetic field deflects the driven current. The physical-current diagrams independently reproduce Eq.~\eqref{eq:orbital-conductivity} after summing triangles and contacts; their momentum-dependent minimal coupling encodes both forces (SM Secs.~SVI and SVIII). We plot the orbital magnetization in Fig.~\ref{fig:channels}(b).

Absorption supplies optical torque $\mathcal T_z=\allowbreak 2\operatorname{Re}\sigma\,\mathcal{H}/\omega$. Its balance with impurity relaxation gives the stationary angular momentum $L_z=\mathcal T_z/\Gamma$ and moment $M_{{\mathrm{orb}},z}=eL_z/(2mc_0)$. The conductivity contains the same optical vertex as spin IFE:
\begin{equation}
\begin{aligned}
 \frac{\sigma(\omega)}{\sigma_D}
 &=w(1+2\delta^2)-\delta^2s(w-u), \quad
 \sigma_D&=\frac{e^2\nu_0v_F^2}{\Gamma}.
\end{aligned}
\label{eq:sigma}
\end{equation}
The Drude term reflects the two-branch density; the interbranch term carries $s(\omega)$, explicitly linking SOC modification of orbital IFE to spin dynamics.

Since $\operatorname{Re}\sigma\ge0$ in a passive metal, the orbital sign remains fixed in this model despite resonant enhancement. For Fig.~\ref{fig:channels}, orbital dominance prevents total-moment reversal when spin reverses; multiband orbital sign reversals remain possible~\cite{Berritta2016,Adamantopoulos2025}.

The numerics retain both Fermi contours and field-induced spectral changes across the transition width $4m\alpha_{\mathrm{so}}^2$. When this rivals $\Gamma$, analytical poles alone do not fix peak heights or linewidths. Spin density and curl current are evaluated independently (SM Appendix~D). The full orbital $\omega^{-3}$ tail is distinct from a density-matched SOC correction (SM Sec.~SIX).\newline

\textit{Discussion and experimental implications.---}%
Our results show why spin polarization reversal need not imply the total magnetization reversal: SOC reshapes both channels, but their distinct frequency dependences allow orbital circulation to dominate while spin polarization reverses (Fig.~\ref{fig:channels}). The total moment therefore depends on the angular momentum retained in both channels. Spin-only descriptions~\cite{Taguchi:2011,Taguchi:2012,Titov2016,Tanaka2020,Tanaka2024} leave this orbital contribution undetermined.

The coherent limit clarifies the role of time-reversal symmetry. Circular driving supplies time-reversal breaking through $i(\mathbf E\times\mathbf E^*)$, whereas static scalar disorder is time-reversal invariant. The detuned clean response [Eq.~\eqref{eq:spin-clean}] extends beyond Edelstein's dissipative hierarchy $\Delta_R\ll\Gamma\ll\omega$~\cite{Edelstein:1998,Edelstein:2000}: disorder controls relaxation, optical-vertex dressing, and resonance broadening, rather than being a universal prerequisite for spin IFE.

Orbital magnetization survives without SOC~\cite{Hertel:2006,Zhang2009,Hertel2015,Yoshino2011,Balatsky2024}. The underlying orbital angular momentum accumulates through the balance of optical torque and relaxation. The magnetic force supplies half this response [Eq.~\eqref{eq:orbital-conductivity}], making the optical magnetic field essential to nonlinear magnetization despite its nominally weak coupling. Both full moments have $\omega^{-3}$ tails for $\max(\Gamma,\Delta_R)\ll\omega\ll\varepsilon_F$, whereas a density-matched SOC correction can decay as $\omega^{-5}$. Comparison with Ref.~\cite{Maxim:2025} therefore requires matching the observable and density convention, as detailed in the SM.

The transition-metal measurements~\cite{Ortiz2023} reveal comparable Kerr responses in Cu, Au, Pd, and Pt despite their different SOC strengths, together with opposite Kerr signs and pronounced structural-phase dependence in Ta and W. Our results suggest a possible interpretation of this weak correlation with SOC: orbital circulation survives without SOC and can dominate the induced moment, masking substantial variations in spin generation. SOC then modifies both channels, so material-dependent changes in their relative magnitudes and signs can reinforce or oppose this orbital background. Free-electron orbital estimates~\cite{Hertel:2006} and interband spin calculations~\cite{Mishra2023,Mishra2025} separately capture only parts of this balance. An oppositely directed spin contribution could reverse the total moment if it becomes sufficiently large, but the observed Kerr signs alone do not establish spin dominance. Multiband orbital responses can themselves change sign~\cite{Berritta2016,Adamantopoulos2025}, and the conversion from induced moments to Kerr rotation depends on the material and probe frequency~\cite{Ortiz2023}. Our theory therefore identifies how competing magnetic channels can produce weak or nonmonotonic SOC trends, while their assignment in individual metals requires a material-specific calculation of both moments and their optical readout.

Ultrafast TGG measurements provide complementary motivation: Faraday rotation exceeds the static-Verdet estimate by about two orders of magnitude~\cite{Mikhaylovskiy2012}, highlighting the need to assess orbital angular momentum alongside spin polarization. Although its localized-electron response differs from the itinerant circulation considered here, both settings raise the question of how optical angular momentum is distributed between the two magnetic channels.

Gate-tunable quantum wells offer a direct test~\cite{Nitta1997,Winkler2003,Matsuda2020,Bennett2019}. For $\alpha_{\mathrm{so}}p_F/\varepsilon_F=0.13$ and $\varepsilon_F=125\,\mathrm{meV}$, the splitting is approximately $32\,\mathrm{meV}$ ($8\,\mathrm{THz}$). Optical orientation~\cite{MeierZakharchenya1984,Ivchenko2005}, photogalvanic spectroscopy~\cite{IvchenkoPikus1978,Ganichev2001CPGE,GanichevPrettl2003,GanichevGolub2014,Golub2003}, spin-galvanic conversion~\cite{Ganichev2002SGE}, and orbital photocurrents~\cite{Tarasenko2007} provide complementary probes, although their transport currents differ from our magnetization current. These measurements would complement studies of current-induced spin and orbital polarizations~\cite{Yoda2015,Yoda2018,Johansson2021,Salemi2021} and orbitronics~\cite{Go2021Orbitronics}. Combining conductivity, spin-sensitive detection, and calibrated magneto-optics across the tunable resonance can test the predicted separation between dispersive spin reversal and absorptive orbital accumulation, identifying when reversal of spin polarization leaves the total magnetic moment unchanged.\newline

\textit{Acknowledgments.---}We thank Jigang Wang for discussions. J.H. and C.S. are supported by Ames National Laboratory and Iowa State University start--up funds. M. D. acknowledges the support of the NSF grant No. DMR-2400484.
\bibliography{biblio}

\end{document}


\title{\texorpdfstring{Supplemental Material\\}{Supplemental Material: }Unified Theory of Spin and Orbital Inverse Faraday Effects in Rashba Metals}
\author{Jaglul Hasan}
\affiliation{Department of Physics and Astronomy, Iowa State University, Ames, Iowa 50011, USA}
\affiliation{Ames National Laboratory, U.S. Department of Energy, Ames, Iowa 50011, USA}
\author{Maxim Dzero}
\affiliation{Department of Physics, Kent State University, Kent, Ohio 44242, USA}
\author{Chandan Setty}
\affiliation{Department of Physics and Astronomy, Iowa State University, Ames, Iowa 50011, USA}
\affiliation{Ames National Laboratory, U.S. Department of Energy, Ames, Iowa 50011, USA}
\maketitle
\onecolumngrid
\setcounter{secnumdepth}{2}
\setcounter{tocdepth}{2}
\begingroup
\makeatletter
\def\l@f@section{\addpenalty{\@secpenalty}\addvspace{0.55em plus 1pt}}
\makeatother
\tableofcontents
\endgroup

\section*{Purpose and organization}
This Supplemental Material supplies the derivations supporting the Letter's spin and orbital inverse Faraday effects (IFEs) in a disordered Rashba two-dimensional electron gas. Sections~\ref{sec:model}--\ref{sec:normal-total} define the conventions and establish the SOC-free orbital benchmark, including the electric and magnetic force contributions. Sections~\ref{sec:phase-space-framework}--\ref{sec:optical-transport} develop the conserving kinetic equation, impurity ladders, and linear optical response. The spin and orbital kinetic calculations are followed by their independent diagrammatic derivations, including the differentiated measurement vertices and physical-current contacts. Section~\ref{sec:scope} collects the frequency and relaxation limits used in the Letter and the convention-matched comparisons needed to interpret them. The appendices specify the quadratic spectral approximation, provide independent scalar checks, and document the full-momentum numerical calculation.

The optical vertices, static measurement vertices, and current contacts are distinct. Their definitions and momentum derivatives are retained explicitly. Figure~\ref{new:contact-diagrams} displays the two measurement-current contacts and the separate quadratic Hamiltonian contact; individual diagram topologies are not identified with the electric and magnetic halves of the kinetic force balance.

\section{Model and zero-temperature conventions}
\label{sec:model}

We consider a 2DEG with Rashba spin--orbit coupling (SOC) and spin-independent, short-range disorder. The unit vector $\mathbf{c} $ is normal to the electron plane. With $e<0$ the signed electron charge, the single-particle Hamiltonian is
\begin{equation}
 \hat H=\frac{\boldsymbol\pi^2}{2m}-\varepsilon_F
 +\alpha_{\mathrm{so}}(\boldsymbol\pi\times\mathbf{c} )\cdot\boldsymbol{\sigma} 
 +e\varphi+U(\mathbf r),\qquad
 \boldsymbol\pi=-\mathrm{i} \boldsymbol\nabla-\frac{e}{c_0}\mathbf A .
 \label{eq:model}
\end{equation}
Here $c_0$ denotes the speed of light, to distinguish it from $\mathbf{c} $, and $\alpha_{\mathrm{so}}$ is the Rashba velocity. We use Gaussian electromagnetic units and set $\hbar=k_B=1$. Energies $\epsilon$ are measured from the equilibrium chemical potential; consequently occupation functions are given as
\begin{equation}
 f_0(\epsilon)=\Theta(-\epsilon),\qquad
 F(\epsilon)\equiv 1-2f_0(\epsilon)=\operatorname{sgn}\epsilon,
 \qquad F'(\epsilon)=2\delta(\epsilon).
 \label{eq:fermi}
\end{equation}
All results in this supplement are evaluated at zero temperature. The symbol $f_0$ denotes equilibrium occupation; $w(\mathbf{R} ,\mathbf{p} ,\epsilon,T)$ denotes the SOC-free energy-resolved Wigner density, and $f(\mathbf r,\mathbf{p} ,t)$ denotes its energy-integrated occupation. The center time is initially denoted by $T$ and subsequently by $t$. A hat denotes a matrix in physical spin space. Hats are suppressed where the matrix structure is unambiguous.

The field-free Hamiltonian and velocity are
\begin{equation}
 \hat H_0(\mathbf{p} )=\xi_p\mathbf{1} +\mathbf b_p\cdot\boldsymbol{\sigma} ,
 \qquad \xi_p=\frac{p^2}{2m}-\varepsilon_F,
 \qquad \mathbf b_p=\alpha_{\mathrm{so}}\mathbf{p} \times\mathbf{c} ,
 \qquad \hat v_i=\frac{p_i}{m}\mathbf{1} +\alpha_{\mathrm{so}}(\mathbf{c} \times\boldsymbol{\sigma} )_i .
 \label{eq:velocity}
\end{equation}
We assume that both Rashba branches are occupied and that the system is metallic,
\begin{equation}
 \varepsilon_F\tau\gg1,\qquad
 \delta\equiv\frac{\alpha_{\mathrm{so}}}{v_F}\ll1,\qquad
 |\omega|,\ \tau^{-1},\ \Delta_R\ll\varepsilon_F,
 \qquad
 p_F=\sqrt{2m\varepsilon_F},\quad v_F=\frac{p_F}{m},\quad
 \Delta_R=2\alpha_{\mathrm{so}} p_F .
 \label{eq:regime}
\end{equation}
The response is evaluated in the single-splitting metallic, noncrossing impurity-ladder approximation: the leading branch asymmetry of the phase-space measure and velocity is retained, while $\omega\tau$ and $\Delta_R\tau$ are not expanded. Radial integrals are extended to the full real axis after the metallic expansion. We retain the energy moments generated by this expansion before setting the energy to the Fermi level. The collision block and spectral reconstruction used at quadratic order are specified together in Appendix~\ref{app:nonlinear}.

\subsection{Observables, conserved moments, and notation}
We choose $\mathbf{c} =\mathbf e_z$, write $\sigma_z=\mathbf{c} \cdot\boldsymbol{\sigma} $, and set
\begin{equation}
 \Gamma=\tau^{-1},\qquad z=\Gamma-\mathrm{i} \omega,\qquad
 \hat{\mathsf R}_i=(\mathbf{c} \times\boldsymbol{\sigma} )_i,\qquad \hat{\mathsf R}_x=-\sigma_y,\quad \hat{\mathsf R}_y=\sigma_x,
 \qquad \varepsilon_{xyz}=1.
 \label{eq:shared-shortcuts}
\end{equation}
The matrices $\hat{\mathsf R}_i$ are the components of $\hat{\boldsymbol{\mathsf R}}=\mathbf{c} \times\boldsymbol{\sigma} $. We use $\varepsilon_{ijk}$ for the three-dimensional Levi-Civita symbol. Its two-dimensional restriction is $\varepsilon^{(2)}_{ij}\equiv\varepsilon_{ijk}c_k=\varepsilon_{ijz}$ for $i,j\in\{x,y\}$, so $\varepsilon^{(2)}_{xy}=1$, $\varepsilon^{(2)}_{yx}=-1$, and $\varepsilon^{(2)}_{xx}=\varepsilon^{(2)}_{yy}=0$.
$\Gamma$ is the relaxation rate of an anisotropic charge distribution, whereas the one-particle propagator has width $\Gamma/2$. Neither is generally a spin relaxation rate. The out-of-plane spin density is $S_z=\langle\sigma_z/2\rangle$ per unit area, and its magnetic moment is $M_{\mathrm{spin},z}=-g\mu_B S_z$, with $\mu_B>0$. The orbital quantity below is the coefficient of a local magnetization current,
\begin{equation}
 \mathbf j^M=c_0\boldsymbol\nabla\times(M_{\mathrm{orb},z}\mathbf{c} ),\qquad
 j_i^M(\mathbf Q)=\mathrm{i}  c_0\varepsilon^{(2)}_{i\ell}Q_\ell M_{\mathrm{orb},z}(\mathbf Q).
 \label{orb:eq:curl}
\end{equation}
It is a sheet magnetic moment: no layer thickness is introduced. A three-dimensional comparison must replace the sheet density by the volume density, or specify a thickness.

At the retained local-response order the induced moment is
$M_{\mathrm{tot},z}=M_{\mathrm{spin},z}+M_{\mathrm{orb},z}$, with each component measured by its own vertex. Spin angular momentum, orbital angular momentum, magnetization, and optical helicity are distinct quantities. They will not be denoted by the same symbol.

\begin{center}
\begin{tabular}{p{.20\linewidth}p{.73\linewidth}}
\toprule
Symbol & Definition and role\\\midrule
$\epsilon,\xi_p$ & Electronic energies relative to $\varepsilon_F$, not kinetic energy measured from the band bottom.\\
$\omega,\Omega,\mathbf Q$ & Optical frequency, output frequency, and output momentum. The orbital calculation sets $\Omega=0$ before evaluating its coefficient linear in $Q$.\\
$w(\mathbf{R} ,\mathbf{p} ,\epsilon,T)$ & Energy-resolved Wigner density per spin.\\
$\Delta_R,\delta$ & Rashba splitting $2\alpha_{\mathrm{so}} p_F$ and small contour-asymmetry parameter $\alpha_{\mathrm{so}}/v_F$.\\
$w,u,v,s$ & Charge, out-of-plane-spin, and in-plane-spin ladder eigenvalues, and anomalous-velocity dressing, respectively. Their formulas are repeated at the start of each calculation; $\hat v_i$ is a velocity matrix.\\
$\mathcal{H} $ & $\mathrm{i} \mathbf{c} \cdot(\mathbf E\times\mathbf E^*)$ in the positive-frequency Fourier-amplitude convention.\\
\bottomrule
\end{tabular}
\end{center}
\newpage
\section{Normal metal without SOC: from the energy-resolved Wigner equation to orbital IFE}
\label{sec:normal-total}
This section derives the normal-metal result before introducing spin matrices or diagrammatic vertices. The starting point is the distribution resolved in the center coordinates and in the energy and momentum conjugate to the relative coordinates. Its gauge-covariant Wigner equation establishes the force terms before an energy integral is taken. The distribution remains energy resolved through the linear harmonics, quadratic source, and current moments; the energy integral is taken only after those moments have been evaluated. The resulting scalar calculation provides a consistency limit for the later Rashba calculation. Set $\alpha_{\mathrm{so}}=0$ throughout this section. The two spin states have the same scalar occupation, and spin enters only as a factor of two in the density and current.

\subsection{The distribution resolved in center coordinates, momentum, and energy}
The starting object is the two-point lesser Green function per spin,
\begin{equation}
G^<(x_1,x_2)=\mathrm{i} \langle\psi^\dagger(x_2)\psi(x_1)\rangle,
\qquad x_a=(t_a,\mathbf r_a).
\end{equation}
Introduce the center coordinates and the relative coordinates by
\begin{align}
X=(T,\mathbf{R} ),\qquad
T&=\frac{t_1+t_2}{2},\quad \mathbf{R} =\frac{\mathbf r_1+\mathbf r_2}{2},\nonumber\\
s=(s_t,\mathbf s),\qquad
s_t&=t_1-t_2,\quad \mathbf s=\mathbf r_1-\mathbf r_2,
\qquad x_{1,2}=X\pm s/2.
\label{eq:nm-center-relative}
\end{align}
Here $T$ is the center time, not temperature. The energy $\epsilon$ and momentum $\mathbf{p} $ are Fourier conjugate to the relative time $s_t$ and relative position $\mathbf s$. They are independent Wigner variables; no condition $\epsilon=\xi_p$ is imposed when taking their partial derivatives.

In an electromagnetic field the transform must include the straight-line gauge links
\begin{equation}
U(a,b)=\exp\!\left[\mathrm{i}  e\int_b^a
\left(\frac{\mathbf A\cdot\mathrm{d} \mathbf r}{c_0}-\varphi\,\mathrm{d}  t\right)\right].
\end{equation}
The energy-resolved Wigner density is
\begin{align}
w(\mathbf{R} ,\mathbf{p} ,\epsilon,T)
={}&\frac{1}{2\pi\mathrm{i} }\int\mathrm{d}  s_t\,\mathrm{d} ^2s\,
e^{\mathrm{i} \epsilon s_t-\mathrm{i} \mathbf{p} \cdot\mathbf s}\nonumber\\
&\times U(X,x_1)G^<(x_1,x_2)U(x_2,X).
\label{eq:nm-wigner-density}
\end{align}
Here $w(\mathbf{R} ,\mathbf{p} ,\epsilon,T)$ is the energy-resolved one-particle density per spin defined by Eq.~\eqref{eq:nm-wigner-density}. The gauge links make $\mathbf{p} $ the kinetic momentum. The equilibrium density is $w^{(0)}=f_0(\epsilon)\mathcal A_0(\mathbf{p} ,\epsilon)/(2\pi)$, with $\mathcal A_0=\mathrm{i} (G_0^R-G_0^A)$. In the quasiparticle limit it becomes $f_0(\epsilon)\delta(\epsilon-\xi_p)$, where $\xi_p=p^2/(2m)-\varepsilon_F$.

The streaming equation is obtained by subtracting the left and right Dyson equations and expanding their gauge-covariant Wigner products. It is not obtained by first assuming an occupation depending only on $\mathbf{R} $, $\mathbf{p} $, and $T$. For the scalar inverse free propagator $L_0=\epsilon-\xi_p$, the first-gradient commutator is
\begin{align}
\mathrm{i} [w,L_0]_\circ
={}&\partial_T w+\mathbf v\cdot\boldsymbol\nabla_{\mathbf R}w
+e E_i(\partial_{p_i}+v_i\partial_\epsilon)w\nonumber\\
&+\frac{e}{c_0}(\mathbf v\times\mathbf B)\cdot\boldsymbol\nabla_{\mathbf p}w
+\text{higher Wigner gradients},\qquad \mathbf v=\mathbf{p} /m.
\label{eq:nm-wigner-streaming}
\end{align}
This follows by inserting $\partial_\epsilon L_0=1$ and $\partial_{p_i}L_0=-v_i$ into Eq.~\eqref{eq:covariant-product}; the electromagnetic part of that product produces both derivatives in the electric term. The self-energy terms supply the collision operator. In the scalar quasiparticle and isotropic elastic closure used in this section, the retained transport equation is
\begin{equation}
\boxed{\left[\partial_T+\mathbf v\cdot\boldsymbol\nabla_{\mathbf R}
+e\mathbf E\cdot(\boldsymbol\nabla_{\mathbf p}+\mathbf v\partial_\epsilon)
+\frac{e}{c_0}(\mathbf v\times\mathbf B)\cdot\boldsymbol\nabla_{\mathbf p}\right]w
=\mathcal C_W[w].}
\label{eq:nm-wigner-kinetic}
\end{equation}
Here $\mathcal C_W[w]=\Gamma(\langle w\rangle_\phi-w)$ at fixed $p$ and $\epsilon$; its gain and loss are discussed below. Equation~\eqref{eq:nm-wigner-kinetic} is the retained first-gradient scalar transport equation, not the untruncated quantum Dyson equation. The finite-frequency Wigner shifts and the spin-matrix collision structure are derived separately in the subsequent quantum sections.

At this transport order the streaming operator admits characteristics in the extended phase space:
\begin{equation}
\dot{\mathbf{R} }=\mathbf v,\qquad
\dot{\mathbf{p} }=e\left(\mathbf E+\frac{\mathbf v\times\mathbf B}{c_0}\right),\qquad
\dot\epsilon=e\mathbf E\cdot\mathbf v.
\label{eq:nm-characteristics}
\end{equation}
Dots here denote derivatives with respect to the center time $T$. During a short interval $\mathrm{d}  T$, the extended phase-space point moves to $(\mathbf{R} +\dot{\mathbf{R} }\mathrm{d}  T,\mathbf{p} +\dot{\mathbf{p} }\mathrm{d}  T,\epsilon+\dot\epsilon\mathrm{d}  T,T+\mathrm{d}  T)$. Taylor expansion gives the complete chain rule:
\begin{align}
\frac{\mathrm{d}  w}{\mathrm{d}  T}
={}&\lim_{\mathrm{d}  T\to0}\frac{
w(\mathbf{R} +\dot{\mathbf{R} }\mathrm{d}  T,\mathbf{p} +\dot{\mathbf{p} }\mathrm{d}  T,\epsilon+\dot\epsilon\mathrm{d}  T,T+\mathrm{d}  T)
-w(\mathbf{R} ,\mathbf{p} ,\epsilon,T)}{\mathrm{d}  T}\nonumber\\
={}&\partial_T w+\dot{\mathbf{R} }\cdot\boldsymbol\nabla_{\mathbf R}w
+\dot{\mathbf{p} }\cdot\boldsymbol\nabla_{\mathbf p}w
+\dot\epsilon\,\partial_\epsilon w\nonumber\\
={}&\mathcal C_W[w].
\label{eq:nm-total-derivative}
\end{align}
In the absence of collisions, $w$ is constant along this characteristic at the retained transport order. Impurity scattering changes it by transferring particles between momentum directions. Each term has a distinct origin: the time derivative follows the change at a fixed phase-space point; spatial streaming follows the displacement of the electron; the momentum derivatives follow the electric and magnetic changes of kinetic momentum; and $\dot\epsilon\partial_\epsilon w$ follows the electric work in the independent energy coordinate. Keeping only a frequency or energy derivative would not reproduce this total derivative. Every momentum derivative here holds $\epsilon$ fixed, and every energy derivative holds $\mathbf{p} $ fixed. The energy derivative is present from the outset. For the parabolic dispersion,
\begin{equation}
\dot\xi_p=\mathbf v\cdot\dot{\mathbf{p} }=e\mathbf E\cdot\mathbf v=\dot\epsilon,
\qquad \mathbf v\cdot(\mathbf v\times\mathbf B)=0.
\label{eq:nm-work}
\end{equation}
Thus the scalar streaming preserves $\epsilon-\xi_p$; this is a consequence of the extended equation and does not identify the two independent variables before differentiation. The electric field changes energy, while the magnetic field changes momentum direction. Both can contribute to a nonlinear transverse current even though only the electric field does work.

\subsection{Elastic gain, loss, and spectral current moments}
The gain and loss have a direct Pauli-blocking interpretation. Let $n_{\mathbf{p} }$ and $n_{\mathbf{p} '}$ be the dimensionless occupations per spin of an outgoing and an incoming one-particle state at the same elastic energy. A transition into $\mathbf{p} $ requires an occupied incoming state and an empty final state, giving $n_{\mathbf{p} '}(1-n_{\mathbf{p} })$; scattering out of $\mathbf{p} $ gives $n_{\mathbf{p} }(1-n_{\mathbf{p} '})$. For a static spin-independent impurity potential the reverse elastic transition has the same rate, $W_{\mathbf{p} \mathbf{p} '}=W_{\mathbf{p} '\mathbf{p} }$. With $\int_{\mathbf{p} '}\equiv\int\mathrm{d} ^2p'/(2\pi)^2$, the same collision form follows from the Pauli factors:
\begin{align}
C[n]_{\mathbf{p} }
&=\int_{\mathbf{p} '} W_{\mathbf{p} \mathbf{p} '}
\bigl[n_{\mathbf{p} '}(1-n_{\mathbf{p} })-n_{\mathbf{p} }(1-n_{\mathbf{p} '})\bigr]\nonumber\\
&=\int_{\mathbf{p} '} W_{\mathbf{p} \mathbf{p} '}(n_{\mathbf{p} '}-n_{\mathbf{p} }).
\label{eq:nm-pauli-collision}
\end{align}
where the elastic-energy delta function is included in $W$. The terms quadratic in the occupations cancel exactly; the linear gain-minus-loss form is therefore consistent with Fermi statistics and does not assume a dilute gas. Pauli factors apply to these dimensionless occupations. In the scalar quasiparticle closure, multiplying by the common direction-independent spectral weight $\mathcal A_0/(2\pi)$ gives the corresponding equation for the physical energy-resolved density $w$. Elastic scattering preserves the momentum magnitude and redistributes its direction. For isotropic scattering with total rate $\Gamma=\tau^{-1}$, the full angular collision operator consequently acts at fixed $p$ and $\epsilon$:
\begin{align}
C[w](p,\phi,\epsilon)
&=\Gamma\int_0^{2\pi}\frac{\mathrm{d} \phi'}{2\pi}
[w(p,\phi',\epsilon)-w(p,\phi,\epsilon)]\nonumber\\
&=\underbrace{\Gamma\langle w\rangle_\phi}_{\text{gain}}
-\underbrace{\Gamma w}_{\text{loss}},\qquad \Gamma=\tau^{-1}.
\label{eq:nm-collision}
\end{align}
Both terms are retained. With $P_0X=\langle X\rangle_\phi$, Eq.~\eqref{eq:nm-wigner-kinetic} can be written in the source form used for the perturbative expansion:
\begin{equation}
\boxed{[\partial_T+\mathbf v\cdot\boldsymbol\nabla_{\mathbf R}+\Gamma(1-P_0)]w
=-e\mathbf E\cdot(\boldsymbol\nabla_{\mathbf p}+\mathbf v\partial_\epsilon)w
-\frac{e}{c_0}(\mathbf v\times\mathbf B)\cdot\boldsymbol\nabla_{\mathbf p}w.}
\label{eq:nm-Boltzmann}
\end{equation}
Define the moments before integrating over energy:
\begin{align}
N(\epsilon)&=2\int\!\frac{\mathrm{d}^2p}{(2\pi)^2}  w_{\mathbf{p} \epsilon},\qquad
\mathcal P_i(\epsilon)=2\int\!\frac{\mathrm{d}^2p}{(2\pi)^2}  v_iw_{\mathbf{p} \epsilon},\nonumber\\
\Pi_{ij}(\epsilon)&=2\int\!\frac{\mathrm{d}^2p}{(2\pi)^2}  v_iv_jw_{\mathbf{p} \epsilon},\qquad
j_i(\epsilon)=e\mathcal P_i(\epsilon),\nonumber\\
n&=\int\mathrm{d} \epsilon\,N(\epsilon),\qquad
j_i=\int\mathrm{d} \epsilon\,j_i(\epsilon).
\label{eq:nm-moments}
\end{align}
The factor of two counts the two spin states. The symbol $n$ is number density; $en$ is charge density. $\mathcal P_i$ is a particle-current moment; the mechanical momentum density is $m\int\mathrm{d} \epsilon\,\mathcal P_i(\epsilon)$. Conservation and angular parity give
\begin{equation}
\langle C[w]\rangle_\phi=0,\qquad
2\int\!\frac{\mathrm{d}^2p}{(2\pi)^2}  v_iC[w]=-\Gamma\mathcal P_i(\epsilon).
\label{eq:nm-collision-moments}
\end{equation}
The gain cancels the loss in the density channel but has zero first angular moment because $\langle v_i\rangle_\phi=0$, equivalently $\langle p_i\rangle_\phi=0$. The density is conserved, while the anisotropic first harmonic carrying current decays at $\Gamma$. Replacing $C[w]$ by $-\Gamma w$ would incorrectly relax the density as well.

Energy is measured from the chemical potential. The kinetic energy measured from the band bottom is consequently
\begin{equation}
K(\epsilon)=\epsilon+\varepsilon_F,\qquad
v^2\big|_{\epsilon=\xi_p}=\frac{2K(\epsilon)}m.
\label{eq:nm-radial-energy}
\end{equation}
For $K>0$ the radial integrals needed below are
\begin{equation}
\int\!\frac{\mathrm{d}^2p}{(2\pi)^2}  v^2\delta(\epsilon-\xi_p)=\frac{K}{\pi},\qquad
\int\!\frac{\mathrm{d}^2p}{(2\pi)^2}  v^4\delta(\epsilon-\xi_p)=\frac{2K^2}{\pi m}.
\label{eq:nm-current-integral}
\end{equation}
They vanish below the band bottom. The zero-temperature derivatives are localized at $\epsilon=0$, where $K=\varepsilon_F>0$. The equilibrium sheet density and density of states per spin are
\begin{equation}
n_0=2\int_0^{p_F}\frac{p\,\mathrm{d}  p}{2\pi}=\frac{p_F^2}{2\pi},\qquad
\nu_0=\frac{m}{2\pi},\qquad n_0=m\nu_0v_F^2.
\label{eq:nm-density}
\end{equation}
Here $\nu_0$ is the density of states per spin and $v_F=p_F/m$.

\subsection{Independent optical sources and the output limit}
The magnetization is extracted from its current,
\begin{equation}
\mathbf j^M=c_0\boldsymbol\nabla\times(M_z\mathbf{c} ),\qquad
j_i^M(\mathbf Q)=\mathrm{i}  c_0\varepsilon^{(2)}_{i\ell}Q_\ell M_z,\qquad \varepsilon^{(2)}_{xy}=1.
\label{eq:nm-curl}
\end{equation}
Use independent optical amplitudes $\mathbf E_+$ and $\mathbf E_-$ with
\begin{equation}
\omega_s=s\omega+\frac\Omega2,\qquad
\mathbf q_s=s\mathbf k+\frac{\mathbf Q}{2},\qquad s=\pm.
\label{eq:nm-source-routing}
\end{equation}
Their mixed response has output $(\Omega,\mathbf Q)$. In the local optical response set $\mathbf k=0$ while retaining the output momentum. The coefficient linear in $\mathbf Q$ describes the derivative of the optical envelope. At finite output the two source amplitudes remain independent arguments of the response kernel; the physical condition $\mathbf E_- =\mathbf E_+^*$ is imposed only after the response coefficient is obtained. The conjugate response at $(-\Omega,-\mathbf Q)$ restores a real physical field. Define the source helicity by
\begin{equation}
\mathcal{H}_z=\mathrm{i} \mathbf{c} \cdot(\mathbf E_+\times\mathbf E_-).
\label{eq:nm-helicity}
\end{equation}
Faraday's law gives
\begin{equation}
\mathbf B_s=\frac{c_0}{\omega_s}\mathbf q_s\times\mathbf E_s,
\qquad \boldsymbol\nabla\times\mathbf E=-\frac1{c_0}\partial_T\mathbf B.
\label{eq:nm-Faraday}
\end{equation}
In the dc limit and at zero carrier momentum,
\begin{equation}
\mathbf B_+=\frac{c_0}{2\omega}\mathbf Q\times\mathbf E_+,
\qquad \mathbf B_-=-\frac{c_0}{2\omega}\mathbf Q\times\mathbf E_-.
\label{eq:nm-fields}
\end{equation}
The minus sign in $\mathbf B_-$ follows from $\omega_-=-\omega$, and the factor $1/2$ from the chosen momentum routing. Each magnetic field is already first order in $Q$, exactly the order needed for the magnetization current. The coefficient linear in $\mathbf Q$ must be calculated before $Q\to0$; a uniform magnetization itself has no bulk curl current, so the homogeneous current alone does not determine its coefficient.

\subsection{Equilibrium source and the energy-resolved linear harmonics}
Use the quasiparticle equilibrium density
\begin{equation}
w^{(0)}_{\mathbf{p} \epsilon}=f_0(\epsilon)\delta(\epsilon-\xi_p).
\label{eq:nm-energy-resolved}
\end{equation}
The electric derivative is
\begin{equation}
\mathscr D_i=\left.\partial_{p_i}\right|_\epsilon
+v_i\left.\partial_\epsilon\right|_{\mathbf{p} }.
\label{eq:nm-phase-direction}
\end{equation}
Applying both terms before simplifying gives
\begin{align}
\partial_{p_i}w^{(0)}&=-v_if_0(\epsilon)\delta'(\epsilon-\xi_p),\nonumber\\
v_i\partial_\epsilon w^{(0)}&=v_if_0'(\epsilon)\delta(\epsilon-\xi_p)
+v_if_0(\epsilon)\delta'(\epsilon-\xi_p),\nonumber\\
\mathscr D_iw^{(0)}&=v_if_0'(\epsilon)\delta(\epsilon-\xi_p).
\label{eq:nm-delta-cancel}
\end{align}
The magnetic equilibrium source is zero because $(\mathbf v\times\mathbf B_s)\cdot\mathbf v=0$.

Let $P_0X=\langle X\rangle_\phi$, $d_s=-\mathrm{i} \omega_s$, and $z_s=\Gamma+d_s$. Expanding $w=w^{(0)}+w^{(1)}+w^{(2)}+\cdots$, without factorials, in Eq.~\eqref{eq:nm-Boltzmann} gives at first order
\begin{equation}
\boxed{[d_s+\mathrm{i} \mathbf q_s\cdot\mathbf v+\Gamma(1-P_0)]w_s^{(1)}
=-e\mathbf E_s\cdot\mathbf v\,f_0'(\epsilon)\delta(\epsilon-\xi_p).}
\label{eq:nm-linear}
\end{equation}
This is an equation for $w_s^{(1)}(\mathbf{p} ,\epsilon)$, with no energy integral taken. At zero gradient the source is a first angular harmonic and has zero angular average. Its $\ell=1$ solution is
\begin{equation}
w_{s,1}^{(1;0)}=-\frac{e}{z_s}(\mathbf E_s\cdot\mathbf v)
f_0'(\epsilon)\delta(\epsilon-\xi_p).
\label{eq:nm-drift}
\end{equation}
The uniform drift has zero number-density moment. The first gradient produces an isotropic density harmonic. Applying $P_0$ to Eq.~\eqref{eq:nm-linear} eliminates the entire gain-minus-loss collision operator and gives
\begin{equation}
d_sw_{s,0}^{(1)}+\mathrm{i} \langle(\mathbf q_s\cdot\mathbf v)w_s^{(1)}\rangle_\phi=0.
\label{eq:nm-density-harmonic}
\end{equation}
The collision operator cancels in this projection. Inserting Eq.~\eqref{eq:nm-drift} on the right to first order in $q_s$ gives
\begin{equation}
w_{s,0}^{(1;1)}=\frac{\mathrm{i}  e v^2}{2z_sd_s}(\mathbf q_s\cdot\mathbf E_s)
f_0'(\epsilon)\delta(\epsilon-\xi_p)
=-\frac{ev^2}{2z_s\omega_s}(\mathbf q_s\cdot\mathbf E_s)
f_0'(\epsilon)\delta(\epsilon-\xi_p).
\label{eq:nm-density-distribution}
\end{equation}
The denominator $\omega_s$ in the density harmonic comes from conserved charge. If the gain were omitted, its angular average would instead decay at $\Gamma$, and the electric density force at quadratic order would be wrong.

The quadrupole follows by applying $P_2$ to the same equation:
\begin{equation}
w_{s,2}^{(1;1)}=\frac{\mathrm{i}  e}{z_s^2}
\left[(\mathbf q_s\cdot\mathbf v)(\mathbf E_s\cdot\mathbf v)
-\frac{v^2}{2}(\mathbf q_s\cdot\mathbf E_s)\right]
f_0'(\epsilon)\delta(\epsilon-\xi_p).
\label{eq:nm-second-harmonic}
\end{equation}
The expression in brackets has zero angular average and is a traceless second harmonic. It describes the optical stress. Its radial derivatives must be retained before taking a current moment; below, the corresponding spectral current is evaluated before its energy integral vanishes. Thus $w_s^{(1)}=w_{s,1}^{(1;0)}+w_{s,0}^{(1;1)}+w_{s,2}^{(1;1)}+O(q_s^2)$.

The spectral current from the drift harmonic is
\begin{equation}
j_{s,i}^{(1)}(\epsilon)=-\frac{e^2K}{\pi z_s}f_0'(\epsilon)E_{s,i},\qquad
N_s^{(1)}(\epsilon)=-\frac{eK}{\pi\omega_sz_s}
(\mathbf q_s\cdot\mathbf E_s)f_0'(\epsilon).
\label{eq:nm-linear-spectral-moments}
\end{equation}
These obey the charge continuity identity at every energy,
\begin{equation}
\boxed{e\omega_sN_s^{(1)}(\epsilon)=\mathbf q_s\cdot\mathbf j_s^{(1)}(\epsilon).}
\label{eq:nm-continuity}
\end{equation}
For the independent linear conductivity check, the current integral can also be evaluated explicitly at zero temperature:
\begin{align}
j_{s,i}^{(1)}
&=-\frac{2e^2}{z_s}E_{s,j}\int\!\frac{\mathrm{d}^2p}{(2\pi)^2}  v_iv_jf_0'(\xi_p),\nonumber\\
\int\!\frac{\mathrm{d}^2p}{(2\pi)^2}  v_iv_j\delta(\xi_p)
&=\int_0^\infty\frac{p\,\mathrm{d}  p}{2\pi}
\delta\!\left(\frac{p^2}{2m}-\varepsilon_F\right)
\frac{p^2}{2m^2}\delta_{ij}
=\frac{\nu_0v_F^2}{2}\delta_{ij}.
\label{eq:nm-drude-radial-check}
\end{align}
The angular identity is $\langle n_in_j\rangle_\phi=\delta_{ij}/2$, where $\mathbf n=\mathbf{p} /p$, and the radial delta function gives $p\,\mathrm{d}  p=m\,\mathrm{d} \xi_p$. Thus the energy integral of the linear current gives
\begin{equation}
\boxed{\sigma_0(\omega)=\frac{n_0e^2}{m(\Gamma-\mathrm{i} \omega)},\quad
\operatorname{Re} \sigma_0=\frac{n_0e^2\Gamma}{m(\Gamma^2+\omega^2)},\quad
\operatorname{Im} \sigma_0=\frac{n_0e^2\omega}{m(\Gamma^2+\omega^2)}.}
\label{eq:nm-Drude}
\end{equation}
The subscript zero denotes absence of SOC. In particular $\sigma_0(-\omega)=\sigma_0(\omega)^*$. The real part measures the current component in phase with the field and therefore the absorbed power. This independent check does not replace $w_s^{(1)}$ by its energy integral in the quadratic calculation.

\subsection{The quadratic equation before the energy integral}
For the mixed output $(\Omega,\mathbf Q)$, the full force structure gives
\begin{align}
[-\mathrm{i} \Omega+\mathrm{i} \mathbf Q\cdot\mathbf v+\Gamma(1-P_0)]w^{(2)}
={}&-e\sum_{s=\pm}E_{s,j}\mathscr D_jw_{\bar s}^{(1)}\nonumber\\
&-\frac{e}{c_0}\sum_{s=\pm}(\mathbf v\times\mathbf B_s)_j
\partial_{p_j}w_{\bar s}^{(1)},\qquad \bar s=-s.
\label{eq:nm-quadratic}
\end{align}
Both optical orderings and the energy derivative are retained. For the helicity current at first order in $Q$, the electric source acts on the first-gradient distribution, including its density and stress harmonics. The magnetic source needs only the uniform drift because $\mathbf B_s$ is already first order in $Q$. Both sources are order $E^2Q$.

A first velocity moment is enough to determine this current: it selects the $\ell=1$ part of $w^{(2)}$. All other angular harmonics are orthogonal to $v_i$ and have zero current moment. The following integrations therefore evaluate the observable directly without solving every harmonic of the quadratic distribution. Define $z_0=\Gamma-\mathrm{i} \Omega$ for the output current harmonic.

\subsubsection{Electric-force moment at fixed energy}
Integrate by parts only in momentum, holding $\epsilon$ fixed:
\begin{align}
R_{E,i}(\epsilon)
&=-2eE_j\int\!\frac{\mathrm{d}^2p}{(2\pi)^2}  v_i(\partial_{p_j}+v_j\partial_\epsilon)w\nonumber\\
&=2eE_j\int\!\frac{\mathrm{d}^2p}{(2\pi)^2}  (\partial_{p_j}v_i)w
-2eE_j\partial_\epsilon\int\!\frac{\mathrm{d}^2p}{(2\pi)^2}  v_iv_jw\nonumber\\
&=\boxed{\frac{e}{m}E_iN(\epsilon)-eE_j\partial_\epsilon\Pi_{ij}(\epsilon).}
\label{eq:nm-electric-force}
\end{align}
The momentum boundary term vanishes and $\partial_{p_j}v_i=\delta_{ij}/m$. This calculation includes derivatives of the velocity numerator, the spectral delta function, and the radial factors in the first-gradient harmonics. Differentiating only $f_0$, or replacing a radial numerator by its Fermi value before the derivative, would not implement the integration by parts. The momentum derivative transfers from the entire distribution to the measurement velocity and produces the density force. The independent energy derivative produces the second, energy-stress term. That term is still present at finite energy and cannot be removed before the energy integral.

\subsubsection{Magnetic-force moment at fixed energy}
Similarly,
\begin{align}
R_{B,i}(\epsilon)
&=-\frac{2e}{c_0}\int\!\frac{\mathrm{d}^2p}{(2\pi)^2}  v_i(\mathbf v\times\mathbf B)_j\partial_{p_j}w\nonumber\\
&=\frac{2e}{c_0}\int\!\frac{\mathrm{d}^2p}{(2\pi)^2} 
\partial_{p_j}[v_i(\mathbf v\times\mathbf B)_j]w\nonumber\\
&=\boxed{\frac{e}{mc_0}(\boldsymbol{\mathcal P}(\epsilon)\times\mathbf B)_i.}
\label{eq:nm-magnetic-force}
\end{align}
Here $\partial_{p_j}(\mathbf v\times\mathbf B)_j
=\varepsilon_{j\ell k}\delta_{j\ell}B_k/m=0$. Although the magnetic force annihilates $w^{(0)}$, it does not annihilate the anisotropic optical drift. This is the source of its quadratic contribution. The resulting spectral current-moment equation is
\begin{align}
\boxed{z_0\mathcal P_i^{(2)}(\epsilon)+\mathrm{i}  Q_j\Pi_{ij}^{(2)}(\epsilon)}
={}&\sum_{s=\pm}\left[
\frac{e}{m}E_{s,i}N_{\bar s}^{(1)}(\epsilon)
-eE_{s,j}\partial_\epsilon\Pi_{\bar s,ij}^{(1)}(\epsilon)\right.\nonumber\\
&\left.\hspace{10mm}+\frac{e}{mc_0}
(\boldsymbol{\mathcal P}_{\bar s}^{(1)}(\epsilon)\times\mathbf B_s)_i\right].
\label{eq:nm-momentum-balance}
\end{align}
The $\Gamma$ part of $z_0$ is precisely the first collision moment in Eq.~\eqref{eq:nm-collision-moments}. It survives at zero output frequency and sets the size of the stored current response. This finite-energy moment equation retains the spectral energy-stress derivative before integration. The zero-gradient output stress is symmetric in $i,j$; in an isotropic inversion-symmetric system its optical tensors are proportional to $\delta_{ij}(\mathbf E_+\cdot\mathbf E_-)$ and $E_{+,i}E_{-,j}+E_{-,i}E_{+,j}$. Therefore its contraction with $Q_j$ has no helicity-odd magnetization projection. Only the output stress at $Q=0$ is needed in this already first-gradient term. The helicity-even output stress can contribute at order $Q$, so its absence from the magnetization projection follows from symmetry. The superscript $H$ selects the part odd under reversal of circular polarization; it is this part that defines the orbital IFE.

\subsection{The three input harmonics and the spectral cancellations}
The angular moments are evaluated using
\begin{equation}
\langle v_iv_j\rangle_\phi=\frac{v^2}{2}\delta_{ij},\qquad
\langle v_iv_jv_kv_l\rangle_\phi
=\frac{v^4}{8}(\delta_{ij}\delta_{kl}+\delta_{ik}\delta_{jl}+\delta_{il}\delta_{jk}).
\label{eq:nm-angular-moments}
\end{equation}

\subsubsection{Quadrupole: a nonzero spectral current whose integral vanishes}
For the $\ell=2$ input, $N_{s,2}^{(1)}=0$ and $\boldsymbol{\mathcal P}_{s,2}^{(1)}=0$, but
\begin{equation}
\Pi_{s,2,ij}^{(1)}(\epsilon)
=\frac{\mathrm{i}  eK^2f_0'(\epsilon)}{2\pi m z_s^2}
[q_{s,i}E_{s,j}+q_{s,j}E_{s,i}-\delta_{ij}(\mathbf q_s\cdot\mathbf E_s)].
\label{eq:nm-quadrupole-stress}
\end{equation}
Its helicity current is consequently
\begin{equation}
j_{2,i}^{(2),H}(\epsilon)
=-\frac{e^2}{z_0}\sum_s
E_{s,j}\partial_\epsilon\Pi_{\bar s,2,ij}^{(1)}(\epsilon)\big|_H.
\label{eq:nm-quadrupole-current}
\end{equation}
This spectral current is generally nonzero. Only after taking its energy integral does the localized stress boundary term vanish:
\begin{equation}
\int\mathrm{d} \epsilon\,j_{2,i}^{(2),H}(\epsilon)=0,\qquad
\int\mathrm{d} \epsilon\,\partial_\epsilon[K^2f_0'(\epsilon)]
=-2\varepsilon_F+2\varepsilon_F=0.
\label{eq:nm-quadrupole-cancellation}
\end{equation}

\subsubsection{Density harmonic: retain both the density and energy-stress terms}
For the $\ell=0$ input,
\begin{align}
N_{s,0}^{(1)}(\epsilon)
&=-\frac{eKf_0'(\epsilon)}{\pi\omega_sz_s}(\mathbf q_s\cdot\mathbf E_s),\nonumber\\
\Pi_{s,0,ij}^{(1)}(\epsilon)
&=-\frac{eK^2f_0'(\epsilon)}{\pi m\omega_sz_s}
(\mathbf q_s\cdot\mathbf E_s)\delta_{ij},\qquad
\boldsymbol{\mathcal P}_{s,0}^{(1)}=0.
\label{eq:nm-density-spectral-moments}
\end{align}
Substitution into Eq.~\eqref{eq:nm-momentum-balance} gives
\begin{align}
j_{E,i}^{(2),H}(\epsilon)
&=\frac{e^3}{\pi m z_0}\sum_s
\frac{E_{s,i}(\mathbf q_{\bar s}\cdot\mathbf E_{\bar s})}
{\omega_{\bar s}z_{\bar s}}\,
[Kf_0'(\epsilon)+K^2f_0''(\epsilon)]\big|_H.
\label{eq:nm-density-spectral-current}
\end{align}
The combination follows explicitly from $-Kf_0'+\partial_\epsilon(K^2f_0')=Kf_0'+K^2f_0''$. At zero temperature,
\begin{equation}
f_0'=-\delta(\epsilon),\qquad f_0''=-\delta'(\epsilon),\qquad
\int\mathrm{d} \epsilon\,[Kf_0'+K^2f_0'']
=-\varepsilon_F+2\varepsilon_F=\varepsilon_F.
\label{eq:nm-density-energy-integral}
\end{equation}
The energy moment follows from $\int\mathrm{d} \epsilon\,g(\epsilon)\delta'(\epsilon)=-g'(0)$ with the energy reference kept fixed. Taking $\Omega\to0$ after this evaluation, define $z_+=\Gamma-\mathrm{i} \omega$ and $z_-=\Gamma+\mathrm{i} \omega$. Before projecting helicity, the integrated electric density current is
\begin{equation}
\mathbf j_E^{(2)}=\frac{e^3\varepsilon_F}{2\pi m\Gamma\omega}
\left[\frac{\mathbf E_-(\mathbf Q\cdot\mathbf E_+)}{z_+}
-\frac{\mathbf E_+(\mathbf Q\cdot\mathbf E_-)}{z_-}\right].
\label{eq:nm-j-electric}
\end{equation}
In the dc routing, charge continuity gives
$n_-^{(1)}=-\sigma_0(-\omega)(\mathbf Q\cdot\mathbf E_-)/(2e\omega)$.
After imposing the physical amplitudes this is
$n_-^{(1)}=-\sigma_0^*(\mathbf Q\cdot\mathbf E^*)/(2e\omega)$;
the minus sign from the negative optical frequency is essential for the ordering sum.

\subsubsection{Drift harmonic: the magnetic contribution}
For the $\ell=1$ input,
\begin{equation}
N_{s,1}^{(1)}=0,\qquad \Pi_{s,1,ij}^{(1)}=0,\qquad
\boldsymbol{\mathcal P}_{s,1}^{(1)}(\epsilon)
=-\frac{eKf_0'(\epsilon)}{\pi z_s}\mathbf E_s.
\label{eq:nm-drift-spectral-moments}
\end{equation}
Only the magnetic force remains in its helicity current:
\begin{equation}
\mathbf j_B^{(2),H}(\epsilon)
=-\frac{e^3Kf_0'(\epsilon)}{\pi m z_0}
\sum_s\frac{\mathbf E_{\bar s}\times(\mathbf q_s\times\mathbf E_s)}
{\omega_s z_{\bar s}}\bigg|_H.
\label{eq:nm-magnetic-spectral-current}
\end{equation}
Using $\int\mathrm{d} \epsilon\,Kf_0'=-\varepsilon_F$ and then taking $\Omega\to0$ gives
\begin{align}
\mathbf j_B^{(2)}
&=\frac{e^3\varepsilon_F}{2\pi m\Gamma\omega}
\left[\frac{\mathbf E_-\times(\mathbf Q\times\mathbf E_+)}{z_-}
-\frac{\mathbf E_+\times(\mathbf Q\times\mathbf E_-)}{z_+}\right]\nonumber\\
&=\mathbf j_E^{(2)}
+\frac{e^3\varepsilon_F}{2\pi m\Gamma\omega}
\left(\frac1{z_-}-\frac1{z_+}\right)
\mathbf Q(\mathbf E_+\cdot\mathbf E_-).
\label{eq:nm-j-magnetic}
\end{align}
The second line follows from
$\mathbf a\times(\mathbf b\times\mathbf d)
=\mathbf b(\mathbf a\cdot\mathbf d)-\mathbf d(\mathbf a\cdot\mathbf b)$.
The additional current is longitudinal and helicity even. It therefore has zero projection onto the orbital magnetization coefficient. The equal helicity parts result from the separately evaluated spectral electric and magnetic sources.

\subsection{The energy-integrated helicity current and magnetization}
Integrating Eq.~\eqref{eq:nm-momentum-balance} only after evaluating its spectral terms, and multiplying by $e$, gives
\begin{equation}
\boxed{\Gamma j_i^{(2),H}
=\frac{e^2}{m}\sum_s[n_{\bar s}^{(1)}E_{s,i}]_H
+\frac{e}{mc_0}\sum_s[(\mathbf j_{\bar s}^{(1)}\times\mathbf B_s)_i]_H.}
\label{eq:nm-current-balance}
\end{equation}
Here $n_s^{(1)}=\int\mathrm{d} \epsilon\,N_s^{(1)}(\epsilon)$. The vanished energy-stress boundary term has been established explicitly above; this integrated identity is an outcome of the spectral calculation.

Now impose $\mathbf E_+=\mathbf E$ and $\mathbf E_-=\mathbf E^*$ and use
\begin{equation}
[E_iE_j^*]_H=-\frac{\mathrm{i} }{2}\varepsilon^{(2)}_{ij}\mathcal{H}_z,\qquad
[E_i^*E_j]_H=\frac{\mathrm{i} }{2}\varepsilon^{(2)}_{ij}\mathcal{H}_z,\qquad
\mathcal{H}_z=\mathrm{i} (\mathbf E\times\mathbf E^*)_z.
\label{eq:nm-polarization}
\end{equation}
Since $z_+^{-1}+z_-^{-1}=2\Gamma/(\Gamma^2+\omega^2)$, both force contributions give
\begin{equation}
\boxed{j_{E,i}^{(2),H}=j_{B,i}^{(2),H}
=\mathrm{i} \varepsilon^{(2)}_{i\ell}Q_\ell
\frac{e^3\varepsilon_F}{2\pi m\omega(\Gamma^2+\omega^2)}\mathcal{H}_z
=\mathrm{i} \varepsilon^{(2)}_{i\ell}Q_\ell
\frac{e\operatorname{Re} \sigma_0}{2m\Gamma\omega}\mathcal{H}_z.}
\label{eq:nm-equal-currents}
\end{equation}
Comparing with Eq.~\eqref{eq:nm-curl} before $Q\to0$ gives
\begin{equation}
\boxed{M_{E,z}^{(0)}=M_{B,z}^{(0)}
=\frac{e^3p_F^2}{4\pi m^2c_0\omega(\Gamma^2+\omega^2)}\mathcal{H}_z,\qquad
M_{{\rm orb},z}^{(0)}
=\frac{n_0e^3}{m^2c_0\omega(\Gamma^2+\omega^2)}\mathcal{H}_z
=\frac{e\operatorname{Re} \sigma_0}{mc_0\Gamma\omega}\mathcal{H}_z.}
\label{eq:nm-final}
\end{equation}
The electric and magnetic force contributions give equal helicity magnetizations. The factor of two in their sum comes from these two force contributions. The spin degeneracy was already included in $n_0$, and both optical orderings were already included in the sums; neither supplies an additional factor here. Since $e<0$, the odd power carries the negative electronic charge.

For comparison, an energy-integrated occupation can now be defined by
\begin{equation}
f(\mathbf{R} ,\mathbf{p} ,T)=\int\mathrm{d} \epsilon\,w(\mathbf{R} ,\mathbf{p} ,\epsilon,T).
\label{eq:nm-occupation-projection}
\end{equation}
The spectral-delta derivatives cancel only after both parts of $\mathscr D_i$ are included, as shown in Eq.~\eqref{eq:nm-delta-cancel}. Integrating that identity over energy now recovers $\partial_{p_i}f_0(\xi_p)=v_if_0'(\xi_p)$, the familiar occupation source. Appendix~\ref{app:scalar} records additional electric and magnetic moment checks before the energy integral. The later Rashba calculation retains this phase-space structure and impurity gain, with velocity $\mathbf{p} /m+\alpha_{\mathrm{so}}\hat{\boldsymbol{\mathsf R}}$, and derives the additional spin precession, matrix collision, and spectral reconstruction explicitly. The occupation $f$ is not used as the starting distribution or as a replacement for $w_s^{(1)}$ in the nonlinear derivation. In the optical-orbit interpretation below, $t$ denotes the same center time as $T$.

\subsection{Loss-only density relaxation and the low-frequency limit}
The role of scattering-in can be isolated without changing the electric phase-space derivatives. In the SOC-free model, replace the conserving scalar operator by $\mathcal C_W^{\mathrm{loss}}[w]=-\Gamma w$. The spatially uniform drift still gives the Drude current, but the first-order density equation becomes
\begin{equation}
 (\Gamma-i\omega)\delta n^{(1)}
 +\frac{i}{e}\mathbf q\cdot\mathbf j^{(1)}=0
 \label{eq:loss-only-density}
\end{equation}
instead of the conserving equation with $-i\omega$. Evaluating the same electric density force with this spuriously relaxing density gives
\begin{equation}
 M_{E,z}^{\mathrm{loss}}=
 \frac{n_0e^3\omega}{m^2c_0(\Gamma^2+\omega^2)^2}\mathcal{H}
 =\frac{n_0e^3\mathcal{H}}{m^2c_0\Gamma^4}\omega+O(\omega^3).
 \label{eq:loss-only-electric}
\end{equation}
Thus the loss-only electric contribution is regular as $\omega\to0$, whereas the conserving calculation gives $M_{E,z}=M_{\mathrm{orb},z}^{(0)}/2\propto\omega^{-1}$. The optical magnetic force must be considered separately. Its SOC-free contribution is unchanged because the uniform first angular harmonic already relaxes at $\Gamma$:
\begin{equation}
 M_{B,z}^{\mathrm{loss}}=
 \frac{n_0e^3}{2m^2c_0\omega(\Gamma^2+\omega^2)}\mathcal{H}.
 \label{eq:loss-only-magnetic}
\end{equation}
When both forces are retained, their loss-only sum is therefore
\begin{equation}
 M_{E,z}^{\mathrm{loss}}+M_{B,z}^{\mathrm{loss}}=
 \frac{n_0e^3(\Gamma^2+3\omega^2)}
 {2m^2c_0\omega(\Gamma^2+\omega^2)^2}\mathcal{H},
 \label{eq:loss-only-total}
\end{equation}
which still has a $1/\omega$ low-frequency term. Missing gain regularizes the electric contribution; it does not, by itself, regularize the total electric-plus-magnetic response. This diagnostic isolates scattering-in while retaining the complete scalar phase-space force. It is not a separate test of truncating momentum or energy derivatives.

\subsection{Optical torque and the angular momentum stored in the Drude orbit}
With $\mathbf E_{\rm phys}=\mathbf E e^{-\mathrm{i} \omega t}+\mathbf E^*e^{\mathrm{i} \omega t}$ and $\mathbf j_+=\sigma_0\mathbf E$, the cycle-averaged absorbed power is
\begin{equation}
\mathcal P_{\rm abs}=\overline{\mathbf j\cdot\mathbf E_{\rm phys}}
=2\operatorname{Re} \sigma_0|\mathbf E|^2.
\label{eq:nm-power}
\end{equation}
The dipole density $\mathbf d$ obeys $\dot{\mathbf d}=\mathbf j$, so $\mathbf d_+=\mathrm{i} \sigma_0\mathbf E/\omega$. Its electric torque per area is
\begin{align}
\mathcal T_z^{\rm opt}
&=\overline{(\mathbf d\times\mathbf E_{\rm phys})_z}\nonumber\\
&=\left[\frac{\mathrm{i} \sigma_0}{\omega}\mathbf E\times\mathbf E^*
-\frac{\mathrm{i} \sigma_0^*}{\omega}\mathbf E^*\times\mathbf E\right]_z
=\frac{2\operatorname{Re} \sigma_0}{\omega}\mathcal{H}_z.
\label{eq:nm-torque}
\end{align}
Writing $\eta_{\rm pol}=\mathcal{H}_z/|\mathbf E|^2$, this is
$\mathcal T_z^{\rm opt}=\eta_{\rm pol}\mathcal P_{\rm abs}/\omega$.
For circular polarization $\eta_{\rm pol}=\pm1$; for linear polarization it is zero. Torque has the units of angular momentum per area per time.

Let $\boldsymbol\xi$ be displacement about the orbit center and $\mathbf u$ the coherent optical velocity. The Drude equation gives
\begin{equation}
\mathbf u_+=\frac{e\mathbf E}{m(\Gamma-\mathrm{i} \omega)},\qquad
\boldsymbol\xi_+=\frac{\mathbf u_+}{-\mathrm{i} \omega}.
\label{eq:nm-orbit}
\end{equation}
The local orbital angular momentum per area is
\begin{align}
L_z&=n_0m\,\overline{(\boldsymbol\xi\times\mathbf u)_z}\nonumber\\
&=\frac{2n_0e^2}{m\omega(\Gamma^2+\omega^2)}\mathcal{H}_z
=\frac{\mathcal T_z^{\rm opt}}{\Gamma}.
\label{eq:nm-stored-L}
\end{align}
Equivalently, crossing $m\dot{\mathbf u}=e\mathbf E_{\rm phys}-m\Gamma\mathbf u$ with $\boldsymbol\xi$ gives
\begin{equation}
\dot L_z=\mathcal T_z^{\rm opt}-\Gamma L_z,
\qquad
M_{{\rm orb},z}^{(0)}=\frac{e}{2mc_0}L_z
=\frac{e}{2mc_0}\frac{\mathcal T_z^{\rm opt}}{\Gamma}.
\label{eq:nm-gyro}
\end{equation}
The gyromagnetic factor $e/(2mc_0)$ converts orbital angular momentum into magnetic moment, $\mathcal T_z^{\rm opt}$ is its injection rate, and $\Gamma^{-1}$ is its survival time in this Drude balance. This reproduces Eq.~\eqref{eq:nm-final} independently. The equal electric and magnetic pieces above describe the decomposition of the slowly varying magnetization current; the dipole torque describes the local circulation stored in the same response.

At fixed nonzero $\omega$, the limit $\Gamma\to0$ leaves $M_{{\rm orb},z}^{(0)}=n_0e^3\mathcal{H}_z/(m^2c_0\omega^3)$. In that limit $\mathcal T_z^{\rm opt}\propto\Gamma$ and $\Gamma^{-1}$ compensates it. The periodic optical orbit, rather than a finite dissipative rate, determines the limiting circulation.

\section{Gauge-covariant phase space and a conserving collision operator}
\label{sec:phase-space-framework}
\subsection{Keldysh distribution and physical density}
\label{sec:keldysh}

We now introduce the Keldysh component $\hat G^K$ and the distribution kernel $\hat h$. A check denotes a matrix in Keldysh space whose entries are spin matrices. The disorder-averaged Green function has the triangular representation
\begin{equation}
 \check G=\begin{pmatrix}\hat G^R&\hat G^K\\0&\hat G^A\end{pmatrix},
 \qquad
 \hat G^K=\hat G^R\circ\hat h-\hat h\circ\hat G^A .
 \label{eq:param}
\end{equation}
The symbol $\circ$ denotes a space-time convolution, or its corresponding Wigner product. The distribution kernel is a general spin matrix,
\begin{equation}
 \hat h=h_0\mathbf{1} +\mathbf h\cdot\boldsymbol{\sigma}  .
 \label{eq:spin-decomp}
\end{equation}
All three spin components are retained to describe spin precession in the momentum-dependent Rashba field.

We use the lesser-function convention $G^<_{\alpha\beta}(1,2)=\mathrm{i} \langle\psi^\dagger_\beta(2)\psi_\alpha(1)\rangle$. The identities
\begin{equation}
 \hat G^K=\hat G^>+\hat G^<,\qquad
 \hat G^R-\hat G^A=\hat G^>-\hat G^<
\end{equation}
give
\begin{equation}
 \hat G^<=\frac12\left[\hat G^K-\hat G^R+\hat G^A\right],\qquad
 \hat{\mathcal A}=\mathrm{i} (\hat G^R-\hat G^A),\qquad
 \hat w_{\mathbf{p} \epsilon}=\frac{\hat G^<_{\mathbf{p} \epsilon}}{2\pi\mathrm{i} }.
 \label{eq:lesser}
\end{equation}
Thus $\hat w$ is an energy-resolved density, whereas $\hat h$ is a distribution kernel. In the present convention, the momentum density matrix is $\hat\rho_{\mathbf{p} }=\int\mathrm{d} \epsilon\,\hat w_{\mathbf{p} \epsilon}$; there is no additional factor of $1/(2\pi)$ in this energy integral.

In stationary equilibrium the fluctuation--dissipation relation reads
\begin{equation}
 \hat h^{(0)}=F(\epsilon)\mathbf{1} ,\qquad
 \hat G_0^K=F(\epsilon)(\hat G_0^R-\hat G_0^A),\qquad
 \hat G_0^<=\mathrm{i}  f_0(\epsilon)\hat{\mathcal A}_0,
 \qquad \hat w^{(0)}=\frac{f_0(\epsilon)}{2\pi}\hat{\mathcal A}_0 .
 \label{eq:equilibrium}
\end{equation}
Define $\mathbf n=\mathbf{p} /p$, $\boldsymbol\eta_{\phi}=\mathbf n\times\mathbf{c} $, and the helicity projectors
\begin{equation}
 \hat P_\lambda(\phi)=\frac12(\mathbf{1} +\lambda\boldsymbol\eta_{\phi}\cdot\boldsymbol{\sigma} ),\qquad
 \xi_{p\lambda}=\xi_p+\lambda\alpha_{\mathrm{so}} p,\qquad \lambda=\pm1 .
 \label{eq:projectors}
\end{equation}
They satisfy $P_\lambda P_{\lambda'}=\delta_{\lambda\lambda'}P_\lambda$ and $P_++P_-=\mathbf{1} $. In the clean limit,
\begin{equation}
 \hat{\mathcal A}_0(\mathbf{p} ,\epsilon)=2\pi\sum_\lambda
 \hat P_\lambda(\phi)\delta(\epsilon-\xi_{p\lambda}),\qquad
 \hat w^{(0)}_{\mathbf{p} \epsilon}=f_0(\epsilon)\sum_\lambda
 \hat P_\lambda(\phi)\delta(\epsilon-\xi_{p\lambda}).
 \label{eq:clean-density}
\end{equation}
SOC therefore changes the spectral density even though the equilibrium distribution kernel remains spin scalar.

\subsection{Gauge-covariant Wigner representation}
\label{sec:wigner}

With the center time $T$ of Eq.~\eqref{eq:nm-center-relative} now denoted by $t$, set $X=(t,\mathbf{R} )$ and $s=(s_t,\mathbf s)$. The matrix gauge-covariant Wigner transform is
\begin{equation}
 \widetilde G(X;\mathbf{p} ,\epsilon)=\int\mathrm{d}  s_t\,\mathrm{d} ^2s\,
 e^{\mathrm{i} \epsilon s_t-\mathrm{i} \mathbf{p} \cdot\mathbf s}
 U(X,X+s/2)G(X+s/2,X-s/2)U(X-s/2,X),
 \label{eq:wigner}
\end{equation}
where the straight-line phase factor is
\begin{equation}
 U(a,b)=\exp\!\left[\mathrm{i}  e\int_b^a
 \left(\frac{\mathbf A\cdot\mathrm{d} \mathbf r}{c_0}-\varphi\,\mathrm{d}  t\right)\right].
 \label{eq:wilson}
\end{equation}
The endpoint gauge phases cancel in Eq.~\eqref{eq:wigner}, and $\mathbf{p} $ is the kinetic momentum. Tildes will be suppressed below. Gauge-covariant phase-space methods organize electric and magnetic force terms without introducing gauge-dependent distributions~\cite{Kita:2001}.

For a convolution $C(1,2)=\int\mathrm{d} 3\,A(1,3)B(3,2)$, the two endpoint phase factors of each kernel combine into the phase around the triangle formed by $1,2,3$. Stokes' theorem converts this phase to the electromagnetic flux through the triangle. Expanding the fields about its center gives the field-strength contribution to the product, while expanding the center coordinates of $A$ and $B$ gives the ordinary Wigner gradients. Relative-coordinate factors are converted to derivatives by
\begin{equation}
 s_i e^{\mathrm{i} \epsilon s_t-\mathrm{i} \mathbf{p} \cdot\mathbf s}
 =\mathrm{i} \partial_{p_i}e^{\mathrm{i} \epsilon s_t-\mathrm{i} \mathbf{p} \cdot\mathbf s},\qquad
 s_t e^{\mathrm{i} \epsilon s_t-\mathrm{i} \mathbf{p} \cdot\mathbf s}
 =-\mathrm{i} \partial_\epsilon e^{\mathrm{i} \epsilon s_t-\mathrm{i} \mathbf{p} \cdot\mathbf s}.
 \label{eq:relative-derivatives}
\end{equation}
The resulting phase-space brackets obey
\begin{equation}
 [R_i,p_j]_\circ=\mathrm{i} \delta_{ij},\quad
 [t,\epsilon]_\circ=-\mathrm{i} ,\quad
 [p_i,p_j]_\circ=\frac{\mathrm{i}  e}{c_0}\varepsilon_{ijk}B_k,\quad
 [\epsilon,p_i]_\circ=\mathrm{i}  eE_i,
 \label{eq:covariant-brackets}
\end{equation}
which also fix the electric and magnetic signs below. Here the brackets denote the star commutator of the indicated scalar phase-space coordinates.

The field-free Wigner product is
\begin{equation}
 A\circ_0 B=A\exp\!\left\{\frac{\mathrm{i} }{2}\left[
 \overleftarrow\nabla_R\cdot\overrightarrow\nabla_p
 -\overleftarrow\nabla_p\cdot\overrightarrow\nabla_R
 -\overleftarrow\partial_t\overrightarrow\partial_\epsilon
 +\overleftarrow\partial_\epsilon\overrightarrow\partial_t\right]\right\}B .
 \label{eq:star0}
\end{equation}
If $A_a\propto e^{\mathrm{i} \mathbf q_a\cdot\mathbf{R} -\mathrm{i} \omega_a t}$ and $B_b\propto e^{\mathrm{i} \mathbf q_b\cdot\mathbf{R} -\mathrm{i} \omega_b t}$, its exact harmonic rule is
\begin{equation}
 (A_a\circ_0 B_b)_{a+b}(\mathbf{p} ,\epsilon)=
 A_a\!\left(\mathbf{p} +\frac{\mathbf q_b}{2},\epsilon+\frac{\omega_b}{2}\right)
 B_b\!\left(\mathbf{p} -\frac{\mathbf q_a}{2},\epsilon-\frac{\omega_a}{2}\right).
 \label{eq:shifts}
\end{equation}
Matrix order is unchanged. The half-frequency shifts follow from the Wigner transformation and should be retained before a frequency expansion is chosen.

To first order in the gauge-covariant gradient expansion, the product is
\begin{align}
 A\circ B={}&AB+\frac{\mathrm{i} }{2}\bigl[
 (\partial_{R_i}A)(\partial_{p_i}B)-(\partial_{p_i}A)(\partial_{R_i}B)
 -(\partial_tA)(\partial_\epsilon B)+(\partial_\epsilon A)(\partial_tB)\bigr]
 \nonumber\\
 &+\frac{\mathrm{i}  e}{2}\left[
 E_i\bigl((\partial_\epsilon A)(\partial_{p_i}B)-(\partial_{p_i}A)(\partial_\epsilon B)\bigr)
 +\frac{1}{c_0}\varepsilon_{ijk}B_k(\partial_{p_i}A)(\partial_{p_j}B)\right]
 +\cdots .
 \label{eq:covariant-product}
\end{align}
Repeated Cartesian indices are summed. We retain the displayed first gradients and the local electric and magnetic forces. Derivatives act on the complete matrix products before any radial expansion or energy integration.

\subsection{Self-consistent Born collision integral}
\label{sec:collision}

The disorder average is specified by
\begin{equation}
 \langle U(\mathbf r)\rangle=0,\qquad
 \langle U(\mathbf r)U(\mathbf r')\rangle
 =\gamma_{\rm imp}\delta^{(2)}(\mathbf r-\mathbf r'),\qquad
 \gamma_{\rm imp}=\frac{1}{2\pi\nu_0\tau},\qquad
 \nu_0=\frac{m}{2\pi}.
 \label{eq:disorder}
\end{equation}
Here $\nu_0$ is the density of states per spin of the spin-degenerate parabolic band. In the self-consistent Born approximation (SCBA),
\begin{equation}
 \hat\Sigma^j(X,\epsilon)=\gamma_{\rm imp}\int\!\frac{\mathrm{d}^2k}{(2\pi)^2} \hat G^j(X;\mathbf{k} ,\epsilon),
 \qquad j=R,A,K,
 \label{eq:scba}
\end{equation}
The self-energy is independent of momentum.

In the stationary response with adiabatic switching, the Dyson equations give $\hat G^K=\hat G^R\circ\hat\Sigma^K\circ\hat G^A$. Combining this identity with Eq.~\eqref{eq:param} and multiplying by the inverse propagators yields
\begin{equation}
 \hat h\circ(\hat G^A)^{-1}-(\hat G^R)^{-1}\circ\hat h=\hat\Sigma^K.
 \label{eq:dyson-h}
\end{equation}
Initial-state boundary terms are encoded by the equilibrium state and the switching prescription. Let $\mathcal G^{R,A}$ denote the disorder-free propagators. With $(G^{R,A})^{-1}=(\mathcal G^{R,A})^{-1}-\Sigma^{R,A}$, Eq.~\eqref{eq:dyson-h} becomes
\begin{equation}
 \mathrm{i} \bigl[\hat h\circ(\hat{\mathcal G}^A)^{-1}-(\hat{\mathcal G}^R)^{-1}\circ\hat h\bigr]
 =\hat I_{\mathrm{SCBA} }[\hat h],\qquad
 \hat I_{\mathrm{SCBA} }[\hat h]=\mathrm{i} \hat\Sigma^K
 -\mathrm{i} \bigl(\hat\Sigma^R\circ\hat h-\hat h\circ\hat\Sigma^A\bigr).
 \label{eq:collision-exact}
\end{equation}
This expression contains both scattering into a state and scattering out of it. In particular,
\begin{equation}
 \hat\Sigma^K=\gamma_{\rm imp}\int\!\frac{\mathrm{d}^2k}{(2\pi)^2} 
 \bigl(\hat G^R_{\mathbf{k} }\circ\hat h_{\mathbf{k} }
 -\hat h_{\mathbf{k} }\circ\hat G^A_{\mathbf{k} }\bigr).
 \label{eq:gain-full}
\end{equation}
Dropping this term gives a relaxation-time loss term rather than the full SCBA collision operator. Keeping it generates the impurity vertex corrections that couple charge and spin~\cite{MSH:2004}.

\subsubsection{Equilibrium self-energy and lifetime normalization}

Within the metallic approximation, the real scalar self-energy is absorbed into $\varepsilon_F$, and
\begin{equation}
 \hat G_0^{R,A}(\mathbf{p} ,\epsilon)=\sum_\lambda\hat P_\lambda(\phi)g_\lambda^{R,A}(p,\epsilon),\qquad
 g_\lambda^{R,A}=\frac{1}{\epsilon-\xi_{p\lambda}\pm\mathrm{i} /(2\tau)}.
 \label{eq:propagators}
\end{equation}
The exact Fermi momenta and branch densities of states for the two occupied branches are
\begin{equation}
 p_\lambda=p_0-\lambda m\alpha_{\mathrm{so}},\qquad p_0=\sqrt{p_F^2+m^2\alpha_{\mathrm{so}}^2},\qquad
 v_\lambda=\frac{p_0}{m},\qquad
 \nu_\lambda=\frac{p_\lambda}{2\pi v_\lambda}
 =\nu_0\left(1-\lambda\frac{m\alpha_{\mathrm{so}}}{p_0}\right),\qquad
 \sum_\lambda\nu_\lambda=2\nu_0.
 \label{eq:dos}
\end{equation}
Near each contour, $\int\!\frac{\mathrm{d}^2p}{(2\pi)^2} \simeq\nu_\lambda\int\mathrm{d} \xi_{p\lambda}\int\mathrm{d} \phi/(2\pi)$. Since $\left\langle P_\lambda\right\rangle_{\phi}=\mathbf{1} /2$ and $\int\mathrm{d} \xi\,g^{R,A}=\mp\mathrm{i} \pi$ in the wide-band radial approximation,
\begin{equation}
 \hat\Sigma_0^{R,A}=\gamma_{\rm imp}\sum_\lambda
 \nu_\lambda\left\langle \hat P_\lambda\right\rangle_{\phi}(\mp\mathrm{i} \pi)
 =\mp\frac{\mathrm{i} }{2\tau}\mathbf{1}  .
 \label{eq:sigma0}
\end{equation}
The one-particle width is $1/(2\tau)$; a retarded--advanced pair has width $1/\tau$. Equilibrium satisfies $\Sigma_0^K=F(\Sigma_0^R-\Sigma_0^A)$, so Eq.~\eqref{eq:collision-exact} vanishes for $\hat h^{(0)}=F\mathbf{1} $.

\subsection{Transport equation and its linearization}
\label{sec:transport}

Applying Eq.~\eqref{eq:covariant-product} to the left-hand side of Eq.~\eqref{eq:collision-exact} gives
\begin{align}
 \partial_t\hat h+\frac12\{\hat v_i,\partial_{R_i}\hat h\}
 +\mathrm{i} [\mathbf b_p\cdot\boldsymbol{\sigma} ,\hat h]
 &+e E_i\left(\partial_{p_i}\hat h+\frac12\{\hat v_i,\partial_\epsilon\hat h\}\right)
 +\frac{e}{2c_0}\{(\hat{\mathbf v}\times\mathbf{B} )_i,\partial_{p_i}\hat h\}
 =\hat I_{\mathrm{SCBA} }[\hat h].
 \label{eq:transport}
\end{align}
The anticommutators retain the spin-dependent velocity. The electric term contains an energy derivative because $\hat h$ is energy resolved. Its equilibrium source is
\begin{equation}
 \mathcal D_a\hat h^{(0)}=e\mathbf{E} _a\cdot\hat{\mathbf v}\,F'(\epsilon),
 \qquad \partial_{p_i}\hat h^{(0)}=0,
 \label{eq:linear-drive}
\end{equation}
where $\mathcal D_a$ denotes the field-dependent part of the transport operator. The magnetic force has no direct first-order action on $\hat h^{(0)}$. A direct Zeeman coupling is not included in Eq.~\eqref{eq:model}.

We use the optical convention
\begin{equation}
 \mathbf{E} (\mathbf{R} ,t)=\sum_{a=\pm}\mathbf{E} _a e^{\mathrm{i} \mathbf q_a\cdot\mathbf{R} -\mathrm{i} \omega_a t},\qquad
 \omega_\pm=\pm\omega+\frac{\Omega}{2},\qquad
 \mathbf q_\pm=\pm\mathbf k+\frac{\mathbf Q}{2}.
 \label{eq:optical}
\end{equation}
The mixed second-order response has output $(\mathbf Q,\Omega)$. For a real monochromatic field at zero output frequency, $\mathbf{E} _- =\mathbf{E} _+^*$. There is no factor of $1/2$ in Eq.~\eqref{eq:optical}. Write $\hat h=\hat h^{(0)}+\hat h^{(1)}+\hat h^{(2)}+\cdots$, with no factorials in the field expansion.

\subsubsection{Cancellation of the first-order spectral self-energies}

Let $\Phi_a[A,B]$ be the term linear in field $a$ in the covariant product, at the order displayed in Eq.~\eqref{eq:covariant-product}. Then
\begin{equation}
 (A\circ B)^{(1)}_a=A_0\circ_0 B_a^{(1)}+A_a^{(1)}\circ_0 B_0+\Phi_a[A_0,B_0].
 \label{eq:product-linear}
\end{equation}
Before any cancellation, the first-order collision integral is
\begin{align}
 \hat I_a^{(1)}={}&\mathrm{i} \hat\Sigma_a^{K,(1)}
 -\mathrm{i} \bigl(\hat\Sigma_0^R\circ_0\hat h_a^{(1)}-\hat h_a^{(1)}\circ_0\hat\Sigma_0^A\bigr)
 -\mathrm{i} \bigl(\hat\Sigma_a^{R,(1)}\circ_0\hat h^{(0)}-\hat h^{(0)}\circ_0\hat\Sigma_a^{A,(1)}\bigr)
 \nonumber\\
 &-\mathrm{i} \bigl(\Phi_a[\hat\Sigma_0^R,\hat h^{(0)}]-\Phi_a[\hat h^{(0)},\hat\Sigma_0^A]\bigr),
 \label{eq:collision-linear-full}\\
 \hat\Sigma_a^{K,(1)}={}&\gamma_{\rm imp}\int\!\frac{\mathrm{d}^2k}{(2\pi)^2} \bigl[
 \hat G_0^R\circ_0\hat h_a^{(1)}-\hat h_a^{(1)}\circ_0\hat G_0^A
 +\hat G_a^{R,(1)}\circ_0\hat h^{(0)}-\hat h^{(0)}\circ_0\hat G_a^{A,(1)}
 \nonumber\\
 &\hspace{22mm}+\Phi_a[\hat G_0^R,\hat h^{(0)}]-\Phi_a[\hat h^{(0)},\hat G_0^A]\bigr].
 \label{eq:sigmak-linear}
\end{align}
The terms involving $G_a^{R,A,(1)}$ cancel the corresponding self-energy terms in Eq.~\eqref{eq:collision-linear-full} by Eq.~\eqref{eq:scba}; this does not require setting the spectral corrections to zero. Moreover,
\begin{equation}
 \Phi_a[G_0^R,h^{(0)}]-\Phi_a[h^{(0)},G_0^A]
 =-\frac{\mathrm{i}  e}{2}E_{a,i}F'(\epsilon)
 \partial_{k_i}(G_0^R+G_0^A).
 \label{eq:total-derivative}
\end{equation}
Its full momentum integral vanishes when the boundary term vanishes. The analogous terms with $\Sigma_0^{R,A}$ vanish because both $\Sigma_0$ and $h^{(0)}$ are momentum independent. These cancellations must be performed before a Fermi-surface projection that could obscure the total derivative.

Using Eq.~\eqref{eq:sigma0}, the linear collision operator therefore reduces to
\begin{equation}
 \hat I_a^{(1)}=-\frac{\hat h_a^{(1)}}{\tau}+\hat K_{g,a},\qquad
 \hat K_{g,a}=\mathrm{i} \gamma_{\rm imp}\int\!\frac{\mathrm{d}^2k}{(2\pi)^2} 
 \left[\hat G_0^R\!\left(\mathbf{k} +\frac{\mathbf q_a}{2},\epsilon+\frac{\omega_a}{2}\right)\hat h_{a,\mathbf{k} }^{(1)}
 -\hat h_{a,\mathbf{k} }^{(1)}\hat G_0^A\!\left(\mathbf{k} -\frac{\mathbf q_a}{2},\epsilon-\frac{\omega_a}{2}\right)\right].
 \label{eq:linear-gain}
\end{equation}
The first-order kinetic equation is
\begin{equation}
 \left(\tau^{-1}-\mathrm{i} \omega_a\right)\hat h_a^{(1)}
 +\frac{\mathrm{i}  q_{a,i}}{2}\{\hat v_i,\hat h_a^{(1)}\}
 +\mathrm{i} [\mathbf b_p\cdot\boldsymbol{\sigma} ,\hat h_a^{(1)}]
 =-e F'(\epsilon)\mathbf{E} _a\cdot\hat{\mathbf v}+\hat K_{g,a}.
 \label{eq:linear-kinetic}
\end{equation}
This is the local-force, leading-gradient equation. Retaining the shifts in its collision kernel resolves the impurity ladder at finite $\omega_a\tau$. The field derivative and impurity gain are kept together in solving for the optical response.

\section{Linear optical response and conductivity}
\label{sec:optical-transport}
\subsection{Homogeneous response and impurity ladder}
\label{sec:ladder}

For the electric-dipole spin response, set $\mathbf q_a=0$ and define
\begin{equation}
 \hat K_a(\mathbf{p} ,\epsilon)=-e F'(\epsilon)\mathbf{E} _a\cdot\hat{\mathbf v}(\mathbf{p} )+\hat K_{g,a}(\epsilon),\qquad
 z_a=\tau^{-1}-\mathrm{i} \omega_a.
 \label{eq:source}
\end{equation}
Equation~\eqref{eq:linear-kinetic} is equivalent to
\begin{equation}
 \mathrm{i} \left[\hat h_a^{(1)}(\hat G_{0,-}^A)^{-1}
 -(\hat G_{0,+}^R)^{-1}\hat h_a^{(1)}\right]=\hat K_a,
 \qquad
 \hat G_{0,\pm}^{R,A}=\hat G_0^{R,A}(\mathbf{p} ,\epsilon\pm\omega_a/2).
 \label{eq:inverse-identity}
\end{equation}
Multiplication by $G_{0,+}^R$ and $G_{0,-}^A$ gives the useful identity
\begin{equation}
 \mathrm{i} \bigl(\hat G_{0,+}^R\hat h_a^{(1)}-\hat h_a^{(1)}\hat G_{0,-}^A\bigr)
 =\hat G_{0,+}^R\hat K_a\hat G_{0,-}^A .
 \label{eq:ra-identity}
\end{equation}
The gain is consequently determined by the closed matrix equation
\begin{equation}
 \hat K_{g,a}=\mathscr T_a\!\left[-e F'(\epsilon)\mathbf{E} _a\cdot\hat{\mathbf v}+\hat K_{g,a}\right],\qquad
 \mathscr T_a[\hat X]=\gamma_{\rm imp}\int\!\frac{\mathrm{d}^2p}{(2\pi)^2} \hat G_{0,+}^R\hat X\hat G_{0,-}^A .
 \label{eq:ladder}
\end{equation}
This representation keeps the radial integral until after the matrix impurity vertex has been determined.

\subsubsection{Radial and angular integrals}

First evaluate the frozen-splitting part of the single-splitting metallic kernel: set $\Delta_p=\Delta_R$ in the propagator denominators and use the common radial measure. A superscript $(0)$ denotes this approximation, before the first radial-energy correction is included. Define
\begin{equation}
 a_a=1-\mathrm{i} \omega_a\tau,\qquad
 w_a=\frac{1}{a_a},\qquad
 u_a^{(0)}=\frac{a_a}{a_a^2+(\Delta_R\tau)^2},\qquad
 v_a^{(0)}=\frac{w_a+u_a^{(0)}}{2}.
 \label{eq:wuv}
\end{equation}
The scalar radial integrals, with the common measure $\nu_0\mathrm{d} \xi$, are
\begin{equation}
 Q_{\lambda\lambda'}=\frac{1}{2\pi\tau}\int_{-\infty}^{\infty}\mathrm{d} \xi\,
 \frac{1}{\epsilon+\omega_a/2-\xi-\lambda\Delta_R/2+\mathrm{i} /(2\tau)}
 \frac{1}{\epsilon-\omega_a/2-\xi-\lambda'\Delta_R/2-\mathrm{i} /(2\tau)}
 =\frac{1}{a_a+\mathrm{i} (\lambda-\lambda')\Delta_R\tau/2}.
 \label{eq:radial-leading}
\end{equation}
The change of variable $\tilde\xi=\xi-\epsilon$ removes the mean energy from this frozen-splitting integral. Hence $\partial_\epsilon u_a^{(0)}=\partial_\epsilon v_a^{(0)}=0$. Thus $Q_{\lambda\lambda}=w_a$ and $\tfrac12\sum_\lambda Q_{-\lambda,\lambda}=u_a^{(0)}$. For a momentum-independent spin matrix $\hat X=X_0\mathbf{1} +\mathbf X\cdot\boldsymbol{\sigma} $, the angular identities
\begin{equation}
 \left\langle \eta_i\right\rangle_{\phi}=0,\qquad
 \left\langle \eta_i\eta_j\right\rangle_{\phi}=\frac12(\delta_{ij}-c_i c_j),\qquad
 (\boldsymbol\eta\cdot\boldsymbol{\sigma} )(\mathbf X\cdot\boldsymbol{\sigma} )(\boldsymbol\eta\cdot\boldsymbol{\sigma} )
 =\bigl[2\boldsymbol\eta(\boldsymbol\eta\cdot\mathbf X)-\mathbf X\bigr]\cdot\boldsymbol{\sigma} 
 \label{eq:angular}
\end{equation}
yield
\begin{equation}
 \mathscr T_a[\hat X]=w_a X_0\mathbf{1} +v_a^{(0)}\mathbf X_{\parallel}\cdot\boldsymbol{\sigma} +u_a^{(0)} X_z\,\mathbf{c} \cdot\boldsymbol{\sigma} ,
 \qquad X_z=\mathbf{c} \cdot\mathbf X,\quad
 \mathbf X_{\parallel}=\mathbf X-X_z\mathbf{c}  .
 \label{eq:map}
\end{equation}
The scalar, in-plane-spin, and out-of-plane-spin channels therefore have different ladder eigenvalues.

The bare velocity contains both $p_i/m$ and the anomalous term proportional to $\alpha_{\mathrm{so}}$. Its angular projection is
\begin{equation}
 \left\langle \hat P_\lambda\hat v_i\hat P_{\lambda'}\right\rangle_{\phi}
 =\frac14\left[\alpha_{\mathrm{so}}+\frac{\lambda+\lambda'}{2}\frac{p}{m}\right](\mathbf{c} \times\boldsymbol{\sigma} )_i .
 \label{eq:angular-velocity}
\end{equation}
The branch dependence of the radial measure must be kept in the momentum-dependent numerator. In particular,
\begin{equation}
 \sum_\lambda\nu_\lambda\left(\alpha_{\mathrm{so}}+\lambda\frac{p_\lambda}{m}\right)
 =\frac{\nu_0}{m}\sum_\lambda\lambda p_\lambda=-2\nu_0\alpha_{\mathrm{so}}.
 \label{eq:measure-cancellation}
\end{equation}
Freezing both $p_\lambda$ and $\nu_\lambda$ before this step would lose a contribution of the same order as the anomalous velocity. Combining the intrabranch and interbranch terms gives
\begin{equation}
 \mathscr T_a[\hat v_i]=\frac{\alpha_{\mathrm{so}}}{2}(u_a^{(0)}-w_a)(\mathbf{c} \times\boldsymbol{\sigma} )_i,
 \qquad
 \mathscr T_a[-e F'\mathbf{E} _a\cdot\hat{\mathbf v}]
 =\frac{e\alpha_{\mathrm{so}} F'}{2}(w_a-u_a^{(0)})(\mathbf{E} _a\times\mathbf{c} )\cdot\boldsymbol{\sigma}  .
 \label{eq:bare-source-ladder}
\end{equation}

\subsubsection{Dressed source and distribution}

There is no scalar or out-of-plane-spin driving term in Eq.~\eqref{eq:bare-source-ladder}. The retarded response, with no externally imposed density or spin mode, gives
\begin{equation}
 K_{g,a,0}=K_{g,a,z}=0,\qquad
 \hat K_{g,a}=\frac{e\alpha_{\mathrm{so}} F'}{2}\frac{w_a-u_a^{(0)}}{1-v_a^{(0)}}
 (\mathbf{E} _a\times\mathbf{c} )\cdot\boldsymbol{\sigma}  .
 \label{eq:gain-solution}
\end{equation}
The unforced scalar mode at strictly zero frequency is fixed by the density boundary condition, not by dividing by $1-w_a$. At the same frozen-splitting order, the total source is
\begin{equation}
 \hat K_a=-e F'(\epsilon)\left[
 \frac{\mathbf{p} \cdot\mathbf{E} _a}{m}\mathbf{1} +
 \alpha_{\mathrm{so}} s_a^{(0)}(\mathbf{E} _a\times\mathbf{c} )\cdot\boldsymbol{\sigma} \right],\qquad
 s_a^{(0)}=\frac{1-w_a}{1-v_a^{(0)}}.
 \label{eq:dressed-source}
\end{equation}
The factor $s_a^{(0)}$ is the impurity correction to the spin-dependent source at this order, and $\partial_\epsilon s_a^{(0)}=0$. The energy dependence of this source so far comes from $F'(\epsilon)$ only. An energy-dependent ladder cannot be inferred from Eq.~\eqref{eq:radial-leading}; it requires the radial correction derived next.

\subsubsection{First radial-energy correction to the optical ladder}
\label{sec:radial}

To retain the first radial-energy correction needed by the quadratic calculation, return to the propagators before freezing their splitting. The metallic expansion starts from $p=p_F+\xi/v_F+\cdots$, giving
\begin{equation}
 \xi_{p\lambda}=(1+\lambda\delta)\xi+\lambda\Delta_R/2+\cdots,
 \qquad \delta=\alpha_{\mathrm{so}}/v_F .
 \label{eq:linear-dispersion}
\end{equation}
Using this linearized dispersion with the common measure $\nu_0\mathrm{d} \xi$, return explicitly to the opposite-helicity radial integral:
\begin{align}
 Q_{-\lambda,\lambda}(\epsilon,\omega_a)
 &\simeq\frac{1}{2\pi\tau}\int_{-\infty}^{\infty}\mathrm{d} \xi\,
 \frac{1}{\epsilon+\omega_a/2-(1-\lambda\delta)\xi
 +\lambda\Delta_R/2+\mathrm{i} /(2\tau)}\nonumber\\
 &\hspace{12mm}\times
 \frac{1}{\epsilon-\omega_a/2-(1+\lambda\delta)\xi
 -\lambda\Delta_R/2-\mathrm{i} /(2\tau)}.
 \label{eq:radial-linearized-optical}
\end{align}
The two propagators now have unequal radial slopes. A common shift of $\xi$ therefore no longer removes $\epsilon$. Evaluating this integral gives
\begin{equation}
 Q_{-\lambda,\lambda}(\epsilon,\omega_a)
 \simeq\frac{1}{a_a-\mathrm{i} \lambda(\Delta_R+2\delta\epsilon)\tau}
 =\frac{1}{a_a-\mathrm{i} \lambda\Delta_R\tau}
 +\frac{2\mathrm{i} \lambda\delta\epsilon\tau}{(a_a-\mathrm{i} \lambda\Delta_R\tau)^2}+\cdots .
 \label{eq:q-radial}
\end{equation}
For example, closing at the advanced pole gives
\begin{equation}
 Q_{-\lambda,\lambda}
 \simeq\frac{\mathrm{i} /\tau}{\omega_a+\lambda\Delta_R+2\lambda\delta\epsilon+\mathrm{i} /\tau}.
 \label{eq:q-residue}
\end{equation}
The Jacobian from the pole and the unequal slopes in Eq.~\eqref{eq:linear-dispersion} must be included together. Their combination removes a spurious term proportional to $\delta\omega_a$.

The symmetric interbranch kernel is therefore
\begin{align}
 u_a(\epsilon)&\equiv\frac12\sum_\lambda Q_{-\lambda,\lambda}
 \simeq\frac{a_a}{a_a^2+[(\Delta_R+2\delta\epsilon)\tau]^2}
 \nonumber\\
 &=u_a^{(0)}
 -\frac{4\Delta_R\delta\epsilon\tau^2 a_a}{[a_a^2+(\Delta_R\tau)^2]^2}+\cdots .
 \label{eq:u-radial}
\end{align}
The same linearized model gives $Q_{\lambda\lambda}=w_a/(1+\lambda\delta)$. Its symmetric half-sum is $w_a+O(\delta^2)$, so there is no first-order energy correction to the charge eigenvalue in this calculation. The branch weight must nevertheless be retained in each velocity-weighted integral before the helicity sum.

Only after Eq.~\eqref{eq:u-radial} has been derived do we extend the optical ladder factors to
\begin{equation}
 v_a(\epsilon)=\frac{w_a+u_a(\epsilon)}2,\qquad
 s_a(\epsilon)=\frac{1-w_a}{1-v_a(\epsilon)}.
 \label{eq:energy-dependent-optical-ladder}
\end{equation}
For explicit order bookkeeping, write
\begin{align}
 u_a(\epsilon)&=u_a^{(0)}+\epsilon u_a^{(1)}+\cdots,\nonumber\\
 u_a^{(1)}&=-\frac{4a_a\Delta_R\delta\tau^2}
 {[a_a^2+(\Delta_R\tau)^2]^2},\nonumber\\
 v_a(\epsilon)&=v_a^{(0)}+\frac{\epsilon}{2}u_a^{(1)}+\cdots,\nonumber\\
 s_a(\epsilon)&=s_a^{(0)}+\epsilon s_a^{(1)}+\cdots,\qquad
 s_a^{(1)}=\frac{(1-w_a)u_a^{(1)}}{2[1-v_a^{(0)}]^2}.
 \label{eq:optical-ladder-energy-expansion}
\end{align}
The coefficient $u_a^{(1)}$ denotes $\partial_\epsilon u_a|_{\epsilon=0}$, not an additional optical harmonic. The displayed rational form in Eq.~\eqref{eq:u-radial} is a convenient expression within the linearized dispersion; subsequent products are expanded consistently through the stated radial order. At this order $u_a(0)=u_a^{(0)}$, $v_a(0)=v_a^{(0)}$, and $s_a(0)=s_a^{(0)}$. The source entering the distribution below is Eq.~\eqref{eq:dressed-source} with $s_a^{(0)}$ replaced by this explicitly defined $s_a(\epsilon)$.

In the asymmetric routing used for the triangles, the two energies are $\epsilon$ and $\epsilon-\omega$. Their mean is $\bar\epsilon=\epsilon-\omega/2$, and hence
\begin{equation}
 u^{\rm as}(\omega,\epsilon)=u(\omega,\epsilon-\omega/2)
 =\frac{a}{a^2+\Delta_R^2\tau^2}
 -\frac{2\Delta_R\delta\tau^2a(2\epsilon-\omega)}{(a^2+\Delta_R^2\tau^2)^2}.
 \label{eq:routing}
\end{equation}
The factor $2\epsilon-\omega$ is a routing change, not a change of the disorder model.

When no energy argument is printed below, $u_a$, $v_a$, and $s_a$ denote their Fermi-level values; the energy-dependent extension is always written with $(\epsilon)$. The local electron momentum and the mean energy are independent Wigner variables. Thus the impurity ladder $s_a(\epsilon)$ is independent of the local $\mathbf{p} $ in the precession inverse, although it depends on the mean energy through the radial integration.

\subsubsection{Distribution with the energy-dependent source}

For completeness, the local precession operator can be inverted without choosing spin components. Let $\hat K=K_0\mathbf{1} +\mathbf K\cdot\boldsymbol{\sigma} $ and $\Delta_p=2\alpha_{\mathrm{so}} p$. The solution of $z_a\hat h+\mathrm{i} [\mathbf b_p\cdot\boldsymbol{\sigma} ,\hat h]=\hat K$ is
\begin{equation}
 h_0=\frac{K_0}{z_a},\qquad
 \mathbf h=\frac{\mathbf K_{\eta}}{z_a}
 +\frac{z_a\mathbf K_{\perp}+\Delta_p\boldsymbol\eta\times\mathbf K_{\perp}}
 {z_a^2+\Delta_p^2},\qquad
 \mathbf K_{\eta}=\boldsymbol\eta(\boldsymbol\eta\cdot\mathbf K),\quad
 \mathbf K_{\perp}=\mathbf K-\mathbf K_{\eta}.
 \label{eq:inverse-precession}
\end{equation}
For the quadratic calculation it is essential to keep the radial momentum in this inverse. Define
\begin{equation}
 D_{a,p}=z_a^2+4\alpha_{\mathrm{so}}^2p^2,\qquad r_a(\epsilon)=-e\alpha_{\mathrm{so}} F'(\epsilon)s_a(\epsilon).
\end{equation}
The scalar and vector components of the first-order solution are
\begin{align}
 h_{0,a}^{(1)}&=-\frac{e F'}{m z_a}\mathbf{p} \cdot\mathbf{E} _a,\label{eq:h1scalar}\\
 \mathbf h_a^{(1)}&=r_a\left[
 \frac{z_a}{D_{a,p}}(\mathbf{E} _a\times\mathbf{c} )
 +\frac{4\alpha_{\mathrm{so}}^2(\mathbf{p} \cdot\mathbf{E} _a)(\mathbf{p} \times\mathbf{c} )}{z_aD_{a,p}}
 +\frac{2\alpha_{\mathrm{so}}\,\mathbf{c} \cdot(\mathbf{p} \times\mathbf{E} _a)}{D_{a,p}}\mathbf{c} \right].
 \label{eq:h1}
\end{align}
The last term originates from spin precession. Its angular average vanishes at first order, but it contributes to the quadratic out-of-plane-spin response. In particular, $\boldsymbol\eta\times(\mathbf{E} \times\mathbf{c} )=-\mathbf{c} (\boldsymbol\eta\cdot\mathbf{E} )$ fixes its sign. The momentum derivatives in the quadratic calculation act on $D_{a,p}$ and the explicit momentum factors in Eq.~\eqref{eq:h1}; the ladder $s_a(\epsilon)$ is momentum independent.

\subsection{Two occupied contours and density normalization}
At fixed chemical potential, Eq.~\eqref{eq:dos} gives
\begin{equation}
 n=\frac{p_+^2+p_-^2}{4\pi}
 =\frac{m(\varepsilon_F+m\alpha_{\mathrm{so}}^2)}{\pi},\qquad
 \frac nm=\nu_0(v_F^2+2\alpha_{\mathrm{so}}^2).
 \label{orb:eq:density}
\end{equation}
\begin{equation}
 \sigma_D=\frac{e^2\nu_0v_F^2}{\Gamma}.
 \label{orb:eq:sigmaD}
\end{equation}
Here $\sigma_D$ is a reference dc scale for the unsplit band at the same chemical potential. Holding total density fixed instead shifts $\varepsilon_F$ and rearranges the explicit $2\delta^2$ term below. A spin trace includes both components; $\nu_0$ remains the density of states per spin.
\subsection{Independent evaluation of the optical conductivity}
\subsubsection{Why calculate the bubble separately?}
The aim is to identify a nonlinear triangle contraction with an independently computed two-current response. Defining that bubble first avoids assuming the orbital IFE identity as an input. In the single-splitting metallic Kubo normalization,
\begin{equation}
\sigma(\omega)=\frac{e^2}{\omega}
\int\frac{d\epsilon}{2\pi}W_\omega(\epsilon)\int\!\frac{\mathrm{d}^2p}{(2\pi)^2} 
\operatorname{Tr} [v_xG^R_\epsilon V_x^+G^A_{\epsilon-\omega}],
\quad W_\omega=f_0(\epsilon-\omega)-f_0(\epsilon).\label{orb:eq:Kubo}
\end{equation}
The common Kubo/contact subtraction is understood in \eqref{orb:eq:Kubo}. At the retained order, the energy-window integral gives $\omega$ after linear mean-energy terms have been integrated, so $\sigma=e^2\Pi/(2\pi)$ for the radially integrated bubble $\Pi$.

\subsubsection{The four current moments, including the SOC terms}
Define $a_x=p_x/m$ for this subsection only, and use the notation
\begin{equation}
I_{AB}=\gamma_{\rm imp}\int\!\frac{\mathrm{d}^2p}{(2\pi)^2} \operatorname{Tr} [A G^R B G^A_\omega].
\end{equation}
The necessary moments are
\begin{align}
I_{aa}&=w(v_F^2+4\alpha_{\mathrm{so}}^2),\label{orb:eq:Iaa}\\
I_{a\mathsf R}=I_{\mathsf R a}&=-2\alpha_{\mathrm{so}} w,\label{orb:eq:Iat}\\
I_{\mathsf R\mathsf R}&=2v.\label{orb:eq:Itt}
\end{align}
For \eqref{orb:eq:Iaa}, the angle average is $\left\langle p_x^2\right\rangle_{\phi}=p^2/2$, and the branch sum is explicitly
\begin{align}
\frac12\sum_\lambda\frac{\nu_\lambda}{\nu_0}\frac{p_\lambda^2}{m^2}
&=\frac1{2m^2}\sum_\lambda
\left(1-\lambda\frac{m\alpha_{\mathrm{so}}}{p_0}\right)(p_0-\lambda m\alpha_{\mathrm{so}})^2\\
&=p_0^2/m^2+3\alpha_{\mathrm{so}}^2=v_F^2+4\alpha_{\mathrm{so}}^2.
\end{align}
The factor $w$ is the intrabranch RA integral. For \eqref{orb:eq:Iat}, take the trace of $\hat{\mathsf R}_x\mathcal T_\omega[a_x]=-\alpha_{\mathrm{so}} w \hat{\mathsf R}_x^2$ and use $\operatorname{Tr}  \hat{\mathsf R}_x^2=2$. The reversed insertion gives the same trace because the scalar $a_x$ commutes with both Green functions. For \eqref{orb:eq:Itt}, use $\operatorname{Tr} [\hat{\mathsf R}_x\mathcal T_\omega[\hat{\mathsf R}_x]]=v\operatorname{Tr}  \hat{\mathsf R}_x^2=2v$.

Now insert $v_x=a_x+\alpha_{\mathrm{so}} \hat{\mathsf R}_x$ and $V_x^+=a_x+\alpha_{\mathrm{so}} s \hat{\mathsf R}_x$ into the bubble. Every term is
\begin{align}
\gamma_{\rm imp}\Pi
&=I_{aa}+\alpha_{\mathrm{so}} s I_{a\mathsf R}+\alpha_{\mathrm{so}} I_{\mathsf R a}+\alpha_{\mathrm{so}}^2s I_{\mathsf R\mathsf R}\\
&=w(v_F^2+4\alpha_{\mathrm{so}}^2)-2\alpha_{\mathrm{so}}^2w(s+1)+2\alpha_{\mathrm{so}}^2sv\\
&=wv_F^2+2\alpha_{\mathrm{so}}^2w+2\alpha_{\mathrm{so}}^2s(v-w).\label{orb:eq:Pi}
\end{align}
Since $e^2/(2\pi\gamma_{\rm imp})=e^2\nu_0/\Gamma$, division by \eqref{orb:eq:sigmaD} gives
\begin{align}
\frac\sigma{\sigma_D}
&=w(1+2\delta^2)+2\delta^2s(v-w)\\
&=w(1+2\delta^2)
+\delta^2\frac{(1-w)(u-w)}{1-v}\\
&=\boxed{w(1+2\delta^2)
+\delta^2\frac{(w-1)(w-u)}{1-v}.}\label{orb:eq:image}
\end{align}
The last line is the conductivity used in both response calculations. The two minus signs in $(1-w)(u-w)=(w-1)(w-u)$ explain the apparent change of form. All four moments were necessary: retaining only $I_{aa}$ would give the incorrect coefficient $4\delta^2$ instead of the final $2\delta^2$ term and would miss the optical spin ladder.

\subsection{The same conductivity from the linear kinetic solution}
Within the equilibrium-spectral projection used for transport, the optical density correction is
\begin{equation}
 \hat w_a^{(1)}=\frac{G_{0,+}^Rh_a^{(1)}-h_a^{(1)}G_{0,-}^A}{4\pi\mathrm{i} }
 =-\frac{G_{0,+}^RK_aG_{0,-}^A}{4\pi}
 =\frac{eF'}{4\pi}G_{0,+}^R(E_{a,j}V_j)G_{0,-}^A.
 \label{eq:kinetic-linear-density}
\end{equation}
The second equality is Eq.~\eqref{eq:ra-identity}, and the last uses $K_a=-eF'E_{a,j}V_j$. Taking its physical current moment gives
\begin{equation}
 \sigma_{ij}^{\rm kin}(\omega)=\frac{e^2}{4\pi}
 \int\mathrm{d} \epsilon\,F'(\epsilon)\int\!\frac{\mathrm{d}^2p}{(2\pi)^2} 
 \operatorname{Tr} [v_iG_{0,+}^RV_jG_{0,-}^A].
 \label{eq:kinetic-linear-sigma}
\end{equation}
At $T=0$, $\int F'\mathrm{d} \epsilon=2$. The same angular and radial moments in Eqs.~\eqref{orb:eq:Iaa}--\eqref{orb:eq:Itt} therefore give $\sigma=e^2\Pi/(2\pi)$, identical to the optical bubble. In the asymmetric diagram routing the normalized signed window $W_\omega/\omega$ integrates to one, and its linear mean-energy moment vanishes. These two representations yield the same retained metallic coefficient. Both evaluations use the same radial measure, dressed optical velocity, and zero-temperature energy weights.
\subsection{Calculating the real conductivity explicitly and obtaining the final IFE}
\subsubsection{Dimensionless variables and simplification of the kinetic expression}
Introduce
\begin{equation}
X=\omega\tau=\omega/\Gamma,\qquad
B_R=\Delta_R\tau=\Delta_R/\Gamma,\qquad
a=1-iX.
\end{equation}
$B_R$ denotes a dimensionless Rashba splitting; it is unrelated to the normalized conductivity $B$ used in the residue calculation. Then
\begin{equation}
w=1/a,\qquad u=\frac{a}{a^2+B_R^2},\qquad v=(w+u)/2.
\end{equation}
Starting from Eq.~\eqref{orb:eq:image}, calculate each difference:
\begin{align}
w-1&=\frac{iX}{a},\\
w-u&=\frac{B_R^2}{a(a^2+B_R^2)},\\
1-v&=\frac{2a^2(a-1)+B_R^2(2a-1)}{2a(a^2+B_R^2)}.
\end{align}
Define
\begin{equation}
\mathcal P=B_R^2(1-2iX)-2iX(1-iX)^2.
\end{equation}
Because $a-1=-iX$, $2a^2(a-1)+B_R^2(2a-1)=\mathcal P$. Therefore
\begin{equation}
\boxed{\frac\sigma{\sigma_D}
=\frac{1+2\delta^2}{1-iX}
+\frac{2i\delta^2XB_R^2}{(1-iX)\mathcal P}.}\label{orb:eq:Sigmacomplex}
\end{equation}
This is the evaluated conductivity, retaining the impurity optical-spin vertex. Its first term is the density-corrected Drude term; its second is the SOC optical correction.

\subsubsection{Taking the real and imaginary parts without an implicit complex expression}
Write
\begin{equation}
\mathcal P=A_R-iC_R,\qquad
A_R=B_R^2-4X^2,\qquad C_R=2X(1+B_R^2-X^2),
\end{equation}
and define the real nonnegative denominator
\begin{equation}
\mathcal D=A_R^2+C_R^2
=(B_R^2-4X^2)^2+4X^2(1+B_R^2-X^2)^2.\label{orb:eq:Ddimensionless}
\end{equation}
Multiply by the conjugate explicitly:
\begin{align}
\frac1{a\mathcal P}
&=\frac{(1+iX)(A_R+iC_R)}{(1+X^2)\mathcal D}\\
&=\frac{(A_R-XC_R)+i(C_R+XA_R)}{(1+X^2)\mathcal D}.
\end{align}
Thus the real part of the SOC numerator $2i\delta^2XB_R^2/(a\mathcal P)$ is
\begin{equation}
-\frac{2\delta^2XB_R^2(C_R+XA_R)}{(1+X^2)\mathcal D}.
\end{equation}
The remaining real polynomial simplifies to
\begin{equation}
C_R+XA_R=X(2+3B_R^2-6X^2).
\end{equation}
Adding the Drude real part proves
\begin{equation}
\boxed{\operatorname{Re} \sigma(\omega)=\frac{\sigma_D}{1+X^2}
\left[1+2\delta^2
-\frac{2\delta^2X^2B_R^2(2+3B_R^2-6X^2)}{\mathcal D}\right].}\label{orb:eq:ReSigmafinal}
\end{equation}
For completeness, the imaginary part used separately by the triangle and contact calculations is
\begin{equation}
\frac{\operatorname{Im} \sigma}{\sigma_D}
=\frac{X(1+2\delta^2)}{1+X^2}
+\frac{2\delta^2XB_R^2[A_R-XC_R]}{(1+X^2)\mathcal D}.\label{orb:eq:ImSigmafinal}
\end{equation}
Equations \eqref{orb:eq:ReSigmafinal} and \eqref{orb:eq:ImSigmafinal} reproduce the full complex expression \eqref{orb:eq:image} when combined. The real part is even in $\omega$, and the imaginary part is odd, as required for a real causal current response.

\section{Spin IFE from the kinetic equation}
The dimensionless functions used in this section have fixed physical meanings:
\begin{equation}
 w(\omega)=\frac{\Gamma}{\Gamma-\mathrm{i} \omega},\qquad
 u(\omega)=\frac{\Gamma(\Gamma-\mathrm{i} \omega)}{(\Gamma-\mathrm{i} \omega)^2+\Delta_R^2},\qquad
 v(\omega)=\frac{w+u}{2},\qquad s(\omega)=\frac{1-w}{1-v}.
\end{equation}
Here $w$ is the charge-channel ladder eigenvalue, $u$ the out-of-plane-spin eigenvalue, $v$ the in-plane-spin eigenvalue, and $s$ the dressing of the anomalous optical velocity. The scalar $v$ is always a ladder eigenvalue; velocities carry a Cartesian index or a bold symbol. The argument $\omega$ is omitted only within the following calculation. These are the Fermi-level values $u(\omega)=u(\omega,0)$, $v(\omega)=v(\omega,0)$, and $s(\omega)=s(\omega,0)$. Their energy-dependent extensions are defined in Eqs.~\eqref{eq:u-radial}--\eqref{eq:optical-ladder-energy-expansion}.

\subsection{Quadratic kinetic source in the helicity channel}
\label{sec:quadratic}

We now take the electric-dipole limit $\mathbf q_a=0$ and the stationary output $\Omega=0$, with adiabatic switching understood. Write $\mathbf{E} _+=\mathbf{E} $, $\mathbf{E} _-=\mathbf{E} ^*$, and define the real helicity pseudoscalar
\begin{equation}
 \mathcal{H} =\mathrm{i} \mathbf{c} \cdot(\mathbf{E} \times\mathbf{E} ^*),\qquad
 C_{ab}=\mathbf{c} \cdot(\mathbf{E} _a\times\mathbf{E} _b),\qquad C_{+-}=-\mathrm{i} \mathcal{H} .
 \label{eq:helicity}
\end{equation}
There is no factor of $1/2$ in the field amplitudes in Eq.~\eqref{eq:optical}. The helicity projection antisymmetrizes the two Cartesian optical indices while holding their frequencies fixed. Both frequency orderings must subsequently be summed. For a coefficient $A_-=A_+^*$,
\begin{equation}
 \sum_{a=\pm}C_{a,-a}A_{-a}
 =-\mathrm{i} \mathcal{H}  A_+^*+\mathrm{i} \mathcal{H}  A_+
 =-2\mathcal{H} \operatorname{Im}  A_+.
 \label{eq:ordering-identity}
\end{equation}

In the single-splitting metallic kinetic reduction, the quadratic equation has the form
\begin{equation}
 \tau^{-1}\hat h^{(2)}+\mathrm{i} [\mathbf b_p\cdot\boldsymbol{\sigma} ,\hat h^{(2)}]
 =\hat K^{(2)}+\hat K_g^{(2)},\qquad
 \hat K^{(2)}=-e\sum_{a=\pm}E_{a,i}
 \left(\partial_{p_i}\hat h_{-a}^{(1)}
 +\frac12\{\hat v_i,\partial_\epsilon\hat h_{-a}^{(1)}\}\right).
 \label{eq:quadratic-source}
\end{equation}
Here $K_g^{(2)}$ denotes the output impurity gain. Equation~\eqref{eq:quadratic-source} defines the projected kinetic calculation evaluated below; the exact nonlinear SCBA equation and the spectral reconstruction of the lesser function are discussed in Appendix~\ref{app:nonlinear}. In particular, a first-order cancellation of spectral self-energies is not an assertion that every nonlinear spectral term vanishes.

\subsubsection{Scalar and vector decomposition}

Using $\{\mathbf a\cdot\boldsymbol{\sigma} ,\mathbf b\cdot\boldsymbol{\sigma} \}=2\mathbf a\cdot\mathbf b\,\mathbf{1} $, we obtain
\begin{align}
 K_0^{(2)}={}&-e\sum_a\left[
 \mathbf{E} _a\cdot\boldsymbol\nabla_p h_{0,-a}^{(1)}
 +\frac{\mathbf{E} _a\cdot\mathbf{p} }{m}\partial_\epsilon h_{0,-a}^{(1)}
 +\alpha_{\mathrm{so}}(\mathbf{E} _a\times\mathbf{c} )\cdot\partial_\epsilon\mathbf h_{-a}^{(1)}\right],
 \label{eq:Kscalar}\\
 \mathbf K^{(2)}={}&-e\sum_a\left[
 \mathbf{E} _a\cdot\boldsymbol\nabla_p\mathbf h_{-a}^{(1)}
 +\frac{\mathbf{E} _a\cdot\mathbf{p} }{m}\partial_\epsilon\mathbf h_{-a}^{(1)}
 +\alpha_{\mathrm{so}}(\mathbf{E} _a\times\mathbf{c} )\partial_\epsilon h_{0,-a}^{(1)}\right].
 \label{eq:Kvector}
\end{align}
The scalar contractions reduce to $\mathbf{E} _a\cdot\mathbf{E} _{-a}$ and $(\mathbf{p} \cdot\mathbf{E} _a)(\mathbf{p} \cdot\mathbf{E} _{-a})$. They are symmetric in the optical indices, so $K_{0,\mathrm{hel}}^{(2)}=0$. This is a projection on the helicity sector; it does not eliminate the helicity-even quadratic density response.

The necessary momentum derivatives are
\begin{align}
 \mathbf{E} _a\cdot\boldsymbol\nabla_p D_{-a,p}&=8\alpha_{\mathrm{so}}^2\mathbf{E} _a\cdot\mathbf{p} ,\nonumber\\
 \mathbf{E} _a\cdot\boldsymbol\nabla_p[\mathbf{c} \cdot(\mathbf{p} \times\mathbf{E} _b)]&=C_{ab},\nonumber\\
 \mathbf{E} _a\cdot\boldsymbol\nabla_p[(\mathbf{p} \cdot\mathbf{E} _b)(\mathbf{p} \times\mathbf{c} )]
 &=(\mathbf{E} _a\cdot\mathbf{E} _b)(\mathbf{p} \times\mathbf{c} )+(\mathbf{p} \cdot\mathbf{E} _b)(\mathbf{E} _a\times\mathbf{c} ).
 \label{eq:momentum-identities}
\end{align}
The three antisymmetric contractions used in applying these derivatives are
\begin{align}
 [(\mathbf{E} _a\times\mathbf{c} )(\mathbf{p} \cdot\mathbf{E} _b)]_{\rm hel}&=-\tfrac12 C_{ab}\mathbf{p} ,\nonumber\\
 [(\mathbf{p} \cdot\mathbf{E} _a)(\mathbf{E} _b\times\mathbf{c} )]_{\rm hel}&=\tfrac12 C_{ab}\mathbf{p} ,\nonumber\\
 [(\mathbf{p} \cdot\mathbf{E} _a)\,\mathbf{c} \cdot(\mathbf{p} \times\mathbf{E} _b)]_{\rm hel}&=\tfrac12 C_{ab}p^2.
 \label{eq:antisymmetric-identities}
\end{align}
For example, the last identity follows by writing the antisymmetrized expression as $\tfrac12\mathbf{c} \cdot\{\mathbf{p} \times[(\mathbf{p} \cdot\mathbf{E} _a)\mathbf{E} _b-(\mathbf{p} \cdot\mathbf{E} _b)\mathbf{E} _a]\}$ and using that $\mathbf{p} $ lies in the plane.

\subsubsection{The three contributions to the spin source}

For compactness, write $z=\tau^{-1}-\mathrm{i} \omega$, $D_p=z^2+4\alpha_{\mathrm{so}}^2p^2$, and $s=s(\omega,\epsilon)$. The momentum-derivative term in Eq.~\eqref{eq:Kvector}, after Eq.~\eqref{eq:ordering-identity}, is
\begin{equation}
 \mathbf K_{\nabla p}=-4e^2\alpha_{\mathrm{so}}^2F'\mathcal{H} \operatorname{Im} \left[
 \frac{s z^2}{D_p^2}\mathbf{c} 
 -\alpha_{\mathrm{so}} s\left(\frac{1}{zD_p}+\frac{2z}{D_p^2}\right)\mathbf{p} \right].
 \label{eq:Kmomentum}
\end{equation}
The energy derivative multiplying the scalar part of the velocity gives
\begin{equation}
 \mathbf K_{v_0\partial\epsilon}
 =-\frac{2e^2\alpha_{\mathrm{so}}}{m}\mathcal{H} \operatorname{Im} \left[
 \partial_\epsilon(F's)\left(\frac{z}{2D_p}\mathbf{p} +\frac{\alpha_{\mathrm{so}} p^2}{D_p}\mathbf{c} \right)\right].
 \label{eq:Kenergy}
\end{equation}
Finally, the anomalous velocity multiplying the scalar distribution gives
\begin{equation}
 \mathbf K_{v_{\rm so}\partial\epsilon}
 =\frac{e^2\alpha_{\mathrm{so}}}{m}F''\mathcal{H} \operatorname{Im} \left(\frac1z\right)\mathbf{p} .
 \label{eq:Kanomalous}
\end{equation}
Only $e^2=e^2$ remains. Keeping all three terms, the helicity source can be written as
\begin{equation}
 \hat K_{\rm hel}^{(2)}=K_z(p,\epsilon)\,\mathbf{c} \cdot\boldsymbol{\sigma} 
 +K_p(p,\epsilon)\,\mathbf{p} \cdot\boldsymbol{\sigma} ,
 \label{eq:Kcp}
\end{equation}
where
\begin{align}
 K_z={}&-4e^2\alpha_{\mathrm{so}}^2F'\mathcal{H} \operatorname{Im} \frac{s z^2}{D_p^2}
 -\frac{2e^2\alpha_{\mathrm{so}}^2p^2}{m}\mathcal{H} \operatorname{Im} \frac{\partial_\epsilon(F's)}{D_p},
 \label{eq:Kc}\\
 K_p={}&4e^2\alpha_{\mathrm{so}}^3F'\mathcal{H} \operatorname{Im} \left[s\left(\frac1{zD_p}+\frac{2z}{D_p^2}\right)\right]
 -\frac{e^2\alpha_{\mathrm{so}}}{m}\mathcal{H} \operatorname{Im} \left[
 \frac{z\partial_\epsilon(F's)}{D_p}-\frac{F''}{z}\right].
 \label{eq:Kp}
\end{align}
Although the second term in Eq.~\eqref{eq:Kcp} is in plane at each momentum, precession and the output ladder convert it into a spatially uniform normal spin. It must therefore be retained before the angular integral.

\subsection{Output spin ladder and radial phase-space derivative}
\label{sec:output}

The dc counterpart of Eq.~\eqref{eq:ra-identity} yields
\begin{equation}
 \hat K_g^{(2)}=\gamma_{\rm imp}\int\!\frac{\mathrm{d}^2p}{(2\pi)^2} 
 \hat G_0^R(\hat K_{\rm hel}^{(2)}+\hat K_g^{(2)})\hat G_0^A.
 \label{eq:output-ladder}
\end{equation}
The helicity source drives only a momentum-independent out-of-plane-spin gain,
\begin{equation}
 \hat K_g^{(2)}=K_{g,z}^{(2)}\mathbf{c} \cdot\boldsymbol{\sigma} ,\qquad
 K_{g,z}^{(2)}=t_z(\epsilon)+u(0,\epsilon)K_{g,z}^{(2)}
 =\frac{t_z(\epsilon)}{1-u(0,\epsilon)}.
 \label{eq:output-solution}
\end{equation}
The scalar homogeneous mode is fixed to zero in this sector. The angular projection needed for the inhomogeneous term is
\begin{equation}
 \left\langle P_\mu(K_z\mathbf{c} \cdot\boldsymbol{\sigma} +K_p\mathbf{p} \cdot\boldsymbol{\sigma} )P_\nu\right\rangle_{\phi}
 =\frac14\big[(1-\mu\nu)K_z+\mathrm{i} (\mu-\nu)pK_p\big]\mathbf{c} \cdot\boldsymbol{\sigma} .
 \label{eq:output-angle}
\end{equation}
Thus only opposite helicities contribute:
\begin{equation}
 t_z(\epsilon)=\frac1{4\pi\tau}\sum_{\mu=\pm}\int\mathrm{d} \xi\,
 [K_z(p,\epsilon)+\mathrm{i} \mu pK_p(p,\epsilon)]
 g_\mu^R(\epsilon,p)g_{-\mu}^A(\epsilon,p).
 \label{eq:t-integral}
\end{equation}

\subsubsection{Zeroth and first radial moments}

Use the branch variable $x=\xi_{p\mu}$ and expand
\begin{equation}
 p=p_F+\frac{x-\mu\Delta_R/2}{v_F}+\cdots,\qquad
 \frac{\mathrm{d} \xi}{\mathrm{d}  x}=1-\mu\delta+\cdots,\qquad
 d_\mu=1-\mu\delta,\quad B_\mu=\mu\Delta_R-\mathrm{i} /\tau.
 \label{eq:radial-variables}
\end{equation}
For any smooth numerator $N_\mu(p)$,
\begin{equation}
 N_\mu(p)=N_\mu(p_F)
 +\frac{x-\mu\Delta_R/2}{v_F}\partial_pN_\mu(p_F)+\cdots.
 \label{eq:numerator-expansion}
\end{equation}
At the retarded pole $x_R=\epsilon+\mathrm{i} /(2\tau)$, the opposite-helicity advanced denominator is $d_\mu B_\mu+2\mu\delta\epsilon$ to the retained order. Closing the radial contour in the upper half plane gives
\begin{align}
 J_{0\mu}&\equiv\int\mathrm{d}  x\,d_\mu g_\mu^Rg_{-\mu}^A
 =-2\pi\mathrm{i} \left(\frac1{B_\mu}-\frac{2\mu\delta\epsilon}{B_\mu^2}\right)+\cdots,
 \label{eq:radial-J0}\\
 J_{1\mu}&\equiv\operatorname{PV} \int\mathrm{d}  x\,d_\mu
 \frac{x-\mu\Delta_R/2}{v_F}g_\mu^Rg_{-\mu}^A
 =-\frac{2\pi\mathrm{i} \epsilon}{v_FB_\mu}+\cdots.
 \label{eq:radial-J1}
\end{align}
The second integral requires the arc contribution: its integrand behaves as $1/(v_Fx)$. At leading order,
\begin{equation}
 J_{1\mu}=-\frac{\mathrm{i} \pi}{v_F}
 -\frac{2\pi\mathrm{i} }{v_FB_\mu}\left(\epsilon-\frac{B_\mu}{2}\right)
 =-\frac{2\pi\mathrm{i} \epsilon}{v_FB_\mu}.
 \label{eq:arc-cancellation}
\end{equation}
Discarding the arc while retaining the radial numerator gives an erroneous constant. The principal value here is the symmetric wide-band prescription inherited from the metallic expansion.

For reference, the helicity sums are
\begin{align}
 \sum_\mu B_\mu^{-1}&=\frac{2\mathrm{i} \tau}{1+\Delta_R^2\tau^2},&
 \sum_\mu\mu B_\mu^{-1}&=\frac{2\Delta_R\tau^2}{1+\Delta_R^2\tau^2},\nonumber\\
 \sum_\mu B_\mu^{-2}&=\frac{2(\Delta_R^2\tau^2-1)\tau^2}{(1+\Delta_R^2\tau^2)^2},&
 \sum_\mu\mu B_\mu^{-2}&=\frac{4\mathrm{i} \Delta_R\tau^3}{(1+\Delta_R^2\tau^2)^2}.
 \label{eq:helicity-sums}
\end{align}
Substituting $N_\mu=K_z+\mathrm{i} \mu pK_p$ into Eq.~\eqref{eq:numerator-expansion} and using these sums gives
\begin{align}
 t_z={}&\left[\frac{K_z+\Delta_R\tau p_FK_p}{1+\Delta_R^2\tau^2}
 +\frac{\epsilon}{v_F}\frac{\partial_pK_z+\Delta_R\tau(K_p+p_F\partial_pK_p)}{1+\Delta_R^2\tau^2}\right.\nonumber\\
 &\left.\hspace{9mm}+2\delta\epsilon\,
 \frac{-2\Delta_R\tau^2K_z+(1-\Delta_R^2\tau^2)\tau p_FK_p}
 {(1+\Delta_R^2\tau^2)^2}\right]_{p=p_F}.
 \label{eq:t-expanded}
\end{align}

\subsubsection{Energy dependence of the output vertex}

The dc out-of-plane-spin ladder follows from Eq.~\eqref{eq:u-radial}:
\begin{equation}
 \Lambda_z(\epsilon)\equiv\frac1{1-u(0,\epsilon)}
 =\frac{1+\Delta_R^2\tau^2}{\Delta_R^2\tau^2}
 -\frac{4\delta\epsilon}{\Delta_R^3\tau^2}+\cdots.
 \label{eq:spin-ladder-energy}
\end{equation}
Its energy dependence is multiplied by the radial moment before the energy integral is performed. Expanding $K_{g,z}^{(2)}=\Lambda_zt_z$ consistently, one obtains
\begin{equation}
 K_{g,z}^{(2)}=\left[
 \frac{K_z}{\Delta_R^2\tau^2}+\frac{p_FK_p}{\Delta_R\tau}
 +\frac{\epsilon}{v_F}\frac{\partial_pK_z}{\Delta_R^2\tau^2}
 +\frac{\epsilon}{v_F}\frac{p_F\partial_pK_p}{\Delta_R\tau}
 -\frac{4\delta\epsilon K_z}{\Delta_R^3\tau^2}\right]_{p=p_F}.
 \label{eq:gain-expanded}
\end{equation}
The term proportional to $\epsilon K_p$ cancels between the numerator expansion and the vertex correction. Since $\Delta_R=2\alpha_{\mathrm{so}} p_F$ and $\delta=\alpha_{\mathrm{so}}/v_F$, an especially useful form is
\begin{equation}
 \boxed{\displaystyle
 \tau K_{g,z}^{(2)}=
 \frac1{4\alpha_{\mathrm{so}}^2p_F^2\tau}
 \left[K_z+\frac{\epsilon}{v_F}\left(\partial_p-\frac2{p_F}\right)K_z\right]_{p=p_F}
 +\frac1{2\alpha_{\mathrm{so}}}\left[K_p+\frac{\epsilon}{v_F}\partial_pK_p\right]_{p=p_F}.}
 \label{eq:phase-space-operator}
\end{equation}
The $-2K_z/p_F$ term includes the energy dependence of the output ladder. Equation~\eqref{eq:phase-space-operator} is the radial phase-space operation required for the single-splitting metallic spin response; replacing it by its first, energy-independent terms is insufficient.

\subsection{Energy integration and the spin inverse Faraday effect}
\label{sec:kinetic-result}

\subsubsection{Four energy moments}

At $T=0$, $F'=2\delta(\epsilon)$ and $F''=2\delta'(\epsilon)$. Integrating by parts before setting $\epsilon=0$ gives
\begin{align}
 \int\mathrm{d} \epsilon\,F's&=2s(\omega,0),&
 \int\mathrm{d} \epsilon\,\epsilon F's&=0,\nonumber\\
 \int\mathrm{d} \epsilon\,\partial_\epsilon(F's)&=0,&
 \int\mathrm{d} \epsilon\,\epsilon\partial_\epsilon(F's)&=-2s(\omega,0),\nonumber\\
 \int\mathrm{d} \epsilon\,F''&=0,&
 \int\mathrm{d} \epsilon\,\epsilon F''&=-2.
 \label{eq:energy-identities}
\end{align}
These identities retain the derivative of the energy-dependent vertex automatically. In particular, $\epsilon\delta'(\epsilon)=-\delta(\epsilon)$ as a distribution, so a factor of energy cannot be set to zero before integration.

With $s\equiv s(\omega,0)$ in the following four moments, Eqs.~\eqref{eq:Kc} and \eqref{eq:Kp} give
\begin{align}
 I_z(p)&\equiv\int\mathrm{d} \epsilon\,K_z
 =-8e^2\alpha_{\mathrm{so}}^2\mathcal{H} \operatorname{Im} \frac{s z^2}{D_p^2},\label{eq:Ic}\\
 I_p(p)&\equiv\int\mathrm{d} \epsilon\,K_p
 =8e^2\alpha_{\mathrm{so}}^3\mathcal{H} \operatorname{Im} \left[s\left(\frac1{zD_p}+\frac{2z}{D_p^2}\right)\right],\label{eq:Ip}\\
 J_z(p)&\equiv\int\mathrm{d} \epsilon\,\epsilon K_z
 =\frac{4e^2\alpha_{\mathrm{so}}^2p^2}{m}\mathcal{H} \operatorname{Im} \frac{s}{D_p},\label{eq:Jc}\\
 J_p(p)&\equiv\int\mathrm{d} \epsilon\,\epsilon K_p
 =\frac{2e^2\alpha_{\mathrm{so}}}{m}\mathcal{H} \operatorname{Im} \left(\frac{s z}{D_p}-\frac1z\right).
 \label{eq:Jp}
\end{align}
The momentum derivatives are taken at fixed $s$, since the local momentum is distinct from the equilibrium $p_F$ entering the impurity ladder. The required identities are
\begin{equation}
 \left(\partial_p-\frac2p\right)\frac{p^2}{D_p}
 =-\frac{8\alpha_{\mathrm{so}}^2p^3}{D_p^2},\qquad
 \partial_p\left(\frac{s z}{D_p}-\frac1z\right)
 =-\frac{8\alpha_{\mathrm{so}}^2p s z}{D_p^2}.
 \label{eq:moment-derivatives}
\end{equation}

\subsubsection{Combination of the direct and radial terms}

The terms without an explicit radial derivative in Eq.~\eqref{eq:phase-space-operator} give
\begin{equation}
 \frac{I_z}{4\alpha_{\mathrm{so}}^2p_F^2\tau}+\frac{I_p}{2\alpha_{\mathrm{so}}}
 =2e^2\mathcal{H} \operatorname{Im} \left[
 s\left(\frac{2\alpha_{\mathrm{so}}^2}{zD_F}+\frac{4\alpha_{\mathrm{so}}^2z}{D_F^2}
 -\frac{z^2}{p_F^2\tau D_F^2}\right)\right],\quad D_F=z^2+\Delta_R^2.
 \label{eq:direct-moments}
\end{equation}
The two phase-space corrections, using $mv_F=p_F$, are
\begin{align}
 \frac1{4\alpha_{\mathrm{so}}^2p_F^2\tau v_F}
 \left(\partial_p-\frac2{p_F}\right)J_z\bigg|_{p_F}
 &=-\frac{8e^2\alpha_{\mathrm{so}}^2}{\tau}\mathcal{H} \operatorname{Im} \frac{s}{D_F^2},\nonumber\\
 \frac1{2\alpha_{\mathrm{so}} v_F}\partial_pJ_p\bigg|_{p_F}
 &=-8e^2\alpha_{\mathrm{so}}^2\mathcal{H} \operatorname{Im} \frac{s z}{D_F^2}.
 \label{eq:radial-moments}
\end{align}
The $z/D_F^2$ terms cancel, while
$(z^2/p_F^2+4\alpha_{\mathrm{so}}^2)/D_F^2=1/(p_F^2D_F)$. The integrated gain is therefore
\begin{equation}
 \boxed{\displaystyle
 \tau\int\mathrm{d} \epsilon\,K_{g,z}^{(2)}
 =2e^2\mathcal{H} \operatorname{Im} \left[
 s(\omega,0)\frac{2\alpha_{\mathrm{so}}^2/z-1/(p_F^2\tau)}{z^2+\Delta_R^2}\right].}
 \label{eq:integrated-gain}
\end{equation}
The contribution proportional to $1/(p_F^2\tau)$ is fixed by the complete radial and output-vertex expansion. It is not obtained by solving a local relaxation-time equation on a frozen Fermi circle.

\subsubsection{Spin density, magnetic moment, and final coefficient}

The physical spin and its magnetic moment per unit area are defined by
\begin{equation}
 S_z=-\frac{\mathrm{i} }{2}\int\!\frac{\mathrm{d}^2p}{(2\pi)^2} \int\frac{\mathrm{d} \epsilon}{2\pi}
 \operatorname{Tr} [(\mathbf{c} \cdot\boldsymbol{\sigma} )G^<],\qquad M_{\mathrm{spin},z}=-g\mu_B S_z,
 \label{eq:observables}
\end{equation}
where $\mu_B>0$ and $g$ is the effective spin $g$ factor. In the equilibrium-spectral projection underlying the kinetic reduction,
\begin{equation}
 G_{\rm hel}^{<,(2)}=\tfrac12(G_0^Rh_{\rm hel}^{(2)}-h_{\rm hel}^{(2)}G_0^A),\qquad
 \int\!\frac{\mathrm{d}^2p}{(2\pi)^2}  G_{\rm hel}^{<,(2)}=-\mathrm{i} \pi\nu_0\tau K_{g,z}^{(2)}\mathbf{c} \cdot\boldsymbol{\sigma} .
 \label{eq:projected-lesser}
\end{equation}
It follows that
\begin{equation}
 S_z^{(2)}=-\frac{\nu_0\tau}{2}\int\mathrm{d} \epsilon\,K_{g,z}^{(2)},\qquad
 M_{\mathrm{spin},z}^{(2)}=\frac{g\mu_B\nu_0\tau}{2}\int\mathrm{d} \epsilon\,K_{g,z}^{(2)}.
 \label{eq:gain-observable}
\end{equation}
The diagrammatic calculation below independently fixes this normalization and reproduces the same single-splitting metallic coefficient without identifying a general nonlinear density with $h^{(2)}$ alone.

Define the response function
\begin{equation}
 \mathcal C(\omega)=\operatorname{Im} \left[
 \frac{1-w}{1-v}\frac{2\alpha_{\mathrm{so}}^2/z-1/(p_F^2\tau)}{z^2+\Delta_R^2}\right],\qquad
 w=\frac1{z\tau},\quad u=\frac{z}{\tau(z^2+\Delta_R^2)},\quad v=\frac{u+w}{2}.
 \label{eq:Cdefinition}
\end{equation}
The spin inverse Faraday response in the stated approximation is
\begin{equation}
 \boxed{\displaystyle
 \mathbf S_{\rm IFE}^{(2)}=-e^2\nu_0\mathcal C(\omega)\,
 \mathrm{i} (\mathbf{E} \times\mathbf{E} ^*),\qquad
 \mathbf M_{\mathrm{spin},\rm IFE}^{(2)}=g\mu_B e^2\nu_0\mathcal C(\omega)\,
 \mathrm{i} (\mathbf{E} \times\mathbf{E} ^*).}
 \label{eq:final-spin}
\end{equation}
Equivalently, in terms of the dimensionless ladder functions,
\begin{equation}
 \mathbf M_{\mathrm{spin},\rm IFE}^{(2)}
 =\frac{e^2g\mu_B\nu_0\tau}{p_F^2}
 \operatorname{Im} \left[\frac{1-w}{1-v}\,wu
 \left(\frac{\Delta_R^2\tau^2}{2}w-1\right)\right]
 \mathrm{i} (\mathbf{E} \times\mathbf{E} ^*).
 \label{eq:final-wuv}
\end{equation}
For $\mathbf{E} =E_0(\mathbf e_x+\mathrm{i} \ell\mathbf e_y)/\sqrt2$, with $\mathbf e_x\times\mathbf e_y=\mathbf{c} $, $\mathcal{H} =\ell|E_0|^2$. If instead the optical amplitude is defined by $\mathbf E(t)=\operatorname{Re} [\boldsymbol{\mathcal E}e^{-\mathrm{i} \omega t}]$, then $\mathbf{E} =\boldsymbol{\mathcal E}/2$ and the right-hand side contains $\mathrm{i} (\boldsymbol{\mathcal E}\times\boldsymbol{\mathcal E}^*)/4$.

\section{Orbital IFE from the kinetic equation}
\label{sec:orb-kinetics}
\subsection{Why the output momentum must be retained}
The spin density is a local observable whose helicity response exists at $\mathbf Q=0$. Orbital magnetization is extracted here from Eq.~\eqref{orb:eq:curl}; its bulk current vanishes at $\mathbf Q=0$. Its coefficient is therefore a first-gradient response. Set $\Omega=0$, keep $\mathbf Q$ through the calculation, and extract the transverse coefficient linear in $Q$. For this local coefficient choose
\begin{equation}
 \omega_s=s\omega,\qquad \mathbf q_s=\mathbf Q/2,\qquad
 \mathbf E_+=\mathbf E,\quad\mathbf E_- =\mathbf E^*,\qquad s=\pm.
 \label{eq:orb-routing}
\end{equation}
This is a convenient routing of the slowly varying optical product, not a claim that two conjugate plane waves have identical wavevectors. For a real envelope, the response kernel at each output $Q$ pairs a positive-frequency Fourier component with a negative-frequency component whose sum is $Q$. The relative optical momentum in Eq.~\eqref{eq:optical} is set to zero for the constitutive coefficient.

\subsection{Charge conservation and the optical magnetic field}
The first-order density and current obey
\begin{equation}
 -\mathrm{i} \omega_s e n_s^{(1)}+\mathrm{i} \mathbf q_s\cdot\mathbf j_s^{(1)}=0,
 \qquad \mathbf j_s^{(1)}=\sigma(s\omega)\mathbf E_s,
 \qquad \sigma(-\omega)=\sigma(\omega)^*.
 \label{eq:linear-continuity-orb}
\end{equation}
Thus $n_s^{(1)}=\mathbf q_s\cdot\mathbf j_s^{(1)}/(e\omega_s)$. Its charge ladder must conserve the isotropic mode; replacing the full collision operator by $-\Gamma\hat w$ would violate this identity.

Faraday's law in the Fourier convention of Eq.~\eqref{eq:optical} is
\begin{equation}
 \boldsymbol\nabla\times\mathbf E=-\frac1{c_0}\partial_t\mathbf B
 \quad\Longrightarrow\quad
 \mathrm{i} \mathbf q_s\times\mathbf E_s=\frac{\mathrm{i} \omega_s}{c_0}\mathbf B_s,
 \qquad \mathbf B_s=\frac{c_0}{\omega_s}\mathbf q_s\times\mathbf E_s.
 \label{eq:Faraday-explicit}
\end{equation}
For Eq.~\eqref{eq:orb-routing},
\begin{equation}
 \mathbf B_+=\frac{c_0}{2\omega}\mathbf Q\times\mathbf E,
 \qquad \mathbf B_-=-\frac{c_0}{2\omega}\mathbf Q\times\mathbf E^*.
 \label{eq:Bpm}
\end{equation}
The minus sign is the negative optical frequency; the factor $1/2$ is the momentum routing. $B$ is of first order in $Q$, exactly the order needed to extract the magnetization. Setting it to zero because it disappears at $Q=0$ would remove a finite part of that coefficient. It enters through the Lorentz force, although the Hamiltonian has no direct Zeeman term.

\subsection{A momentum moment that retains anomalous Rashba velocity}
Define the physical moments of the lesser density of Eq.~\eqref{eq:lesser}:
\begin{align}
 n&=\int\mathrm{d} \epsilon\int\!\frac{\mathrm{d}^2p}{(2\pi)^2} \operatorname{Tr} \hat w,&
 P_i&=\int\mathrm{d} \epsilon\int\!\frac{\mathrm{d}^2p}{(2\pi)^2}  p_i\operatorname{Tr} \hat w,\nonumber\\
 T_{ij}&=\int\mathrm{d} \epsilon\int\!\frac{\mathrm{d}^2p}{(2\pi)^2}  p_i\operatorname{Tr} (\hat v_j\hat w),&
 j_i&=e\int\mathrm{d} \epsilon\int\!\frac{\mathrm{d}^2p}{(2\pi)^2} \operatorname{Tr} (\hat v_i\hat w)
 =e(P_i/m+\mathcal V_i),\nonumber\\
 \mathcal V_i&=\int\mathrm{d} \epsilon\int\!\frac{\mathrm{d}^2p}{(2\pi)^2} \operatorname{Tr} (\alpha_{\mathrm{so}} \hat{\mathsf R}_i\hat w).
 \label{eq:orb-moments}
\end{align}
$P_i$ is mechanical momentum density, $T_{ij}$ its flux, and $\mathcal V_i$ the anomalous-velocity contribution. With SOC, $j_i/e$ cannot be replaced by $P_i/m$.

For clarity the dc collision closure used by the moment reduction is specified explicitly. Put
\begin{equation}
 \mathcal A_\epsilon^{\rm on}(\mathbf{p} )=\delta(\epsilon-\hat H_0(\mathbf{p} )),\quad
 \mathcal W_\epsilon=\int\!\frac{\mathrm{d}^2p}{(2\pi)^2} \hat w_{\mathbf{p} \epsilon},\qquad
 C_\epsilon[\hat w]=-\Gamma\hat w_{\mathbf{p} \epsilon}
 +\frac{\Gamma}{2\nu_0}\{\mathcal A_\epsilon^{\rm on}(\mathbf{p} ),\mathcal W_\epsilon\}.
 \label{eq:projected-collision-orb}
\end{equation}
$\mathcal A^{\rm on}$ is normalized without $2\pi$, unlike $\mathcal A$ in Eq.~\eqref{eq:lesser}. This is the equilibrium-spectral, on-shell Born gain/loss block at a static output. It is the static Born collision block used in the metallic spectral calculation below. The diagram calculation retains the matching metallic radial terms.

When both branches are available at energy $\epsilon$, their spectral moments are
\begin{equation}
 \int\!\frac{\mathrm{d}^2p}{(2\pi)^2} \mathcal A_\epsilon^{\rm on}=\nu_0\mathbf{1} ,
 \qquad \int\!\frac{\mathrm{d}^2p}{(2\pi)^2}  p_i\mathcal A_\epsilon^{\rm on}=-m\nu_0\alpha_{\mathrm{so}} \hat{\mathsf R}_i.
 \label{eq:spectral-moments-orb}
\end{equation}
For example, $\langle n_x\boldsymbol\eta\cdot\boldsymbol{\sigma} \rangle=\hat{\mathsf R}_x/2$ gives
\begin{align}
 \int\!\frac{\mathrm{d}^2p}{(2\pi)^2}  p_x\mathcal A_\epsilon^{\rm on}
 &=\frac{\hat{\mathsf R}_x}{4}\sum_\lambda\lambda\nu_\lambda(\epsilon)p_\lambda(\epsilon),\nonumber\\
 \sum_\lambda\lambda\nu_\lambda p_\lambda
 &=\nu_0\sum_\lambda\lambda
 \left(1-\lambda\frac{m\alpha_{\mathrm{so}}}{p_0}\right)(p_0-\lambda m\alpha_{\mathrm{so}})
 =-4m\alpha_{\mathrm{so}}\nu_0.
 \label{eq:spectral-momentum-worked}
\end{align}
The nonzero moment is a contour-weight effect. Equalizing the branch weights prematurely removes it.

The density, spin, and momentum moments of the collision block can now be evaluated without guessing a relaxation time for the current. First integrate over momentum:
\begin{align}
 \int\!\frac{\mathrm{d}^2p}{(2\pi)^2}  C_\epsilon
 &=-\Gamma\mathcal W_\epsilon+
 \frac{\Gamma}{2\nu_0}
 \left\{\int\!\frac{\mathrm{d}^2p}{(2\pi)^2} \mathcal A_\epsilon^{\rm on},\mathcal W_\epsilon\right\}\nonumber\\
 &=-\Gamma\mathcal W_\epsilon+
 \frac{\Gamma}{2}\{\mathbf{1} ,\mathcal W_\epsilon\}=0.
 \label{eq:collision-density-cancel}
\end{align}
Thus loss and gain exactly cancel in the integrated matrix density at this static output. Taking its trace gives number conservation; taking its spin trace gives zero impurity torque on the total spin. Neither statement says that the momentum moment vanishes.

The loss contribution to that moment is
\begin{equation}
 F_{{\rm imp},i}^{\rm loss}
 =-\Gamma\int\mathrm{d} \epsilon\int\!\frac{\mathrm{d}^2p}{(2\pi)^2}  p_i\operatorname{Tr} \hat w
 =-\Gamma P_i.
 \label{eq:collision-loss-momentum}
\end{equation}
The gain contribution is obtained by doing its spectral momentum integral first:
\begin{align}
 F_{{\rm imp},i}^{\rm gain}
 &=\frac{\Gamma}{2\nu_0}\int\mathrm{d} \epsilon\,
 \operatorname{Tr} \left\{\int\!\frac{\mathrm{d}^2p}{(2\pi)^2}  p_i\mathcal A_\epsilon^{\rm on},\mathcal W_\epsilon\right\}\nonumber\\
 &=-\frac{m\Gamma}{2}\int\mathrm{d} \epsilon\,
 \operatorname{Tr} \{\alpha_{\mathrm{so}} \hat{\mathsf R}_i,\mathcal W_\epsilon\}\nonumber\\
 &=-m\Gamma\int\mathrm{d} \epsilon\,\operatorname{Tr} (\alpha_{\mathrm{so}} \hat{\mathsf R}_i\mathcal W_\epsilon)
 =-m\Gamma\mathcal V_i.
 \label{eq:collision-gain-momentum}
\end{align}
Cyclicity gives $\operatorname{Tr} \{X,Y\}=2\operatorname{Tr} (XY)$. The spectral moment in Eq.~\eqref{eq:spectral-moments-orb} supplies its sign and factor of $m$. Combining gain and loss gives
\begin{equation}
 \boxed{F_{{\rm imp},i}
 =-\Gamma(P_i+m\mathcal V_i)
 =-\frac{m\Gamma}{e}j_i.}
 \label{eq:impurity-current-moment}
\end{equation}
The current is $j_i=e(P_i/m+\mathcal V_i)$, with $\mathcal V_i$ the anomalous Rashba velocity moment. A loss-only collision operator would produce $-\Gamma P_i$ and miss the $-m\Gamma\mathcal V_i$ term. This gain contribution is the kinetic counterpart of dressing the output current vertex by its impurity ladder.

The equilibrium-spectral Born block above is the collision closure used in this moment calculation. Appendix~\ref{app:nonlinear} states its relation to the full quadratic spectral bookkeeping. No additional assumption that setting output frequency to zero removes collisions is made.

\subsection{Momentum flux and the stationary spin Ward identity}
Taking the momentum moment of the gauge-covariant equation gives
\begin{equation}
 \partial_tP_i+\partial_jT_{ij}
 =enE_i+\frac{(\mathbf j\times\mathbf B)_i}{c_0}+F_{{\rm imp},i}.
 \label{eq:momentum-balance-orb}
\end{equation}
The momentum-space electric derivative is integrated by parts; its derivative of $p_i$ produces $enE_i$. The magnetic derivative produces the full velocity inside $\mathbf j\times\mathbf B$. Energy-derivative terms are boundary terms after the energy integral. These operations are shown before energy integration in Appendix~\ref{app:scalar}.

At the static quadratic output, the time derivative is zero but the impurity force remains. Fourier transforming $\partial_jT_{ij}$ gives $\mathrm{i}  Q_jT_{ij}^{(2)}$. The complete mixed force has one first-order density or current and the other optical field. Substituting Eq.~\eqref{eq:impurity-current-moment}, moving its negative current term to the left, and multiplying the entire equation by $e/m$ gives
\begin{equation}
 \Gamma j_i^{(2)}+\frac{\mathrm{i}  e}{m}Q_jT_{ij}^{(2)}
 =\frac{e^2}{m}\sum_s n_{-s}^{(1)}E_{s,i}
 +\frac{e}{mc_0}\sum_s(\mathbf j_{-s}^{(1)}\times\mathbf B_s)_i.
 \label{eq:orb-ward-quadratic}
\end{equation}
Only the uniform helicity part of $T_{ij}^{(2)}$ is needed in its term already proportional to $Q$. Rotation and vertical-mirror symmetry restrict it to $t_H\varepsilon^{(2)}_{ij}\mathcal{H} $. Its antisymmetric component obeys
\begin{align}
 T_{xy}-T_{yx}
 &=\alpha_{\mathrm{so}}\int\mathrm{d} \epsilon\int\!\frac{\mathrm{d}^2p}{(2\pi)^2} \operatorname{Tr} [(p_x\sigma_x+p_y\sigma_y)\hat w],\nonumber\\
 \mathrm{i} [H_0,\sigma_z/2]&=\alpha_{\mathrm{so}}(p_x\sigma_x+p_y\sigma_y),\nonumber\\
 \partial_tS_z+\partial_jJ_j^{S_z}&=T_{xy}-T_{yx}.
 \label{eq:spin-stress-ward}
\end{align}
Scalar disorder has no total-spin torque. At uniform stationary output the last left-hand side is zero, so $t_H=0$ and $T_{ij}^{(2),H}(0)=0$. For $\alpha_{\mathrm{so}}=0$ the flux is symmetric directly. This cancellation uses the Rashba spin equation and the specified symmetries; it does not follow merely from $\partial_{p_j}v_i=\delta_{ij}/m$.

\subsection{Electric density force and magnetic Lorentz force: the two halves}
Insert charge continuity and Faraday's law into Eq.~\eqref{eq:orb-ward-quadratic}. With Eq.~\eqref{eq:orb-routing}, the electric part is
\begin{equation}
 \mathbf j_E^{(2)}=\frac{e}{2m\Gamma\omega}
 [\sigma\mathbf E^*(\mathbf Q\cdot\mathbf E)
 -\sigma^*\mathbf E(\mathbf Q\cdot\mathbf E^*)].
 \label{eq:jE-orb}
\end{equation}
The Lorentz part follows from $\mathbf a\times(\mathbf b\times\mathbf d)=\mathbf b(\mathbf a\cdot\mathbf d)-\mathbf d(\mathbf a\cdot\mathbf b)$:
\begin{equation}
 \mathbf j_B^{(2)}=\mathbf j_E^{(2)}
 +\frac{e}{2m\Gamma\omega}(\sigma^*-\sigma)
 \mathbf Q(\mathbf E\cdot\mathbf E^*).
 \label{eq:jB-orb}
\end{equation}
The extra term is longitudinal and helicity even. It is irrelevant to the local orbital moment but has not been silently discarded from the current.

The antisymmetric polarization product is
\begin{equation}
 [E_iE_j^*]_H=-\frac{\mathrm{i} }{2}\varepsilon^{(2)}_{ij}\mathcal{H} ,
 \qquad [E_i^*E_j]_H=\frac{\mathrm{i} }{2}\varepsilon^{(2)}_{ij}\mathcal{H} .
 \label{eq:helicity-tensor-orb}
\end{equation}
Therefore the two forces contribute equally to the transverse helicity current:
\begin{equation}
 j_{E,i}^{(2),H}=j_{B,i}^{(2),H}
 =\mathrm{i} \varepsilon^{(2)}_{i\ell}Q_\ell\frac{e\operatorname{Re} \sigma}{2m\Gamma\omega}\mathcal{H} ,
 \qquad
 M_{E,z}=M_{B,z}=\frac{e}{2mc_0\Gamma\omega}\operatorname{Re} \sigma\,\mathcal{H} .
 \label{eq:EB-halves-orb}
\end{equation}
Summing the two force contributions gives
\begin{equation}
 \boxed{M_{\mathrm{orb},z}=\frac{e}{mc_0\Gamma\omega}
 \operatorname{Re} \sigma(\omega)\,\mathcal{H} .}
 \label{eq:orb-kinetic-final}
\end{equation}
The factor of two is the sum of the electric density force and magnetic Lorentz force. It is not an extra spin degeneracy or an optical-ordering convention. Section~\ref{sec:orb-diagrams} derives the same coefficient from the actual differentiated loop traces, without imposing Eq.~\eqref{eq:orb-kinetic-final} on the diagrams.

\subsection{The SOC-free benchmark}
Putting $\alpha_{\mathrm{so}}=0$ in the independent conductivity gives
\begin{equation}
 \sigma_0(\omega)=\frac{n_0e^2}{m(\Gamma-\mathrm{i} \omega)},\quad
 n_0=\frac{p_F^2}{2\pi},\qquad
 \boxed{M_{\mathrm{orb},z}^{(0)}
 =\frac{n_0e^3}{m^2c_0\omega(\Gamma^2+\omega^2)}\mathcal{H} .}
 \label{eq:orb-normal-final}
\end{equation}
Each kinetic force supplies one half of this result. The scalar phase-space calculation in Appendix~\ref{app:scalar} resolves the density harmonic, drift harmonic, and traceless stress harmonic explicitly. In particular, it shows why relaxing the density at rate $\Gamma$ or freezing a radial numerator can spoil this elementary benchmark.
\section{Spin IFE from diagrams}
The dimensionless functions used in this section have fixed physical meanings:
\begin{equation}
 w(\omega)=\frac{\Gamma}{\Gamma-\mathrm{i} \omega},\qquad
 u(\omega)=\frac{\Gamma(\Gamma-\mathrm{i} \omega)}{(\Gamma-\mathrm{i} \omega)^2+\Delta_R^2},\qquad
 v(\omega)=\frac{w+u}{2},\qquad s(\omega)=\frac{1-w}{1-v}.
\end{equation}
Here $w$ is the charge-channel ladder eigenvalue, $u$ the out-of-plane-spin eigenvalue, $v$ the in-plane-spin eigenvalue, and $s$ the dressing of the anomalous optical velocity. The scalar $v$ is always a ladder eigenvalue; velocities carry a Cartesian index or a bold symbol. The argument $\omega$ is omitted only within the following calculation. These are the Fermi-level values $u(\omega)=u(\omega,0)$, $v(\omega)=v(\omega,0)$, and $s(\omega)=s(\omega,0)$. Their energy-dependent extensions are defined in Eqs.~\eqref{eq:u-radial}--\eqref{eq:optical-ladder-energy-expansion}.

\label{sec:diagrammatic}

We evaluate the same single-splitting metallic response using equilibrium propagators and noncrossing impurity ladders. The two analytic sectors and the optical exchange are kept separately until the last step. The scalar disorder broadening, radial expansion, and field amplitudes are identical to those in the kinetic calculation.

\subsection{Electromagnetic vertices and magnetic-moment normalization}

It is convenient to use $\mathbf a=\mathbf A/c_0$. Expanding Eq.~\eqref{eq:model} at zero scalar potential gives
\begin{equation}
 H[\mathbf a]=H_0-e a_i v_i+\frac{e^2}{2m}\mathbf a^2\mathbf{1} ,
 \qquad \mathbf E_\omega=\mathrm{i} \omega\mathbf a_\omega.
 \label{eq:em-vertices}
\end{equation}
Thus the linear Hamiltonian vertex is $-e\hat v_i$, and the second derivative is $e^2\delta_{ij}\mathbf{1} /m$. The latter is symmetric in the two optical indices and does not contribute to the helicity-odd spin response. In Gaussian units,
\begin{equation}
 \frac{A_{+,j}A_{-,k}}{c_0^2}
 =a_{+,j}a_{-,k}=\frac{E_jE_k^*}{\omega^2}.
 \label{eq:AtoE}
\end{equation}
There is therefore no residual $c_0^{-2}$ in a susceptibility expressed in terms of the physical electric field. The spin measurement vertex is $\boldsymbol{\sigma} /2$, while the magnetic-moment vertex is $-g\mu_B\boldsymbol{\sigma} /2$.

\begin{figure}[tb]
 \centering
 \includegraphics[width=0.65\textwidth]{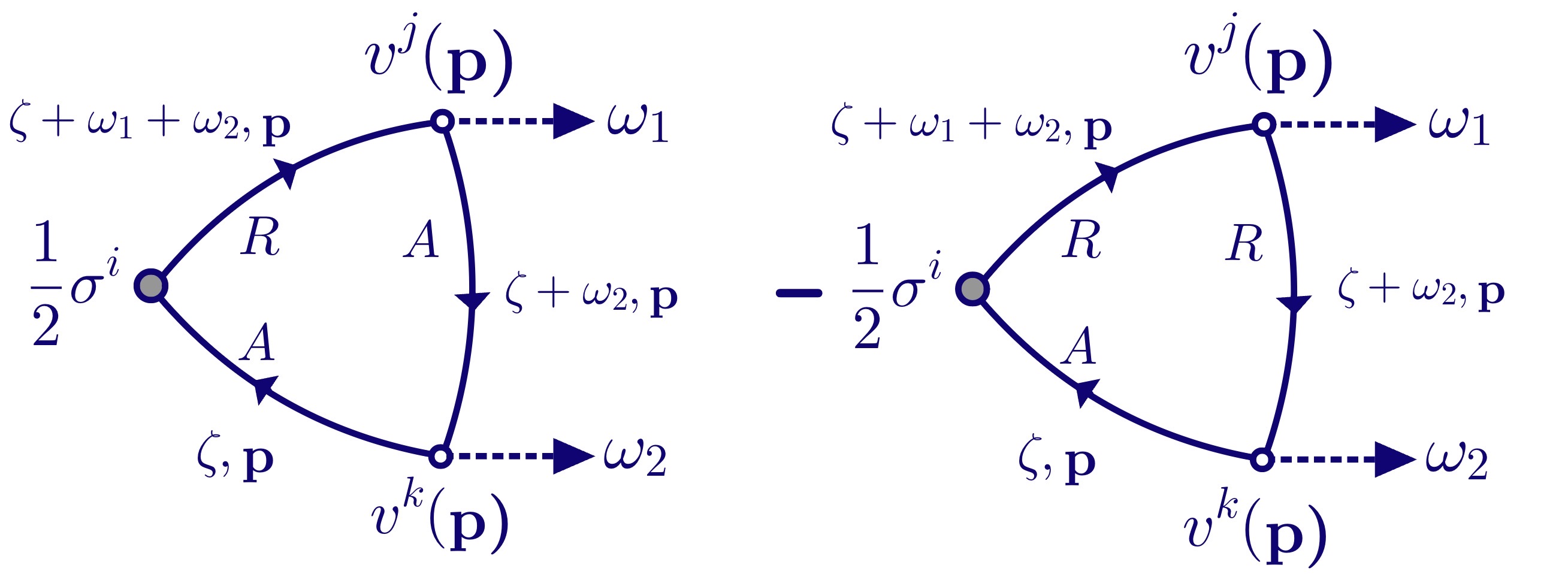}
 \caption{The $RAA$ and $RRA$ sectors for one optical ordering of the spin response. Their relative minus sign is fixed by analytic continuation. The displayed $\sigma^i/2$ is the spin vertex; multiplication by $-g\mu_B$ gives the magnetic-moment vertex. The dc limit has $\omega_1+\omega_2=0$. The exchanged optical ordering must also be included. In the calculation the optical vertex joining an $RA$ pair and the output spin vertex acquire the ladders derived below.}
 \label{fig:spin-triangles}
\end{figure}

\subsection{Matsubara continuation and the two mixed analytic sectors}

For clarity, first suppress the matrix vertices. Let $\zeta_n=(2n+1)\pi T$ be a fermionic Matsubara frequency and $\Omega_{1,2}$ bosonic Matsubara frequencies. Consider
\begin{equation}
 \mathscr S(\mathrm{i} \Omega_1,\mathrm{i} \Omega_2)
 =T\sum_nG_1(\mathrm{i} \zeta_n)G_2(\mathrm{i} \zeta_n-\mathrm{i} \Omega_1)
 G_3(\mathrm{i} \zeta_n-\mathrm{i} \Omega_1-\mathrm{i} \Omega_2).
 \label{eq:matsubara-triangle}
\end{equation}
The contour is initially written at an auxiliary positive temperature, with $f_{\rm F}(\zeta)=(e^{\zeta/T}+1)^{-1}$ to specify its Fermi poles; after continuation we take $T\to0$ and use $f_0(\epsilon)=\Theta(-\epsilon)$ in every real-energy integral. The contour surrounding the Fermi poles is deformed onto the three Green-function cuts. With the contour orientation chosen so that $T\sum_nH(\mathrm{i} \zeta_n)=-\oint\mathrm{d} \zeta\,f_{\rm F}(\zeta)H(\zeta)/(2\pi\mathrm{i} )$, this gives
\begin{align}
 \mathscr S={}&-\int\frac{\mathrm{d} \epsilon}{2\pi\mathrm{i} }f_0(\epsilon)
 [G_1^R(\epsilon)-G_1^A(\epsilon)]G_2(\epsilon-\mathrm{i} \Omega_1)G_3(\epsilon-\mathrm{i} \Omega_1-\mathrm{i} \Omega_2)\nonumber\\
 &-\int\frac{\mathrm{d} \epsilon}{2\pi\mathrm{i} }f_0(\epsilon)
 G_1(\epsilon+\mathrm{i} \Omega_1)[G_2^R(\epsilon)-G_2^A(\epsilon)]G_3(\epsilon-\mathrm{i} \Omega_2)\nonumber\\
 &-\int\frac{\mathrm{d} \epsilon}{2\pi\mathrm{i} }f_0(\epsilon)
 G_1(\epsilon+\mathrm{i} \Omega_1+\mathrm{i} \Omega_2)G_2(\epsilon+\mathrm{i} \Omega_2)
 [G_3^R(\epsilon)-G_3^A(\epsilon)].
 \label{eq:three-cuts}
\end{align}
Continue $\mathrm{i} \Omega_{1,2}\to\omega_{1,2}+\mathrm{i} 0^+$ before imposing $\omega_1=\omega$, $\omega_2=-\omega$. Shifting the middle-cut energy by $\omega$, and writing all products in their original matrix order, yields
\begin{align}
 \mathscr S^R={}&-\int\frac{\mathrm{d} \epsilon}{2\pi\mathrm{i} }f_0(\epsilon)
 \big[G_1^R(\epsilon)G_2^R(\epsilon-\omega)G_3^R(\epsilon)
 -G_1^A(\epsilon)G_2^A(\epsilon-\omega)G_3^A(\epsilon)\big]\nonumber\\
 &+\int\frac{\mathrm{d} \epsilon}{2\pi\mathrm{i} }[f_0(\epsilon)-f_0(\epsilon-\omega)]
 \big[G_1^R(\epsilon)G_2^R(\epsilon-\omega)G_3^A(\epsilon)
 -G_1^R(\epsilon)G_2^A(\epsilon-\omega)G_3^A(\epsilon)\big].
 \label{eq:continued-triangle}
\end{align}
The $RRR$ and $AAA$ radial integrals vanish in the extended-band approximation when the contour can be closed in the empty half plane and the integrand has no arc contribution. This step is a property of the present radial approximation, not a statement about an arbitrary finite band. The mixed sectors remain, with their relative minus sign.

The overall observable sign can be checked directly using $G^<_0=f(G^A_0-G^R_0)$. For a fixed optical ordering, the mixed part of the lesser correction is
\begin{equation}
 G^{<,(2)}_{\rm mixed}
 =[f_0(\epsilon-\omega)-f_0(\epsilon)]
 G^R_\epsilon V_+\big(G^A_{\epsilon-\omega}-G^R_{\epsilon-\omega}\big)V_-G^A_\epsilon.
 \label{eq:lesser-triangle}
\end{equation}
Equations~\eqref{eq:observables} and \eqref{eq:lesser-triangle} fix the signs in the formulas below independently of the diagram labels.

\subsection{Impurity tensor, optical vertex, and output spin vertex}

The same $RA$ map as in Eq.~\eqref{eq:ladder} appears in the Bethe--Salpeter equation. Its action on spin matrices is diagonal in the scalar, out-of-plane-spin, and in-plane-spin sectors. To exhibit the spin-index structure, introduce the projectors
\begin{equation}
 \mathscr P_0[X]=\tfrac12\operatorname{Tr}  X\,\mathbf{1} ,\qquad
 \mathscr P_z[X]=\tfrac12\operatorname{Tr} [(\mathbf{c} \cdot\boldsymbol{\sigma} )X]\,\mathbf{c} \cdot\boldsymbol{\sigma} ,\qquad
 \mathscr P_\parallel[X]=\tfrac12\sum_{i=x,y}\operatorname{Tr} [\sigma_iX]\sigma_i.
 \label{eq:super-projectors}
\end{equation}
They obey $\mathscr P_a\mathscr P_b=\delta_{ab}\mathscr P_a$ and sum to the identity map on $2\times2$ matrices. The elementary impurity rung and its geometric sum are
\begin{equation}
 \mathscr T=w\mathscr P_0+u\mathscr P_z+v\mathscr P_\parallel,\qquad
 \mathscr D=(1-\mathscr T)^{-1}
 =\frac{\mathscr P_0}{1-w}+\frac{\mathscr P_z}{1-u}
 +\frac{\mathscr P_\parallel}{1-v}.
 \label{eq:full-ladder-map}
\end{equation}
For example, $(\mathscr P_z)_{ab;cd}=\tfrac12(\mathbf{c} \cdot\boldsymbol{\sigma} )_{ab}(\mathbf{c} \cdot\boldsymbol{\sigma} )_{dc}$; the reversed indices in the second factor specify the trace contraction. This removes any ambiguity associated with tensor-product conventions.

The angular factor for one rung is
\begin{equation}
 \left\langle P_\mu\otimes P_\nu\right\rangle_{\phi}
 =\frac14\left[\mathbf{1} \otimes\mathbf{1} 
 +\frac{\mu\nu}{2}\sum_{i=x,y}\sigma_i\otimes\sigma_i\right]
 \label{eq:rung-angle}
\end{equation}
in the direct index pairing. Changing to the input--output pairing uses the Pauli completeness (Fierz) identity
\begin{equation}
 \delta_{ab}\delta_{cd}=\frac12\sum_{\rho=0,x,y,z}(\sigma_\rho)_{ad}(\sigma_\rho)_{cb}.
 \label{eq:fierz}
\end{equation}
Together with $Q_{++}+Q_{--}=2w$ and $Q_{+-}+Q_{-+}=2u$ at the retained order, this reproduces Eq.~\eqref{eq:full-ladder-map}, with $v=(u+w)/2$. The energy in $u$ is the mean energy of the pair, as in Eq.~\eqref{eq:routing}.

\begin{figure}[tb]
 \centering
 \includegraphics[width=0.45\textwidth]{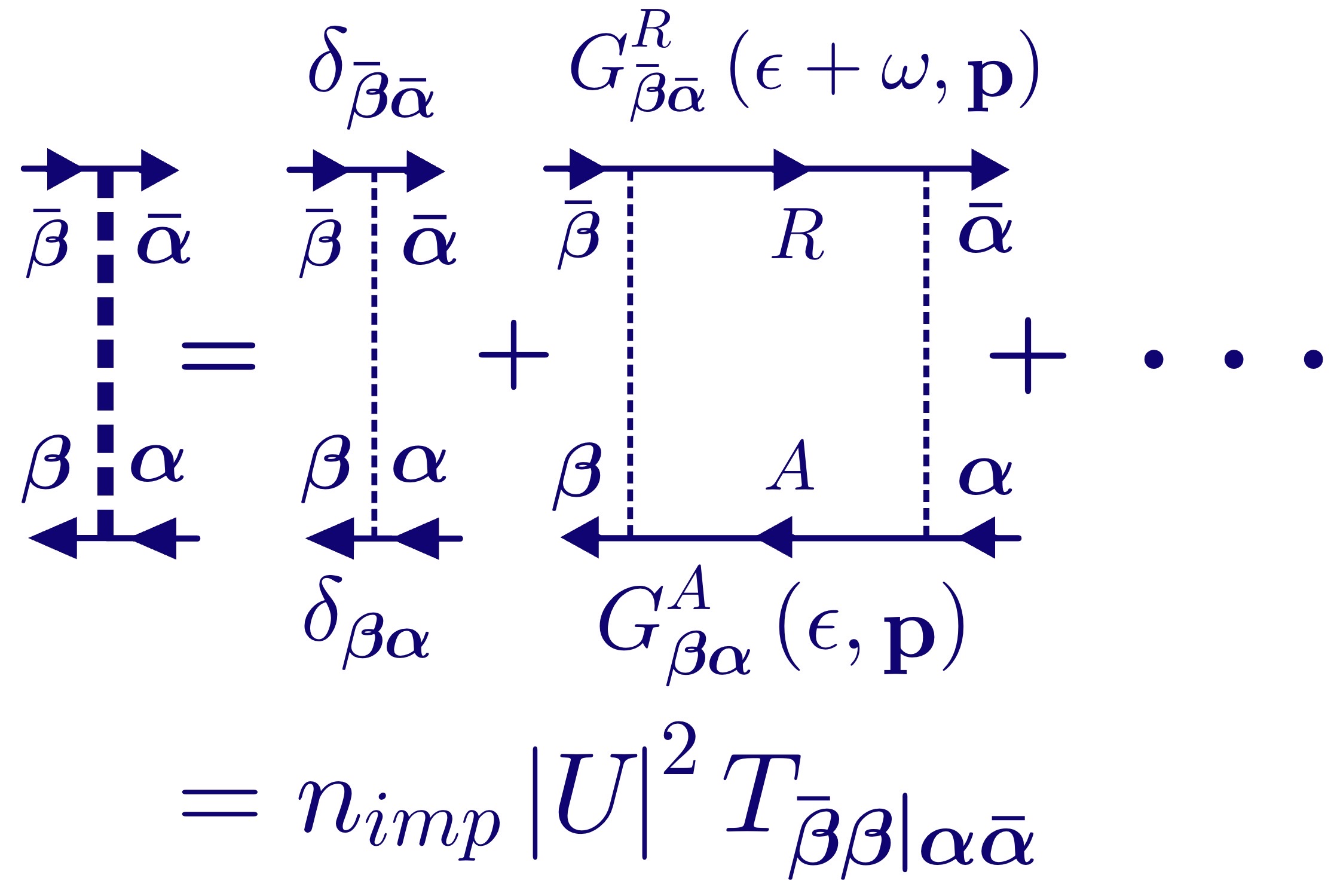}
 \caption{Noncrossing impurity ladder. The thin dashed line is one disorder correlator $n_{\rm imp}|U|^2=\gamma_{\rm imp}$, and the thick dashed line denotes its repeated $RA$ scattering sum. The kernel is diagonal in the three spin sectors of Eq.~\eqref{eq:full-ladder-map}. The two propagator energies shown here are equivalent to the symmetric routing after an energy shift.}
 \label{fig:impurity-ladder}
\end{figure}

The optical vertex obeys
\begin{equation}
 \widetilde v_i=v_i+\gamma_{\rm imp}\int\!\frac{\mathrm{d}^2p}{(2\pi)^2}  G_+^R\widetilde v_iG_-^A.
 \label{eq:optical-BS}
\end{equation}
Its momentum-independent correction has only in-plane spin structure. From Eq.~\eqref{eq:bare-source-ladder},
\begin{equation}
 \widetilde v_i-v_i=\frac{\alpha_{\mathrm{so}}(u-w)}{2(1-v)}(\mathbf{c} \times\boldsymbol{\sigma} )_i,
 \qquad
 \boxed{\widetilde v_i=\frac{p_i}{m}\mathbf{1} +\alpha_{\mathrm{so}}\frac{1-w}{1-v}(\mathbf{c} \times\boldsymbol{\sigma} )_i.}
 \label{eq:dressed-optical}
\end{equation}
This is precisely the kinetic source factor $s$. In an $RAA$ triangle only the first optical vertex joins an $RA$ pair; the other joins two advanced propagators. In the $RRA$ triangle the $RA$ optical vertex is the second one and carries $s(-\omega)=s(\omega)^*$ at the corresponding mean energy. Dressing both optical vertices by $s(\omega)$ in a single mixed triangle would change the specified diagram set.

The output out-of-plane-spin vertex obeys
\begin{equation}
 \widetilde\sigma_z=\sigma_z+\gamma_{\rm imp}\int\!\frac{\mathrm{d}^2p}{(2\pi)^2}  G_0^R\widetilde\sigma_zG_0^A
 =\Lambda_z(\epsilon)\sigma_z,\qquad\sigma_z=\mathbf{c} \cdot\boldsymbol{\sigma} .
 \label{eq:spin-BS}
\end{equation}
Since $\mathscr T_0[\sigma_z]=u(0,\epsilon)\sigma_z$, this is exactly Eq.~\eqref{eq:spin-ladder-energy}. In the variables used below,
\begin{equation}
 \Lambda_z(\epsilon)=1+\frac{\Gamma^2}{\Delta_R^2}
 -\frac{2m\Gamma^2}{p_F^2\Delta_R^2}\epsilon+\cdots,
 \qquad\Gamma=\tau^{-1}.
 \label{eq:Lambda-Gamma}
\end{equation}
The output frequency is zero; it must not be confused with the nonzero optical frequency in $s$.

\begin{figure}[tb]
 \centering
 \includegraphics[width=0.45\textwidth]{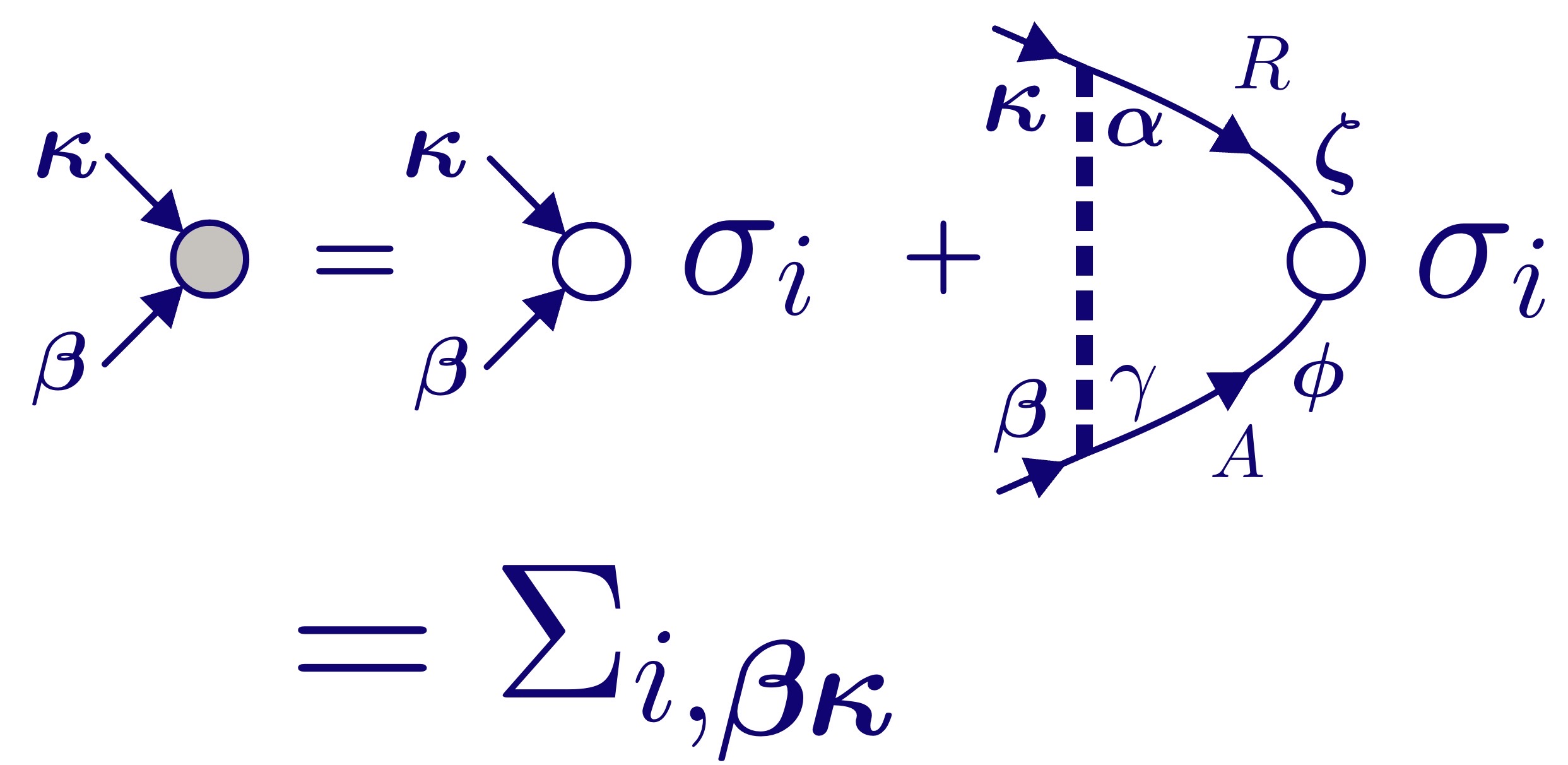}
 \caption{Dressing of the spin measurement vertex. The thick dashed line denotes the impurity ladder of Fig.~\ref{fig:impurity-ladder}. For the out-of-plane-spin component the result is $\widetilde\sigma_z=\sigma_z/[1-u(0,\epsilon)]$. The symbol $\Sigma_i$ in the original artwork denotes the dressed vertex, not the one-particle self-energy $\hat\Sigma^{R,A}$.}
 \label{fig:dressed-spin}
\end{figure}

\subsection{Angular trace and helicity selection}

Set $\widetilde\alpha_{\mathrm{so}}=\alpha_{\mathrm{so}} s$ for the $RAA$ optical vertex. The Pauli identities
\begin{equation}
 \operatorname{Tr} \sigma_i\sigma_j=2\delta_{ij},\quad
 \operatorname{Tr} \sigma_i\sigma_j\sigma_k=2\mathrm{i} \varepsilon_{ijk},\quad
 \operatorname{Tr} \sigma_i\sigma_j\sigma_k\sigma_l
 =2(\delta_{ij}\delta_{kl}-\delta_{ik}\delta_{jl}+\delta_{il}\delta_{jk})
 \label{eq:Pauli-traces}
\end{equation}
and $\left\langle n_in_j\right\rangle_{\phi}=(\delta_{ij}-c_ic_j)/2$ give the helicity-odd out-of-plane-spin trace
\begin{align}
 &\big[\left\langle \operatorname{Tr} [\sigma_zP_\mu\widetilde v_jP_\nu v_kP_\beta]\right\rangle_{\phi}\big]_{\rm hel}
 =\frac{\mathrm{i} \alpha_{\mathrm{so}}}{8}\varepsilon_{jk\ell}c_\ell
 \left[\frac{p}{m}(1+s)(1-\mu\beta)\nu
 +2\alpha_{\mathrm{so}} s(1-\mu\beta)+\frac{p}{m}(1-s)(\mu-\beta)\right].
 \label{eq:triangle-trace}
\end{align}
Both coefficients vanish when $\beta=\mu$. Thus the propagators adjacent to the spin vertex have opposite helicity. Setting $\beta=-\mu$ and separating the two possible middle helicities gives
\begin{align}
 M_z^{(a),RAA}={}&-\frac{g\mu_Be^2\alpha_{\mathrm{so}}\mathcal{H} }{4\omega^2}
 \int\frac{\mathrm{d} \epsilon}{2\pi\mathrm{i} }\,W_\omega(\epsilon)\Lambda_z(\epsilon)
 \sum_\mu\left[I_{1\mu}(\epsilon,\omega)+s\,I_{2\mu}(\epsilon,\omega)\right],
 \label{eq:RAA-radial}\\
 W_\omega(\epsilon)&=f_0(\epsilon-\omega)-f_0(\epsilon),\nonumber\\
 I_{1\mu}&=\int\!\frac{\mathrm{d}^2p}{(2\pi)^2} \left(\mu\frac p m+\alpha_{\mathrm{so}} s\right)
 g_\mu^R(\epsilon)g_\mu^A(\epsilon-\omega)g_{-\mu}^A(\epsilon),\label{eq:I1def}\\
 I_{2\mu}&=\int\!\frac{\mathrm{d}^2p}{(2\pi)^2} \left(-\mu\frac p m+\alpha_{\mathrm{so}}\right)
 g_\mu^R(\epsilon)g_{-\mu}^A(\epsilon-\omega)g_{-\mu}^A(\epsilon).
 \label{eq:I2def}
\end{align}
The angular measure in the last two expressions is unity after Eq.~\eqref{eq:triangle-trace}. In Eqs.~\eqref{eq:RAA-radial}--\eqref{eq:I2def}, $s$ initially means $s(\omega,\epsilon-\omega/2)$ and must be expanded consistently with the energy dependence of the remaining factors.

Before the helicity trace, the two mixed contributions read
\begin{align}
 M_z^{(a),RAA}={}&-\frac{g\mu_B e^2}{2\omega^2}
 E_jE_k^*\int\!\frac{\mathrm{d}^2p}{(2\pi)^2} \int\frac{\mathrm{d} \epsilon}{2\pi\mathrm{i} }W_\omega\Lambda_z
 \operatorname{Tr} [\sigma_zG_\epsilon^R\widetilde v_jG_{\epsilon-\omega}^Av_kG_\epsilon^A],\nonumber\\
 M_z^{(a),RRA}={}&+\frac{g\mu_B e^2}{2\omega^2}
 E_jE_k^*\int\!\frac{\mathrm{d}^2p}{(2\pi)^2} \int\frac{\mathrm{d} \epsilon}{2\pi\mathrm{i} }W_\omega\Lambda_z
 \operatorname{Tr} [\sigma_zG_\epsilon^Rv_jG_{\epsilon-\omega}^R\widetilde v_k^{\,-}G_\epsilon^A],
 \label{eq:two-triangles}
\end{align}
where $\widetilde v^{\,-}$ has the conjugate optical factor. These formulas display the magnetic-moment sign, the spin factor $1/2$, and the relative analytic-sector sign explicitly.

\subsection{Single-splitting metallic radial residues}

Define
\begin{equation}
 B_\mu=\mu\Delta_R-\mathrm{i} \Gamma,\qquad
 d_\mu=1-\mu\delta,\qquad
 L_\mu=d_\mu B_\mu+2\mu\delta\epsilon.
 \label{eq:triangle-residue-variables}
\end{equation}
Both $RAA$ integrals are evaluated by the retarded pole $x_R=\epsilon+\mathrm{i} \Gamma/2$ in $x=\xi_{p\mu}$. The numerator must be evaluated at this complex pole; its imaginary displacement contributes at the same order as the other retained radial terms. Using $p/m=v_F-\alpha_{\mathrm{so}}\mu+x/p_F+\cdots$, the first integral is
\begin{align}
 I_{1\mu}
 &=\frac{2\pi\mathrm{i} \nu_0}{\omega+\mathrm{i} \Gamma}
 \frac{d_\mu[\mu v_F-\alpha_{\mathrm{so}}+\alpha_{\mathrm{so}} s+\mu(\epsilon+\mathrm{i} \Gamma/2)/p_F]}{L_\mu}\nonumber\\
 &=\frac{2\pi\mathrm{i} \nu_0}{\omega+\mathrm{i} \Gamma}
 \left[\frac{\mu v_F-\alpha_{\mathrm{so}}+\alpha_{\mathrm{so}} s+\mu(\epsilon+\mathrm{i} \Gamma/2)/p_F}{B_\mu}
 -\frac{2\alpha_{\mathrm{so}}\epsilon}{B_\mu^2}\right]+\cdots.
 \label{eq:I1residue}
\end{align}
For the second integral one obtains
\begin{align}
 I_{2\mu}
 &=2\pi\mathrm{i} \nu_0\,
 \frac{d_\mu[\mu v_F-2\alpha_{\mathrm{so}}+\mu(\epsilon+\mathrm{i} \Gamma/2)/p_F]}{L_\mu(L_\mu-\omega)}\nonumber\\
 &=2\pi\mathrm{i} \nu_0\left[
 \frac{\mu v_F-2\alpha_{\mathrm{so}}+\mu(\epsilon+\mathrm{i} \Gamma/2)/p_F}{B_\mu(B_\mu-\omega)}
 +\frac{\alpha_{\mathrm{so}}}{(B_\mu-\omega)^2}
 -\frac{2\alpha_{\mathrm{so}}\epsilon(2B_\mu-\omega)}{B_\mu^2(B_\mu-\omega)^2}\right]+\cdots.
 \label{eq:I2residue}
\end{align}
The positive double-pole term in Eq.~\eqref{eq:I2residue} follows by expanding the Jacobian and both denominators together:
\begin{align}
 \frac{d_\mu}{L_\mu(L_\mu-\omega)}
 &=\frac1{B_\mu(B_\mu-\omega)}
 +\mu\delta\left[\frac1{(B_\mu-\omega)^2}
 -\frac{2\epsilon(2B_\mu-\omega)}{B_\mu^2(B_\mu-\omega)^2}\right]+\cdots,
 \label{eq:double-pole-expansion}
\end{align}
and multiplying its correction by the leading numerator $\mu v_F$. Terms quadratic in the small radial corrections are consistently omitted. Freezing the numerator, Jacobian, or opposite-helicity denominator separately would destroy this combination.

\subsection{Energy integral, output ladder, and conjugate sector}

At chemical-potential-referenced energy, the Fermi-window identities are
\begin{equation}
 \int\mathrm{d} \epsilon\,W_\omega=\omega,\qquad
 \int\mathrm{d} \epsilon\,\epsilon W_\omega=\frac{\omega^2}{2},\qquad
 \int\mathrm{d} \epsilon\,(\epsilon-\omega/2)W_\omega=0.
 \label{eq:Fermi-window}
\end{equation}
They hold for either sign of $\omega$. At $T=0$ and $\omega>0$, the integral is over $0<\epsilon<\omega$. The final identity shows why the linear mean-energy correction to the optical ladder has zero integral against an energy-independent leading kernel. Its products with additional radial corrections are beyond the displayed expansion. The output ladder depends on $\epsilon$, not on $\epsilon-\omega/2$, so its linear term in Eq.~\eqref{eq:Lambda-Gamma} survives.

One useful simplification of the $I_2$ terms is
\begin{equation}
 \frac{\alpha_{\mathrm{so}}}{(B_\mu-\omega)^2}
 -\frac{\alpha_{\mathrm{so}}\omega(2B_\mu-\omega)}{B_\mu^2(B_\mu-\omega)^2}
 =\frac{\alpha_{\mathrm{so}}}{B_\mu^2}.
 \label{eq:I2energy-simplification}
\end{equation}
After the energy integration let $s=s(\omega,0)$ and define
\begin{align}
 \mathcal B(\omega)=\sum_{\mu=\pm}\Bigg\{&
 \left(1+\frac{\Gamma^2}{\Delta_R^2}\right)
 \left[\frac{\mu v_F-\alpha_{\mathrm{so}}+\alpha_{\mathrm{so}} s+\mu(\omega+\mathrm{i} \Gamma)/(2p_F)}{(\omega+\mathrm{i} \Gamma)B_\mu}
 +\frac{s[\mu v_F-2\alpha_{\mathrm{so}}+\mu(\omega+\mathrm{i} \Gamma)/(2p_F)]}{(B_\mu-\omega)B_\mu}
 \right.\nonumber\\[-1mm]
 &\left.\hspace{17mm}-\frac{\alpha_{\mathrm{so}}\omega}{(\omega+\mathrm{i} \Gamma)B_\mu^2}
 +\frac{s\alpha_{\mathrm{so}}}{B_\mu^2}\right]
 -\frac{m\omega\Gamma^2\mu v_F}{p_F^2\Delta_R^2}
 \left[\frac1{(\omega+\mathrm{i} \Gamma)B_\mu}
 +\frac{s}{(B_\mu-\omega)B_\mu}\right]\Bigg\}.
 \label{eq:Bfull}
\end{align}
The last line outside the square bracket is the energy correction to the dc spin vertex. It has been inserted before the energy integral, as required by Eq.~\eqref{eq:Fermi-window}.

The resulting $RAA$ contribution and its $RRA$ partner are
\begin{equation}
 M_z^{(a),RAA}=-\frac{g\mu_B e^2\alpha_{\mathrm{so}}\nu_0}{4\omega}\mathcal{H} \mathcal B(\omega),\qquad
 M_z^{(a),RRA}=-\frac{g\mu_B e^2\alpha_{\mathrm{so}}\nu_0}{4\omega}\mathcal{H} \mathcal B(\omega)^*.
 \label{eq:conjugate-sectors}
\end{equation}
To obtain the second equality, reverse the matrix order under conjugation, relabel $\mu\to-\mu$, and use $s(-\omega)=s(\omega)^*$. Every denominator is conjugated: in particular $B_\mu-\omega$ becomes $B_\mu^*-\omega$, not $B_\mu^*+\omega$. The pair is real,
\begin{equation}
 M_z^{(a)}=-\frac{g\mu_B e^2\alpha_{\mathrm{so}}\nu_0}{2\omega}\mathcal{H} \operatorname{Re} \mathcal B(\omega).
 \label{eq:one-ordering}
\end{equation}

\subsection{Helicity sum and equality with the kinetic result}

The sum in Eq.~\eqref{eq:Bfull} can be performed using Eq.~\eqref{eq:helicity-sums}, partial fractions in $B_\mu(B_\mu-\omega)$, and $\Delta_R=2\alpha_{\mathrm{so}} p_F$. With $z=\Gamma-\mathrm{i} \omega$ and $D_F=z^2+\Delta_R^2$, its reduced form is
\begin{equation}
 \mathcal B(\omega)
 =-\frac{2\mathrm{i}  v_F}{\Delta_R z}
 +\frac{2\mathrm{i}  v_F s(\Gamma+z)}{\Delta_R D_F}
 +\frac{2\mathrm{i} \omega}{\Delta_Rp_F}
 \frac{s(\Delta_R^2/(2z)-\Gamma)}{D_F}.
 \label{eq:Breduced}
\end{equation}
The first two terms change to minus their complex conjugates when $\omega\to-\omega$. The third changes to its complex conjugate. The exchanged optical ordering is obtained by $\omega\to-\omega$ and $\mathbf{E} \leftrightarrow\mathbf{E} ^*$, so $\mathcal{H} \to-\mathcal{H} $. Combining both orderings in Eq.~\eqref{eq:one-ordering} gives
\begin{align}
 M_z^{(a)}+M_z^{(b)}
 &=-\frac{g\mu_B e^2\alpha_{\mathrm{so}}\nu_0\mathcal{H} }{2\omega}
 \operatorname{Re} [\mathcal B(\omega)+\mathcal B(-\omega)]\nonumber\\
 &=\frac{2g\mu_B e^2\alpha_{\mathrm{so}}\nu_0\mathcal{H} }{\Delta_Rp_F}
 \operatorname{Im} \left[\frac{s(\Delta_R^2/(2z)-\Gamma)}{D_F}\right]\nonumber\\
 &=g\mu_B e^2\nu_0\mathcal{H} 
 \operatorname{Im} \left[\frac{s(2\alpha_{\mathrm{so}}^2/z-\Gamma/p_F^2)}{z^2+\Delta_R^2}\right].
 \label{eq:diagrammatic-final}
\end{align}
This is Eq.~\eqref{eq:final-spin}, including the sign, both optical orderings, the density of states per spin, and the magnetic-moment conversion. The equality is between the two consistently expanded single-splitting metallic calculations.

\section{Orbital IFE from differentiated current diagrams}
The dimensionless functions used in this section have fixed physical meanings:
\begin{equation}
 w(\omega)=\frac{\Gamma}{\Gamma-\mathrm{i} \omega},\qquad
 u(\omega)=\frac{\Gamma(\Gamma-\mathrm{i} \omega)}{(\Gamma-\mathrm{i} \omega)^2+\Delta_R^2},\qquad
 v(\omega)=\frac{w+u}{2},\qquad s(\omega)=\frac{1-w}{1-v}.
\end{equation}
Here $w$ is the charge-channel ladder eigenvalue, $u$ the out-of-plane-spin eigenvalue, $v$ the in-plane-spin eigenvalue, and $s$ the dressing of the anomalous optical velocity. The scalar $v$ is always a ladder eigenvalue; velocities carry a Cartesian index or a bold symbol. The argument $\omega$ is omitted only within the following calculation. These are the Fermi-level values $u(\omega)=u(\omega,0)$, $v(\omega)=v(\omega,0)$, and $s(\omega)=s(\omega,0)$. Their energy-dependent extensions are defined in Eqs.~\eqref{eq:u-radial}--\eqref{eq:optical-ladder-energy-expansion}.

\label{sec:orb-diagrams}
\subsection{From a uniform moment to a coefficient linear in output momentum}
Equation~\eqref{orb:eq:curl} shows why three current vertices are not enough information until the output momentum is differentiated. Define
\begin{equation}
 j_i^{(2),H}(\mathbf Q)=\mathrm{i} \varepsilon^{(2)}_{i\ell}Q_\ell
 \mathcal K_E(\omega)\mathcal{H} +O(Q^2),\qquad
 M_{\mathrm{orb},z}=\mathcal K_E\mathcal{H} /c_0.
 \label{orb:eq:extractM}
\end{equation}
Keep $\mathbf Q$ in the propagators and all Bethe--Salpeter vertices until the derivative is taken. The output frequency is zero throughout. An isolated triangle at $Q=0$ does not contain this orbital response. We will use $\mathbf a=\mathbf A/c_0$ and $\mathcal{H} _a=\mathcal{H} /\omega^2$; the corresponding current coefficient is $\kappa_a=\omega^2\mathcal K_E$.

\begin{figure}[tb]
\centering
\begin{tikzpicture}[baseline=(current bounding box.center),scale=1.15]
\foreach \x/\txt/\middle/\vj/\vk in {0/RAA/A/V_j^+/v_k,4/RRA/R/v_j/V_k^-}{
\draw (\x,0)--(\x+2,0)--(\x+1,1.6)--cycle;
\fill (\x,0) circle(2pt) (\x+2,0) circle(2pt) (\x+1,1.6) circle(2pt);
\node[below] at (\x,0) {$-e\vk$};
\node[below] at (\x+2,0) {$-e\vj$};
\node[above] at (\x+1,1.6) {$eJ_i$};
\node at (\x+1,.55) {$\txt$};
\node[above left] at (\x+.45,.75) {$G^A$};
\node[above right] at (\x+1.55,.75) {$G^R$};
\node[above] at (\x+1,0) {$G^{\middle}_{\epsilon-\omega}$};}
\end{tikzpicture}
\caption{Current triangles for the orbital response in the vertex conventions of this supplement. The measurement vertex is $eJ_i$, without the spin factor $1/2$. The optical vertex on the $RA$ pair is dressed: $V_j^+$ in $RAA$ and $V_k^-$ in $RRA$. The outer pair carries static transfer $\mathbf Q$, and the middle propagator has energy $\epsilon-\omega$. The response is $RAA-RRA$; both exchanged optical orderings and the physical-current contacts must be added. Detailed momenta are fixed by the routing below.}
\label{fig:orbital-triangles}
\end{figure}
\subsection{Electromagnetic vertices and the current contact}
Minimal coupling gives
\begin{equation}
H[\mathbf a]=H_0-ea_iv_i+\frac{e^2a^2}{2m}\mathbf{1} ,
\qquad j_i[\mathbf a]=-\frac{\partial H}{\partial a_i}
=ev_i-\frac{e^2}{m}a_i\mathbf{1} .\label{orb:eq:minimal}
\end{equation}
There are three distinct objects to keep track of:
\begin{enumerate}
\item The linear optical vertex $-ev_i$ produces the two optical insertions in each triangle.
\item The quadratic Hamiltonian vertex $e^2\delta_{jk}\mathbf{1} /m$ is symmetric in its two optical indices. It gives zero after the helicity projection $\varepsilon^{(2)}_{jk}$.
\item The explicit $-e^2a_i/m$ in the measurement current combines with a first-order density response. It produces two current-contact diagrams. Unlike the quadratic Hamiltonian vertex, their helicity part is nonzero.
\end{enumerate}
We compute the triangles first and add the measurement-current contacts at the end. This separation makes it possible to see which part of the answer actually reduces to the real conductivity.

For momentum-linear SOC,
\begin{equation}
\partial_{p_\ell}v_j=\frac{\delta_{\ell j}}m\mathbf{1} .\label{orb:eq:dv}
\end{equation}
This equation does not discard SOC: the derivative of the momentum-independent anomalous velocity is zero. Momentum derivatives of Green functions still contain the full velocity:
\begin{equation}
\partial_{p_\ell}G^{R,A}=G^{R,A}v_\ell G^{R,A}.\label{orb:eq:dG}
\end{equation}
To prove \eqref{orb:eq:dG}, differentiate $G G^{-1}=\mathbf{1} $ and use $\partial_{p_\ell}G^{-1}=-v_\ell$. This identity explains why differentiating the propagators introduces an additional velocity insertion, including its SOC part.

\subsubsection{Explicit diagrams, vertex definitions, and optical exchange}
\label{new:diagram-guide}
\begin{figure}[tb]
 \centering
 \includegraphics[width=.98\textwidth]{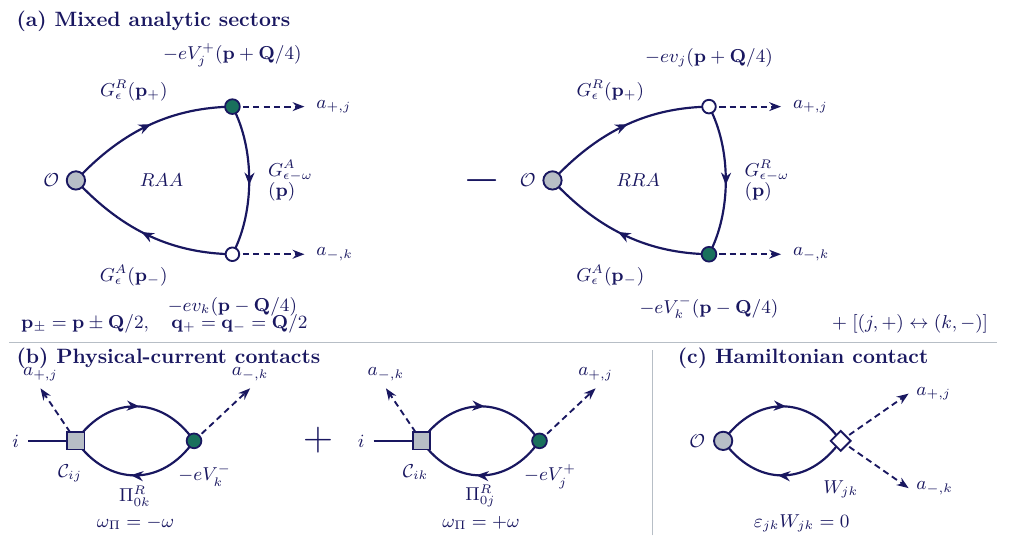}
 \caption{Mixed triangles and the two distinct contact topologies. (a) Curved $RAA-RRA$ loops for one ordering, together with the required optical exchange. Gray circles are measurement insertions. Filled green optical circles denote the impurity dressing of the adjacent $RA$ pair. (b) Two physical-current contacts: one optical leg and the output current meet at the square $\mathcal C_{ij}=-e^2\delta_{ij}\mathbf{1} /m$, while the other field drives a density response. The bubbles denote full retarded density--current responses $\Pi^R$ at $-\omega$ and $+\omega$, respectively. (c) The Hamiltonian contact $W_{jk}=e^2\delta_{jk}\mathbf{1} /m$ has both optical legs at the diamond and vanishes under helicity projection. These diagrams supplement the original figures without identifying a single current contact with the magnetic Lorentz-force half.}
 \label{new:contact-diagrams}
\end{figure}

The mixed triangles of Figs.~\ref{fig:spin-triangles} and \ref{fig:orbital-triangles} and the contact bubbles are displayed together in Fig.~\ref{new:contact-diagrams}. This redraw retains the original $RAA-RRA$ organization while making the vertices and the two different kinds of contact explicit. The positive window $W_\omega=f_0(\epsilon-\omega)-f_0(\epsilon)$ fixes the relative analytic-sector sign as in Eq.~\eqref{orb:eq:triangle}. Both exchanged optical orderings are included. They are distinct from the two analytic sectors of one ordering.

The measurement insertion is $\mathcal O=\widetilde\sigma_z/2$ for spin density and $\mathcal O=eJ_i$ for electrical current. Multiplication of the former observable by $-g\mu_B$ gives the spin magnetic moment. There is no universal factor $1/2$ multiplying the current insertion. In $RAA$, only the upper optical vertex joins an $RA$ pair and carries $V_j^+$. In $RRA$, only the lower optical vertex carries $V_k^-$. The $AA$ or $RR$ optical insertion remains bare at the retained order. Thus drawing both optical vertices as dressed in each mixed triangle would specify a different diagram set.

The finite-transfer current insertion is $J_i(\mathbf{p} ;\mathbf{Q} )$, even though its value at $Q=0$ reduces to $p_i\mathbf{1} /m$ after the static impurity ladder is summed. Its transfer derivative must still be retained. The momenta $\mathbf{p} _\pm=\mathbf{p} \pm\mathbf{Q} /2$ on the outer propagators and $\mathbf{p} \pm\mathbf{Q} /4$ at the optical vertices agree with the routing below. Setting $Q=0$ before differentiating a propagator or a Bethe--Salpeter vertex loses part of the orbital response.

The vertex definitions follow directly from Eq.~\eqref{orb:eq:minimal}, with $\mathbf a=\mathbf A/c_0$:
\begin{equation*}
 \mathcal C_{ij}\equiv\frac{\partial j_i}{\partial a_j}
 =-\frac{e^2}{m}\delta_{ij}\mathbf{1} ,
 \qquad
 W_{jk}\equiv\frac{\partial^2H}{\partial a_j\partial a_k}
 =\frac{e^2}{m}\delta_{jk}\mathbf{1} ,
 \qquad H^{(2)}=\frac12W_{jk}a_ja_k.
\end{equation*}
Although their magnitudes contain the same inverse mass, these are different functional derivatives. The square differentiates the measurement current and has one optical leg. The diamond differentiates the Hamiltonian twice and has two optical legs. The latter is symmetric in $j,k$ and obeys $\varepsilon^{(2)}_{jk}W_{jk}=0$; its vanishing does not eliminate the two current contacts. A spin-density measurement has no explicit $\mathbf a$ dependence and therefore has no analogous measurement-current contact.

The two bubbles in panel (b) are the two terms already present in Eq.~\eqref{orb:eq:contactstart}. Their internal transfer is the nonzero optical frequency: the bubble driven by $a_-$ carries $-\omega$, and that driven by $a_+$ carries $+\omega$. The retarded symbol $\Pi^R$ includes the analytic sectors, density Ward identity, and impurity dressing required by the linear response. It is not an isolated $RA$ product and is not a scalar dc pole. The displayed two contacts must not be doubled again when the optical exchange of the triangle pair is performed.

\subsection{Static output vertex and proof of the RA collapse}
At output frequency zero, $w_0=1$ and $s_0=0$, so
\begin{equation}
\boxed{J_i(\mathbf p;0)=\frac{p_i}{m}\mathbf{1} .}\label{orb:eq:J0}
\end{equation}
This result uses the impurity ladder. It is not a substitution of the bare Rashba velocity by a scalar velocity.

At identical $(\epsilon,\mathbf p)$,
\begin{equation}
(G^R)^{-1}-(G^A)^{-1}=i\Gamma\mathbf{1} .
\end{equation}
Multiplying on the left by $G^A$ and on the right by $G^R$ proves
\begin{align}
i\Gamma G^AG^R
&=G^A[(G^R)^{-1}-(G^A)^{-1}]G^R=G^A-G^R,\\
\boxed{G^AJ_iG^R}&=\boxed{\frac{p_i}{mi\Gamma}(G^A-G^R).}\label{orb:eq:collapse}
\end{align}
No commutation of a spin velocity through a Green function was assumed. The scalar output vertex in \eqref{orb:eq:J0} is what makes the second line possible. We use \eqref{orb:eq:collapse} on undifferentiated factors remaining after the required $Q$ derivative.

\subsection{The intermediate canonical bubble}
The RA collapse of a current triangle initially produces a canonical-current bubble rather than \eqref{orb:eq:Kubo}. Define
\begin{equation}
\bar\sigma=\frac{e^2}{2\omega}
\int\frac{d\epsilon}{2\pi}W_\omega\int\!\frac{\mathrm{d}^2p}{(2\pi)^2} 
\operatorname{Tr} \left[\frac{p_x}{m}G^R(v_x+V_x^+)G^A_\omega\right].\label{orb:eq:barsigma}
\end{equation}
The same moments give
\begin{align}
\bar\sigma&=\frac{e^2\nu_0}{\Gamma}
\left[I_{aa}+\frac{\alpha_{\mathrm{so}}(1+s)}2 I_{a\mathsf R}\right]\\
&=\frac{e^2\nu_0}{\Gamma}
\left[wv_F^2+\alpha_{\mathrm{so}}^2w(3-s)\right].\label{orb:eq:barsigmaeval}
\end{align}
Subtracting this from \eqref{orb:eq:Pi} and using $2v=u+w$ yields
\begin{equation}
\boxed{\sigma-\bar\sigma
=\frac{e^2\nu_0\alpha_{\mathrm{so}}^2}{\Gamma}(su-w).}\label{orb:eq:sigmadiff}
\end{equation}
This difference is the two-vertex bubble that the SOC-dependent outer derivative must supply. It will be obtained from the triangle itself below.

\subsection{Continuation for the orbital measurement vertex}
The continuation in Eqs.~\eqref{eq:three-cuts} and \eqref{eq:continued-triangle} also applies when the spin measurement vertex is replaced by $eJ_i$. The relative sign is therefore $RAA-RRA$ in the positive signed window $W_\omega=f_0(\epsilon-\omega)-f_0(\epsilon)$. Both optical orderings are required. The energy integrals obey Eq.~\eqref{eq:Fermi-window}; the mean energy is $\bar\epsilon=\epsilon-\omega/2$.
\begin{equation}
 \int\mathrm{d} \epsilon\,W_\omega=\omega,\qquad
 \int\mathrm{d} \epsilon\,\epsilon W_\omega=\frac{\omega^2}{2},\qquad
 \int\mathrm{d} \epsilon\,\bar\epsilon W_\omega=0.
 \label{orb:eq:windows}
\end{equation}
Same-analytic radial pieces may be removed only with their correct arc and Kubo subtraction. This qualification matters for velocity-weighted bubbles, even when the differentiated three-line integrals decay fast enough to have no arc.
\subsection{Nonlinear triangles and output-momentum routing}
\subsubsection{Momentum routing that keeps the output frequency zero}
For the coefficient associated with the slowly varying optical intensity, set the relative in-plane optical momentum to zero and choose
\begin{equation}
\mathbf q_+=\mathbf q_-=\mathbf Q/2,\qquad
\omega_+=\omega,\qquad \omega_-=-\omega.
\end{equation}
Then the outer and middle propagators are
\begin{align}
R_+&=G^R(\epsilon,\mathbf p+\mathbf Q/2),&
A_-&=G^A(\epsilon,\mathbf p-\mathbf Q/2),\\
R_\omega&=G^R(\epsilon-\omega,\mathbf p),&
A_\omega&=G^A(\epsilon-\omega,\mathbf p).
\end{align}
The mean optical momenta are $\mathbf p+\mathbf Q/4$ and $\mathbf p-\mathbf Q/4$. Their difference is not their transfer: each optical RA vertex separately has transfer $\mathbf Q/2$. The output vertex has mean momentum $\mathbf p$ and transfer $\mathbf Q$. This distinction explains the factors $1/4$ and $1/2$ in the derivatives below.

Let $J_i$ be the dressed output velocity, with its factor $e$ kept outside. For one ordering define
\begin{align}
K^\omega_{i;jk}(\mathbf Q)=\int\frac{d\epsilon}{2\pi i}W_\omega\int\!\frac{\mathrm{d}^2p}{(2\pi)^2} 
\operatorname{Tr} \Big[ A_-J_iR_+\big(&V_j^+(\mathbf p+\mathbf Q/4;\mathbf Q/2)
A_\omega v_k(\mathbf p-\mathbf Q/4)\\[-2mm]
&-v_j(\mathbf p+\mathbf Q/4)R_\omega
V_k^-(\mathbf p-\mathbf Q/4;\mathbf Q/2)\big)\Big].\label{orb:eq:triangle}
\end{align}
An optical vertex on an RA pair carries its noncrossing ladder. The other optical vertex in that analytic sector is on an AA or RR pair and remains bare at this order. In RAA the upper optical vertex is dressed; in RRA the lower one is dressed. The two triangle contributions to the current are
\begin{equation}
j_i^{\triangle}(\mathbf Q)=\frac{e^3}{\omega^2}E_jE_k^*
\left[K^\omega_{i;jk}(\mathbf Q)+K^{-\omega}_{i;kj}(\mathbf Q)\right].\label{orb:eq:jtriangle}
\end{equation}
This contains one measurement-current insertion and two optical-current insertions. It contains no local spin-magnetization insertion.

\subsubsection{The helicity and curl contraction fixes the factors}
The antisymmetric part of the polarization product is
\begin{equation}
[E_jE_k^*]_H=-\frac i2\varepsilon^{(2)}_{jk}\mathcal{H} .
\end{equation}
Define
\begin{equation}
\mathscr C=\varepsilon^{(2)}_{i\ell}\varepsilon^{(2)}_{jk}
\left.\partial_{Q_\ell}
\left[K^\omega_{i;jk}+K^{-\omega}_{i;kj}\right]\right|_{Q=0}.\label{orb:eq:Cdef}
\end{equation}
The helicity/curl tensor has the form $Q_\ell B_\triangle\varepsilon^{(2)}_{i\ell}\varepsilon^{(2)}_{jk}$. Since $\varepsilon^{(2)}_{i\ell}\varepsilon^{(2)}_{i\ell}=2$ and $\varepsilon^{(2)}_{jk}\varepsilon^{(2)}_{jk}=2$, $\mathscr C=4B_\triangle$. Contracting with the optical product gives
\begin{equation}
j_i^{\triangle,H}=i\varepsilon^{(2)}_{i\ell}Q_\ell\kappa_a^\triangle\mathcal{H} _a,
\qquad \boxed{\kappa_a^\triangle=-e^3\mathscr C/4.}\label{orb:eq:CtoK}
\end{equation}
This is the normalization connecting a differentiated loop trace to the current and, eventually, to magnetization.

\subsection{Every required momentum derivative}
\subsubsection{The product rule and mean-momentum derivatives}
Use $R=G^R_\epsilon$, $A=G^A_\epsilon$ after evaluation at $Q=0$. Define
\begin{equation}
L_i(Q)=A_-J_iR_+,\quad
C_{i\ell}=\left.\partial_{Q_\ell}J_i\right|_0,\quad
D_{j\ell}^\pm=\left.\partial_{q_\ell}V_j^\pm(\mathbf p;\mathbf q)\right|_0.
\end{equation}
The outer product has three derivatives:
\begin{equation}
\boxed{L'_{i\ell}=-\frac12Av_\ell AJ_iR
+AC_{i\ell}R+\frac12AJ_iRv_\ell R.}\label{orb:eq:Lprime}
\end{equation}
The first and last terms differentiate the shifted Green functions. Their signs follow from $\mathbf p\mp\mathbf Q/2$. The middle term differentiates the output impurity ladder; a constant homogeneous vertex does not imply a constant finite-transfer vertex.

For the optical vertices, the derivatives are
\begin{align}
\partial_{Q_\ell}V_j^+(\mathbf p+\mathbf Q/4;\mathbf Q/2)|_0
&=\frac{\delta_{\ell j}}{4m}\mathbf{1} +\frac12D_{j\ell}^+,\\
\partial_{Q_\ell}V_k^-(\mathbf p-\mathbf Q/4;\mathbf Q/2)|_0
&=-\frac{\delta_{\ell k}}{4m}\mathbf{1} +\frac12D_{k\ell}^-,\\
\partial_{Q_\ell}v_j(\mathbf p+\mathbf Q/4)|_0
&=\frac{\delta_{\ell j}}{4m}\mathbf{1} ,\\
\partial_{Q_\ell}v_k(\mathbf p-\mathbf Q/4)|_0
&=-\frac{\delta_{\ell k}}{4m}\mathbf{1} .\label{orb:eq:optderivs}
\end{align}
The middle Green functions have no $Q$ derivative in this routing. Therefore, with
\begin{equation}
S_{jk}=V_j^+A_\omega v_k-v_jR_\omega V_k^-,
\end{equation}
the full differentiated trace is
\begin{align}
\partial_{Q_\ell}\operatorname{Tr} [L_iS_{jk}]|_0
={}&\operatorname{Tr} [L'_{i\ell}S_{jk}]\\
&+\operatorname{Tr} \!\left[L_i\left(
\frac{\delta_{\ell j}}{4m}A_\omega v_k
-\frac{\delta_{\ell k}}{4m}V_j^+A_\omega
-\frac{\delta_{\ell j}}{4m}R_\omega V_k^-
+\frac{\delta_{\ell k}}{4m}v_jR_\omega\right)\right]\\
&+\frac12\operatorname{Tr} [L_i(D_{j\ell}^+A_\omega v_k-v_jR_\omega D_{k\ell}^-)].\label{orb:eq:allterms}
\end{align}
The three lines are, respectively, the outer-pair derivatives, the four explicit optical-velocity derivatives, and the two optical-transfer ladder derivatives. Equation \eqref{orb:eq:allterms}, applied to both optical orderings, is a complete inventory at linear $Q$ in the present diagram set.

\subsubsection{Differentiating the optical Bethe--Salpeter equation}
At finite transfer, the optical ladder has the form
\begin{equation}
V_j(\mathbf p;\mathbf q)=v_j(\mathbf p)
+\gamma_{\rm imp}\int\!\frac{\mathrm{d}^2p}{(2\pi)^2} '
G^R_\epsilon(\mathbf p'+\mathbf q/2)V_j(\mathbf p';\mathbf q)
G^A_{\epsilon-\omega}(\mathbf p'-\mathbf q/2).
\end{equation}
The bare mean-momentum vertex does not depend on $\mathbf q$ in this equation. Differentiate each of the two Green functions and the vertex inside the integral; move the last term to the left. This gives
\begin{equation}
\boxed{(1-\mathcal T_\omega)D_{j\ell}^+
=\frac{\gamma_{\rm imp}}2\int\!\frac{\mathrm{d}^2p}{(2\pi)^2} 
\left[Rv_\ell RV_j^+A_\omega
-RV_j^+A_\omega v_\ell A_\omega\right].}\label{orb:eq:Dsource}
\end{equation}
The $D^-$ equation uses $R_\omega,A$ in place of $R,A_\omega$. This calculation is necessary because spatial variations of the optical field create a first-order density/spin response that is missed by simply inserting the homogeneous optical ladder.

Rotational symmetry allows only
\begin{equation}
D_{j\ell}=d_0\delta_{j\ell}\mathbf{1} +d_z\varepsilon^{(2)}_{j\ell}\sigma_z.\label{orb:eq:Dsectors}
\end{equation}
In particular, $d_0$ is obtained by inverting the scalar factor $1-w$, and $d_z$ by inverting the out-of-plane-spin factor $1-u$.

\subsubsection{Differentiating the output ladder and retaining the SOC commutator}
Use the adjoint output rung $\mathcal T_0^*[X]=\gamma_{\rm imp}\int\!\frac{\mathrm{d}^2p}{(2\pi)^2}  AXR$. The same product-rule derivation gives
\begin{equation}
\boxed{(1-\mathcal T_0^*)C_{i\ell}
=\frac{\gamma_{\rm imp}}2\int\!\frac{\mathrm{d}^2p}{(2\pi)^2} 
[-Av_\ell AJ_iR+AJ_iRv_\ell R].}\label{orb:eq:Csource}
\end{equation}
Only its transverse out-of-plane-spin part is needed. The full Rashba velocity obeys
\begin{equation}
p_yv_x-p_xv_y=-\alpha_{\mathrm{so}}(p_x\sigma_x+p_y\sigma_y)
=\frac{[H_0,\sigma_z]}{2i}.\label{orb:eq:commutator}
\end{equation}
For either analytic sector,
\begin{equation}
G[H_0,\sigma_z]G=[G,\sigma_z].\label{orb:eq:Gcomm}
\end{equation}
To see this, write $H_0=(\epsilon\mathord{\pm}i\Gamma/2)\mathbf{1} -G^{-1}$; the scalar drops out of the commutator.

Take the difference of \eqref{orb:eq:Csource} for $(i,\ell)=(y,x)$ and $(x,y)$, insert $J_i=p_i/m$, and use \eqref{orb:eq:commutator}--\eqref{orb:eq:Gcomm}:
\begin{align}
(1-\mathcal T_0^*)(C_{yx}-C_{xy})
&=\frac{\gamma_{\rm imp}}{4im}\int\!\frac{\mathrm{d}^2p}{(2\pi)^2} 
\left[\{AR,\sigma_z\}-2A\sigma_zR\right]\\
&=\frac{1-u_0}{2im}\sigma_z.\label{orb:eq:Ccurlsource}
\end{align}
The second line uses $\gamma_{\rm imp}\int\!\frac{\mathrm{d}^2p}{(2\pi)^2}  AR=\mathbf{1} $ and $\gamma_{\rm imp}\int\!\frac{\mathrm{d}^2p}{(2\pi)^2}  A\sigma_zR=u_0\sigma_z$, where $u_0=\Gamma^2/(\Gamma^2+\Delta_R^2)$. Division by the out-of-plane-spin ladder factor gives
\begin{equation}
C_{yx}-C_{xy}=\frac{\sigma_z}{2im},
\qquad \boxed{C_{i\ell}^{\perp}=\frac{i}{4m}\varepsilon^{(2)}_{i\ell}\sigma_z.}\label{orb:eq:Canswer}
\end{equation}
The apparent spin-relaxation denominator cancels its source factor $1-u_0$. At exactly zero SOC the spin sector is degenerate; take the spinless diagrams directly or the limit of the complete summed response, not the isolated solution \eqref{orb:eq:Canswer}.

Finally the propagator part of the outer curl is
\begin{equation}
\frac1{4im}\left[\{AR,\sigma_z\}-2A\sigma_zR\right],
\end{equation}
while the vertex part from \eqref{orb:eq:Canswer} is $A\sigma_zR/(2im)$. They cancel the second term, leaving
\begin{equation}
\boxed{\partial_{Q_x}L_y-\partial_{Q_y}L_x
=\frac1{4im}\{AR,\sigma_z\}.}\label{orb:eq:outercurl}
\end{equation}
This is the SOC contribution that would be lost by treating all propagator derivatives as symmetric in the optical indices.

\subsection{First collapse: the four explicit optical-velocity derivatives}
Contract the second line of \eqref{orb:eq:allterms} with $\varepsilon^{(2)}_{i\ell}\varepsilon^{(2)}_{jk}$. Using $\varepsilon^{(2)}_{ij}\varepsilon^{(2)}_{jk}=-\delta_{ik}$ and $\varepsilon^{(2)}_{ik}\varepsilon^{(2)}_{jk}=\delta_{ij}$ gives, for one ordering,
\begin{equation}
\frac1{4m}\sum_i\operatorname{Tr} \!\left[
\frac{p_i}{m}AR\left(-A_\omega v_i-V_i^+A_\omega
+R_\omega V_i^-+v_iR_\omega\right)\right].\label{orb:eq:barecontract}
\end{equation}
Insert $AR=(A-R)/(i\Gamma)$. The mixed parts remaining after the common same-analytic/contact subtraction are
\begin{equation}
\frac1{4mi\Gamma}\sum_i\operatorname{Tr} \!\left[
\frac{p_i}{m}\left(RA_\omega v_i+RV_i^+A_\omega
+AR_\omega V_i^-+Av_iR_\omega\right)\right].\label{orb:eq:barebubble}
\end{equation}
There are only two propagators in every term of \eqref{orb:eq:barebubble}. Its first two terms are the canonical bubble \eqref{orb:eq:barsigma}, and its other two terms are the conjugate bubble. Angular isotropy makes the sum over $i=x,y$ twice the $x$ response. Including the energy factor $1/i$ and the resolvent factor $1/(i\Gamma)$ gives $-2\operatorname{Re} \bar b/(m\Gamma)$ for one ordering, where $\bar b=\omega\bar\sigma/e^2$. The exchanged ordering supplies the same contribution. Hence
\begin{equation}
\mathscr C_{\partial v}=-\frac4{m\Gamma}\operatorname{Re} \bar b,
\qquad \boxed{\kappa_{a,\partial v}=\frac{e\omega}{m\Gamma}\operatorname{Re} \bar\sigma.}\label{orb:eq:bareanswer}
\end{equation}
The result is proportional to $\operatorname{Re} \bar\sigma/\Gamma$, not $\operatorname{Im} \bar\sigma/\omega$. These happen to be equal for a Drude function, but not generally for the SOC-modified bubble.

\subsection{Radial residue machinery for the remaining derivative groups}
\subsubsection{Temporary normalization and the role of the radial expansion}
For the explicit Pauli algebra and residues in this and the next two sections, temporarily set $m=1$ and divide out the overall charge factors: $\sigma/e^2$ for conductivities and $\kappa_a/e^3$ for current coefficients. The signed electron charge remains $e<0$. Define
\begin{equation}
\rho_r=p_F^{-2},\quad h=i\Gamma,\quad d=\omega+i\Gamma,\quad
\bar{d}=d-2h=\omega-i\Gamma,\quad \Delta=\Delta_R,\quad \nu_0=1/(2\pi).
\end{equation}
Here $\rho_r=p_F^{-2}$ is a temporary radial expansion parameter, not an optical momentum. A final coefficient is restored to physical units after combining the groups. The normalized conductivities are $\widetilde\sigma=\sigma/(e^2\nu_0)$ and $\widetilde{\bar\sigma}=\bar\sigma/(e^2\nu_0)$; $\widetilde\sigma_0$ below is the Drude part $i/(\rho_r d)$. The temporary pole variables are $d=\omega+i\Gamma$, $\bar d=\omega-i\Gamma$, and $h=i\Gamma$; $h$ in this residue algebra is a linewidth variable, not the distribution kernel $\hat h$. The independently evaluated bubbles \eqref{orb:eq:Pi} and \eqref{orb:eq:barsigmaeval} become
\begin{align}
{\widetilde\sigma}&=\frac{i}{\rho_r d}+\frac{i\rho_r\Delta^2}{4d}
\left(2+\frac{\Delta^2s}{d^2-\Delta^2}\right),\label{orb:eq:Bnorm}\\
{\widetilde{\bar\sigma}}&=\frac{i}{\rho_r d}+\frac{i\rho_r\Delta^2}{4d}(3-s),\label{orb:eq:Bbarnorm}\\
{\widetilde\sigma}-{\widetilde{\bar\sigma}}&=\frac{i\rho_r\Delta^2}{4}
\left(\frac{s d}{d^2-\Delta^2}-\frac1d\right).\label{orb:eq:Bdiffnorm}
\end{align}
These retain the $\rho_r^{-1}$ charge term and $\rho_r^1$ SOC correction. Intermediate products may have $\rho_r^{-2}$ and $\rho_r^0$ terms, which must be allowed to cancel. Keeping the $\rho_r^1$ orbital correction requires retaining the appropriate higher radial terms in intermediate products; the approximation is imposed on the final response, not by freezing each factor separately.

The contracted traces are rotational scalars, so set $\mathbf p=(p,0)$ after forming the tensor contraction. Put
\begin{equation}
x=\epsilon-\xi+i\Gamma/2,\qquad
p^2=\rho_r^{-1}+2(\epsilon+i\Gamma/2-x),\qquad \alpha_{\mathrm{so}}^2=\rho_r\Delta^2/4.\label{orb:eq:x}
\end{equation}
Write $\eta=\epsilon-\omega/2$. The integrated, compact expressions use $p^2=\rho_r^{-1}+d-2x$, but the expansion is first performed with the $2\eta$ term retained. Equation \eqref{eq:Fermi-window} removes odd mean-energy terms. Quadratic mean-energy terms generated in intermediate products must also be retained until the complete paired identity is formed; they cancel at the displayed response order. Thus the compact mean-energy presentation is not an instruction to freeze individual complex-pole numerators.

\subsubsection{Matrices and the precise residue prescription}
Let
\begin{align}
D_R&=x^2-\alpha_{\mathrm{so}}^2p^2,&D_A&=(x-h)^2-\alpha_{\mathrm{so}}^2p^2,\\
D_B&=(x-d)^2-\alpha_{\mathrm{so}}^2p^2,&D_H&=(x-d+h)^2-\alpha_{\mathrm{so}}^2p^2.
\end{align}
At $\mathbf p=(p,0)$,
\begin{align}
R&=(x\mathbf{1} +\alpha_{\mathrm{so}} p \hat{\mathsf R}_x)/D_R,&
A&=((x-h)\mathbf{1} +\alpha_{\mathrm{so}} p \hat{\mathsf R}_x)/D_A,\\
A_\omega&=((x-d)\mathbf{1} +\alpha_{\mathrm{so}} p \hat{\mathsf R}_x)/D_B,&
R_\omega&=((x-d+h)\mathbf{1} +\alpha_{\mathrm{so}} p \hat{\mathsf R}_x)/D_H.
\end{align}
In particular,
\begin{equation}
AR=\ell_0\mathbf{1} +\ell_1\hat{\mathsf R}_x,\quad
\ell_0=\frac{x(x-h)+\alpha_{\mathrm{so}}^2p^2}{D_RD_A},\quad
\ell_1=\frac{\alpha_{\mathrm{so}} p(2x-h)}{D_RD_A}.\label{orb:eq:c0c1}
\end{equation}
The algebra needed for the traces is
\begin{equation}
\hat{\mathsf R}_x^2=\hat{\mathsf R}_y^2=\mathbf{1} ,\quad \hat{\mathsf R}_x\hat{\mathsf R}_y=i\sigma_z,\quad
\hat{\mathsf R}_y\hat{\mathsf R}_x=-i\sigma_z,\quad \{\hat{\mathsf R}_x,\sigma_z\}=0,\quad
\operatorname{Tr} \mathbf{1} =2,\quad\operatorname{Tr} \sigma_z=\operatorname{Tr}  \hat{\mathsf R}_i=0.\label{orb:eq:paulirules}
\end{equation}

Expand the rational functions in $\rho_r$ before integrating. For example,
\begin{equation}
\frac1{D_R}=\frac1{x^2-\Delta^2/4}
\sum_{n\geq0}\left[
\frac{\rho_r\Delta^2(d+2\eta-2x)}{4(x^2-\Delta^2/4)}\right]^n.\label{orb:eq:radseries}
\end{equation}
Each coefficient has known poles, possibly of higher order, at $x=\pm\Delta/2$. Terms with $R_\omega$ also have retarded poles at $x=\omega\pm\Delta/2$. Since $d\xi=-dx$, define
\begin{equation}
\mathcal I[F]=\int\frac{d\xi}{2\pi}F(x)
=-i\sum_{\text{retarded }x\text{ poles}}\operatorname{Res}F.\label{orb:eq:Iresidue}
\end{equation}
For a pole of order $n$ at $x=x_0$ the explicit rule is
\begin{equation}
\operatorname{Res}_{x_0}F=
\frac1{(n-1)!}\left[\frac{d^{n-1}}{dx^{n-1}}
\big((x-x_0)^nF(x)\big)\right]_{x=x_0}.\label{orb:eq:multipole}
\end{equation}
This evaluates all the double and higher poles created by the momentum derivatives. The derivative traces here decay sufficiently fast to have no arc term.

For a radial contraction $F_\omega$, \eqref{orb:eq:CtoK} and the signed energy window give the normalized current coefficient
\begin{equation}
\boxed{\frac{\kappa_a}{e^3}=\frac{\nu_0 i\omega}{4}
[F_\omega+F_{-\omega}].}\label{orb:eq:FtoK}
\end{equation}
The plus sign includes both the negative signed energy window and the minus sign from exchanging the antisymmetric optical indices.

\subsection{Second collapse: scalar and out-of-plane-spin optical-transfer ladders}
\subsubsection{Projecting and evaluating the derivative source}
The coefficients in \eqref{orb:eq:Dsectors} are extracted by
\begin{equation}
d_0=\frac14\operatorname{Tr} (D_{xx}+D_{yy}),\qquad
d_z=\frac14\operatorname{Tr} [\sigma_z(D_{xy}-D_{yx})].
\end{equation}
In the temporary $m=1$ normalization define the source traces
\begin{align}
S_0(x)=\frac18\operatorname{Tr} \sum_{j=x,y}
\left[Rv_jRV_j^+A_\omega-RV_j^+A_\omega v_jA_\omega\right],\label{orb:eq:S0}\\
S_z(x)=\frac18\operatorname{Tr} \sigma_z\Big[
Rv_yRV_x^+A_\omega-RV_x^+A_\omega v_yA_\omega
-Rv_xRV_y^+A_\omega+RV_y^+A_\omega v_xA_\omega\Big].\label{orb:eq:Sz}
\end{align}
The factors $1/8$ account for the $1/2$ from the shifted propagator and the angular/scalar or spin projection. Since $\gamma_{\rm imp}\nu_0\int d\xi=\Gamma\mathcal I$, the projected Bethe--Salpeter equations are
\begin{equation}
d_0^+=\frac\Gamma{1-w}\mathcal I[S_0],\qquad
d_z^+=\frac\Gamma{1-u}\mathcal I[S_z].\label{orb:eq:projectedD}
\end{equation}
Applying the Pauli rules \eqref{orb:eq:paulirules}, expanding as in \eqref{orb:eq:radseries}, and taking the residues \eqref{orb:eq:multipole} gives
\begin{align}
\mathcal I[S_0]
&=\frac{i}{2\rho_r d^2}
+\frac{i\rho_r\Delta^2[2d^2-2\Delta^2+\Delta^2s]}
{8d^2(d^2-\Delta^2)}
=\frac {\widetilde\sigma}{2d},\label{orb:eq:S0eval}\\
\mathcal I[S_z]
&=\frac{\Delta^2[-d^2+\Delta^2+2d^2s]}
{4d(d^2-\Delta^2)^2}
\quad\hbox{at the order needed in the paired response}.\label{orb:eq:Szeval}
\end{align}
Consequently, because $1-w=\omega/d$,
\begin{equation}
\boxed{d_0^+=\frac{\Gamma {\widetilde\sigma}}{2\omega},\qquad
d_0^-=-\frac{\Gamma {\widetilde\sigma}^*}{2\omega}.}\label{orb:eq:D0bubble}
\end{equation}
The scalar derivative vertex has therefore reduced to a two-current conductivity bubble. The out-of-plane-spin coefficient is
\begin{equation}
d_z^+=-\frac{ih\Delta^2[-d^2+\Delta^2+2d^2s]}
{4d(d^2-\Delta^2)(d^2-\Delta^2-dh)},
\qquad d_z^-=-d_z^{+*}.\label{orb:eq:Dzeval}
\end{equation}

\subsubsection{A worked double pole: why the radial numerator matters}
In the spinless part of $S_0$, at the integrated mean energy,
\begin{equation}
S_0^{(\alpha_{\mathrm{so}}=0)}=-\frac{d(\rho_r^{-1}+d-2x)}{4x^2(x-d)^2}.
\end{equation}
The pole at $x=0$ is double. Equation \eqref{orb:eq:multipole} gives
\begin{align}
\operatorname{Res}_{0}S_0^{(\alpha_{\mathrm{so}}=0)}
&=\left.\frac{d}{dx}
\left[-\frac{d(\rho_r^{-1}+d-2x)}{4(x-d)^2}\right]\right|_{x=0}\\
&=-\frac d4\left[-\frac2{d^2}+\frac{2(\rho_r^{-1}+d)}{d^3}\right]
=-\frac1{2\rho_r d^2},\\
\mathcal I[S_0^{(\alpha_{\mathrm{so}}=0)}]&=\frac{i}{2\rho_r d^2}.
\end{align}
The two pieces independent of $\rho_r^{-1}$ cancel. Freezing $p^2$ before differentiating this double pole would destroy that cancellation. With nonzero mean energy $\eta$, the additional term is $i\eta/d^2$, whose energy integral is zero. The SOC terms in \eqref{orb:eq:S0eval} are obtained by the same rule at the two helicity poles.

\subsubsection{Explicit Pauli traces multiplying the derivative vertices}
At $\mathbf p=(p,0)$, $v_x=p\mathbf{1} +\alpha_{\mathrm{so}} \hat{\mathsf R}_x$ and $v_y=\alpha_{\mathrm{so}} \hat{\mathsf R}_y$. The contraction in the last line of \eqref{orb:eq:allterms} gives
\begin{equation}
F_{\rm transfer}=d_0^+K_++d_0^-K_-+d_z^+Z_++d_z^-Z_-.
\end{equation}
Here the scalar-transfer traces are
\begin{align}
K_+(x)&=-p^2\left[
\frac{\ell_0(x-d)+\alpha_{\mathrm{so}}^2\ell_0}{D_B}
+\frac{\alpha_{\mathrm{so}}^2(2x-h)(p^2+x-d)}{D_RD_AD_B}\right],\label{orb:eq:Kplus}\\
K_-(x)&=-p^2\left[
\frac{\ell_0(x-d+h)+\alpha_{\mathrm{so}}^2\ell_0}{D_H}
+\frac{\alpha_{\mathrm{so}}^2(2x-h)(p^2+x-d+h)}{D_RD_AD_H}\right].\label{orb:eq:Kminus}
\end{align}
For example, $K_+=-p\operatorname{Tr} [AR A_\omega v_x]/2$. Writing $AR=\ell_0+\ell_1\hat{\mathsf R}_x$ and $A_\omega=b_0+b_1\hat{\mathsf R}_x$, its trace is $-p[\ell_0(pb_0+\alpha_{\mathrm{so}} b_1)+\ell_1(pb_1+\alpha_{\mathrm{so}} b_0)]$, which directly gives \eqref{orb:eq:Kplus}. The signs follow from the two Levi-Civita tensors, not from a choice of damping convention.

The out-of-plane-spin traces are
\begin{align}
Z_+(x)&=i\alpha_{\mathrm{so}}^2p^2\left[
\frac{\ell_0}{D_B}-\frac{(2x-h)(x-d)}{D_RD_AD_B}\right],\label{orb:eq:Zplus}\\
Z_-(x)&=i\alpha_{\mathrm{so}}^2p^2\left[
\frac{(2x-h)(x-d+h)}{D_RD_AD_H}-\frac{\ell_0}{D_H}\right].\label{orb:eq:Zminus}
\end{align}
For example, $Z_+=p\operatorname{Tr} [AR\sigma_zA_\omega v_y]/2
=i\alpha_{\mathrm{so}} p(\ell_0b_1-\ell_1b_0)$; this uses $\hat{\mathsf R}_x\hat{\mathsf R}_y=i\sigma_z$. These formulas display the SOC-dependent insertions explicitly.

\subsubsection{Residue table and two worked entries}
Define the density-corrected Drude bubble
\begin{equation}
{\widetilde\sigma_0}=\frac{i}{\rho_r d}+\frac{i\rho_r\Delta^2}{2d}.
\end{equation}
The radial residues through the needed order are
\begin{center}
\begin{tabular}{ll}
\toprule
Trace & Radial integral $\mathcal I$\\
\midrule
$K_+$ & $({\widetilde\sigma_0}+i)/h$\\[2mm]
$K_-$ & $(-{\widetilde\sigma_0}^*+i)/h$\\[2mm]
$Z_+$ & $-\Delta^2/[2h(d^2-\Delta^2)]$\\[2mm]
$Z_-$ & $+\Delta^2/[2h((d-2h)^2-\Delta^2)]$\\
\bottomrule
\end{tabular}
\end{center}
For the spinless part of $K_+$,
\begin{equation}
K_+^{(0)}=-\frac{\rho_r^{-1}+d-2x}{x(x-h)(x-d)}.
\end{equation}
Its single retarded pole at $x=0$ has residue $-(\rho_r^{-1}+d)/(hd)$, and hence
\begin{equation}
\mathcal I[K_+^{(0)}]=\frac{i}{\rho_r hd}+\frac ih.
\end{equation}
The constant $i/h$ is essential. It originates in the radial numerator and ultimately produces the imaginary-conductivity term in the triangle response. SOC adds $i\rho_r\Delta^2/(2hd)$, giving the table entry.

For a worked spin residue, write $b=\Delta/2$. At leading order,
\begin{equation}
Z_+=\frac{ib^2[-x^2+2dx+b^2-hd]}
{(x^2-b^2)[(x-h)^2-b^2][(x-d)^2-b^2]}.
\end{equation}
At $x=\lambda b$, the numerator becomes $ib^2d(2x-h)$, so
\begin{equation}
\operatorname{Res}_{\lambda b}Z_+=-\frac{ib^2}{2xh(d-2x)}\bigg|_{x=\lambda b}.
\end{equation}
Sum the two poles explicitly:
\begin{align}
\sum_\lambda\operatorname{Res}_{\lambda b}Z_+
&=-\frac{ib^2}{2hb}
\left[\frac1{d-2b}-\frac1{d+2b}\right]
=-\frac{2ib^2}{h(d^2-4b^2)},\\
\mathcal I[Z_+]&=-\frac{2b^2}{h(d^2-4b^2)}
=-\frac{\Delta^2}{2h(d^2-\Delta^2)}.
\end{align}
This is the out-of-plane-spin entry in the table. The $K_-,Z_-$ entries require both sets of retarded poles, since their integrands contain $R$ and $R_\omega$.

\subsubsection{Scalar-transfer contribution and cancellation of the large terms}
Using the residue table and \eqref{orb:eq:D0bubble}, the scalar radial contraction for one ordering is
\begin{align}
F_0(\omega)
&=\frac{\Gamma {\widetilde\sigma}}{2\omega}\frac{{\widetilde\sigma_0}+i}{h}
-\frac{\Gamma {\widetilde\sigma}^*}{2\omega}\frac{-{\widetilde\sigma_0}^*+i}{h}\\
&=-\frac i{2\omega}\left[{\widetilde\sigma}{\widetilde\sigma_0}+{\widetilde\sigma}^*{\widetilde\sigma_0}^*+i({\widetilde\sigma}-{\widetilde\sigma}^*)\right]\\
&=-\frac i\omega\left[\operatorname{Re} ({\widetilde\sigma}{\widetilde\sigma_0})-\operatorname{Im}  {\widetilde\sigma}\right].\label{orb:eq:F0}
\end{align}
The first term can scale as $\rho_r^{-2}$. It is an intermediate optical-density term, not a large magnetization. Under optical exchange, $\omega\to-\omega$, ${\widetilde\sigma}\to {\widetilde\sigma}^*$, and ${\widetilde\sigma_0}\to {\widetilde\sigma_0}^*$. Thus
\begin{equation}
F_0(-\omega)=\frac i\omega[\operatorname{Re} ({\widetilde\sigma}{\widetilde\sigma_0})+\operatorname{Im}  {\widetilde\sigma}],
\quad \boxed{F_0(\omega)+F_0(-\omega)=\frac{2i}{\omega}\operatorname{Im}  {\widetilde\sigma}.}
\end{equation}
Equation \eqref{orb:eq:FtoK} then gives
\begin{equation}
\frac{\kappa_{a,0}}{e^3}=-\frac12\operatorname{Im} \left(\frac{\sigma}{e^2}\right)\quad(m=1),
\qquad \boxed{\kappa_{a,0}=-\frac{e}{2m}\operatorname{Im} \sigma}\quad\hbox{in physical units}.\label{orb:eq:scalaranswer}
\end{equation}
This is an explicit cancellation between the two orderings. Keeping just one ordering would leave the spurious $\rho_r^{-2}$ term.

\subsubsection{Normal-spin transfer contribution}
Let $Z=\mathcal I[Z_+]$ and $d_z=d_z^+$. The table and \eqref{orb:eq:Dzeval} give $\mathcal I[Z_-]=Z^*$ and $d_z^-=-d_z^*$. Therefore
\begin{equation}
F_z(\omega)=d_zZ-d_z^*Z^*=2i\operatorname{Im} (d_zZ).
\end{equation}
On optical exchange the two $Z$ traces acquire a minus sign and their roles are exchanged, so $F_z(-\omega)=-F_z(\omega)$. Hence
\begin{equation}
\boxed{F_z(\omega)+F_z(-\omega)=0.}\label{orb:eq:spincancel}
\end{equation}
The cancellation holds at the displayed radial order after the signed energy integration; odd mean-energy corrections vanish by \eqref{eq:Fermi-window}. This is why the out-of-plane-spin optical-transfer group has zero net contribution. It was calculated rather than omitted.

\subsection{Third collapse: the SOC outer derivative becomes the missing bubble}
\subsubsection{Taking the Pauli trace of the outer curl}
Equation \eqref{orb:eq:outercurl} contains the out-of-plane-spin matrix $\sigma_z$. With $AR=\ell_0+\ell_1\hat{\mathsf R}_x$, $\{AR,\sigma_z\}=2\ell_0\sigma_z$. At $\mathbf p=(p,0)$, write $A_\omega=b_0+b_1\hat{\mathsf R}_x$ and $R_\omega=r_0+r_1\hat{\mathsf R}_x$. The antisymmetric optical trace is
\begin{align}
\operatorname{Tr} \sigma_z(V_x^+A_\omega v_y-V_y^+A_\omega v_x)
&=2i\alpha_{\mathrm{so}}[p(1+s)b_1+2\alpha_{\mathrm{so}} s b_0],\\
\operatorname{Tr} \sigma_z(v_xR_\omega V_y^--v_yR_\omega V_x^-)
&=2i\alpha_{\mathrm{so}}[p(1+s^*)r_1+2\alpha_{\mathrm{so}} s^*r_0].
\end{align}
These follow by multiplying $\hat{\mathsf R}_x,\hat{\mathsf R}_y$ using \eqref{orb:eq:paulirules}; terms proportional to $\mathbf{1} $ or $\hat{\mathsf R}_i$ have zero trace against $\sigma_z$.

Combining with the outer curl, including its Levi-Civita sign, gives the scalar integrand
\begin{equation}
\boxed{O(x)=-\ell_0\left[
\alpha_{\mathrm{so}}^2p^2\left(\frac{1+s}{D_B}-\frac{1+s^*}{D_H}\right)
+2\alpha_{\mathrm{so}}^2\left(\frac{s(x-d)}{D_B}-\frac{s^*(x-d+h)}{D_H}\right)
\right].}\label{orb:eq:Ointegrand}
\end{equation}
This equation is the full retained SOC outer trace before radial integration. It explicitly contains both the anomalous velocity and SOC in the propagators.

\subsubsection{Residues, their cancellation, and the bubble identification}
Let $D=d^2-\Delta^2$ and $\bar{D}=(d-2h)^2-\Delta^2$. Expanding \eqref{orb:eq:Ointegrand} with \eqref{orb:eq:radseries} and applying \eqref{orb:eq:Iresidue}--\eqref{orb:eq:multipole} gives the order-$\rho_r^0$ residue sum
\begin{equation}
\mathcal I[O]^{(0)}_\omega
=-\frac{i\Delta^2}{4hDD^*}
\left[D(s-s^*)-4h\omega(1+s)\right].\label{orb:eq:Ozero}
\end{equation}
For the opposite ordering, $d\to-\bar{d}$, $D\leftrightarrow \bar{D}$, $\omega\to-\omega$, and $s\leftrightarrow s^*$. The sum of the square brackets in \eqref{orb:eq:Ozero} is
\begin{equation}
(D-\bar{D})(s-s^*)-4h\omega(s-s^*)=0,
\end{equation}
because $D-\bar{D}=d^2-(d-2h)^2=4h\omega$. Thus the order-$\rho_r^0$ outer contribution cancels before evaluating $s$.

For clarity, the reduced polynomial numerator of the paired order-$\rho_r^1$ residues is
\begin{equation}
\mathcal P_1=-2hDD^*+s^*dD(\bar{d})^2-sd^2\bar{d}\bar{D}.
\end{equation}
The radial result is
\begin{align}
F_{\rm out}(\omega)+F_{\rm out}(-\omega)
&=-\frac{i\rho_r\Delta^2\mathcal P_1}{2d\bar{d}hDD^*}\\
&=\frac{i\rho_r\Delta^2}{2h}
\left[\frac{sd}{D}-\frac{s^*\bar{d}}{\bar{D}}
+\frac{2h}{d\bar{d}}\right].\label{orb:eq:Opaired}
\end{align}
The first line specifies the polynomial obtained from the double/higher poles. The second line follows by canceling the displayed factors in each of its three terms. It avoids hiding the SOC result in an unspecified ``integration by parts.''

Now use the independently calculated bubble \eqref{orb:eq:Bdiffnorm}:
\begin{align}
({\widetilde\sigma}-{\widetilde{\bar\sigma}})+({\widetilde\sigma}-{\widetilde{\bar\sigma}})^*
&=\frac{i\rho_r\Delta^2}{4}
\left[\frac{sd}{D}-\frac{s^*\bar{d}}{\bar{D}}
-\frac1d+\frac1{\bar{d}}\right]\\
&=\frac{i\rho_r\Delta^2}{4}
\left[\frac{sd}{D}-\frac{s^*\bar{d}}{\bar{D}}
+\frac{2h}{d\bar{d}}\right].
\end{align}
Comparison with \eqref{orb:eq:Opaired} proves the loop-to-bubble identity
\begin{equation}
\boxed{F_{\rm out}(\omega)+F_{\rm out}(-\omega)
=\frac2h[({\widetilde\sigma}-{\widetilde{\bar\sigma}})+({\widetilde\sigma}-{\widetilde{\bar\sigma}})^*]
=-\frac{4i}{\Gamma}\operatorname{Re} ({\widetilde\sigma}-{\widetilde{\bar\sigma}}).}\label{orb:eq:Ocollapse}
\end{equation}
Multiplication by \eqref{orb:eq:FtoK} gives
\begin{equation}
\frac{\kappa_{a,\rm out}}{e^3}=\frac\omega\Gamma\operatorname{Re} \left(\frac{\sigma-\bar\sigma}{e^2}\right)
\quad(m=1),
\qquad \boxed{\kappa_{a,\rm out}=\frac{e\omega}{m\Gamma}\operatorname{Re} (\sigma-\bar\sigma)}.\label{orb:eq:outeranswer}
\end{equation}
Equations \eqref{orb:eq:bareanswer} and \eqref{orb:eq:outeranswer} now add to the full conductivity bubble. This is precisely where the extra SOC terms are needed in the triangle reduction.

\subsubsection{Restoring mass and charge}
To justify the restored prefactors, put $\mathbf p=\sqrt m\,\widetilde{\mathbf p}$, $\widetilde\alpha_{\mathrm{so}}=\sqrt m\alpha_{\mathrm{so}}$, and $\widetilde{\mathbf Q}=\mathbf Q/\sqrt m$. Physical velocities are the normalized velocities divided by $\sqrt m$, while $d^2p=m\,d^2\widetilde p$. A three-velocity loop differentiated once with respect to physical $Q$ therefore carries $m/m^2=1/m$. A two-velocity conductivity loop carries $m/m=1$. The triangle has $e^3$, while conductivity has $e^2$, leaving $e/m$ in the group formulas above.

\subsection{Summing the triangles and adding the physical-current contact}
\subsubsection{The complete triangle-to-two-vertex reduction}
The four groups, in the convention $j_i=i\varepsilon^{(2)}_{i\ell}Q_\ell\kappa_a\mathcal{H} _a$, are
\begin{center}
\begin{tabular}{p{0.57\linewidth}l}
\toprule
Derivative group & Contribution to $\kappa_a$\\
\midrule
Four explicit optical-velocity derivatives & $e\omega\operatorname{Re} \bar\sigma/(m\Gamma)$\\[2mm]
Outer propagators and output ladder, including SOC & $e\omega\operatorname{Re} (\sigma-\bar\sigma)/(m\Gamma)$\\[2mm]
Scalar optical-transfer ladder & $-e\operatorname{Im} \sigma/(2m)$\\[2mm]
Normal-spin optical-transfer ladder & $0$ after both orderings\\
\bottomrule
\end{tabular}
\end{center}
Thus
\begin{align}
\kappa_a^\triangle
&=\frac{e\omega}{m\Gamma}
\left[\operatorname{Re} \bar\sigma+\operatorname{Re} (\sigma-\bar\sigma)\right]
-\frac{e}{2m}\operatorname{Im} \sigma\\
&=\boxed{\frac{e}{m}
\left[\frac\omega\Gamma\operatorname{Re} \sigma-\frac12\operatorname{Im} \sigma\right].}\label{orb:eq:triangleresult}
\end{align}
Equivalently, the tensor contraction is
\begin{equation}
\boxed{\mathscr C=-\frac{4\omega}{me^2\Gamma}\operatorname{Re} \sigma
+\frac2{me^2}\operatorname{Im} \sigma.}\label{orb:eq:completeC}
\end{equation}
This is a reduction of the differentiated three-current diagrams to two-current conductivity bubbles. The isolated triangles retain an imaginary-conductivity term. They do not by themselves equal the final boxed IFE formula.

\subsubsection{Evaluating the two measurement-current contacts}
From \eqref{orb:eq:minimal}, the contact current is
\begin{equation}
j_i^{\rm ct}=-\frac{e^2}{m}
\left(a_{+,i}n_-^{(1)}+a_{-,i}n_+^{(1)}\right).\label{orb:eq:contactstart}
\end{equation}
Here $n^{(1)}$ is the first-order number density, not electrical charge density. Its density-current bubble satisfies charge continuity:
\begin{equation}
-i\omega_s en_s^{(1)}+i\mathbf q_s\cdot\mathbf j_s^{(1)}=0,
\qquad \mathbf j_s^{(1)}=\sigma(s\omega)\mathbf E_s
=i\omega_s\sigma(s\omega)\mathbf a_s.
\end{equation}
Therefore
\begin{equation}
n_s^{(1)}=\frac{\mathbf q_s\cdot\mathbf j_s^{(1)}}{e\omega_s}
=\frac{i\sigma(s\omega)}{e}\mathbf q_s\cdot\mathbf a_s.\label{orb:eq:densityWard}
\end{equation}
This linear charge Ward identity evaluates a two-vertex contact diagram. It is not an angular-momentum balance assumption.

With $\mathbf q_s=\mathbf Q/2$, substitute \eqref{orb:eq:densityWard} in \eqref{orb:eq:contactstart}:
\begin{equation}
j_i^{\rm ct}=-\frac{ie}{2m}
\left[\sigma^*\delta_{ij}Q_k+\sigma\delta_{ik}Q_j\right]
a_{+,j}a_{-,k}.
\end{equation}
Insert $[a_{+,j}a_{-,k}]_H=-i\varepsilon^{(2)}_{jk}\mathcal{H} _a/2$. The contraction is
\begin{align}
j_i^{\rm ct,H}
&=-\frac{e}{4m}
\left[\sigma^*\varepsilon^{(2)}_{ik}Q_k+\sigma\varepsilon^{(2)}_{ji}Q_j\right]\mathcal{H} _a\\
&=-\frac{e}{4m}(\sigma^*-\sigma)\varepsilon^{(2)}_{i\ell}Q_\ell\mathcal{H} _a\\
&=\boxed{i\varepsilon^{(2)}_{i\ell}Q_\ell\frac{e}{2m}\operatorname{Im} \sigma\,\mathcal{H} _a.}\label{orb:eq:contactresult}
\end{align}
It cancels exactly the last term of \eqref{orb:eq:triangleresult}. The sum is
\begin{equation}
j_i^{(2),H}=i\varepsilon^{(2)}_{i\ell}Q_\ell
\frac{e\omega}{m\Gamma}\operatorname{Re} \sigma\,\mathcal{H} _a
=i\varepsilon^{(2)}_{i\ell}Q_\ell\frac{e}{m\Gamma\omega}
\operatorname{Re} \sigma\,\mathcal{H} .\label{orb:eq:fullcurrent}
\end{equation}
Comparison with \eqref{orb:eq:curl} proves, at the stated approximation order,
\begin{equation}
\boxed{M_{\mathrm{orb},z}=\frac{e}{mc_0\Gamma\omega}\operatorname{Re} \sigma(\omega)\mathcal{H} .}\label{orb:eq:Midentity}
\end{equation}
The $1/\Gamma$ arose from the static RA resolvent and its impurity vertex. The real part arose from the conjugate conductivity bubbles after cancellation of the imaginary term by the measurement-current contact. No finite output frequency was introduced.

\subsubsection{Relating the drawn bubbles to the evaluated contact term}
The square in Fig.~\ref{new:contact-diagrams}(b) multiplies the first-order \emph{number} density. Its electromagnetic charge factor is therefore $-e^2/m$, rather than the coefficient obtained by interpreting $n^{(1)}$ as charge density. The retarded bubble is evaluated using Eq.~\eqref{orb:eq:densityWard}; the two diagram terms then give precisely Eq.~\eqref{orb:eq:contactresult}. Together with Eq.~\eqref{orb:eq:triangleresult}, their cancellation can be displayed as
\begin{align*}
 \kappa_a^\triangle&=\frac em\left[
 \frac\omega\Gamma\operatorname{Re} \sigma-\frac12\operatorname{Im} \sigma\right],
 &\kappa_a^{\rm ct}&=\frac{e}{2m}\operatorname{Im} \sigma,\\
 \kappa_a^\triangle+\kappa_a^{\rm ct}
 &=\frac{e\omega}{m\Gamma}\operatorname{Re} \sigma,
 &M_{{\rm orb},z}&=\frac{e}{mc_0\Gamma\omega}\operatorname{Re} \sigma\,\mathcal{H} .
\end{align*}
This is a check of the contact signs and of the optical exchange, not an additional assumption about angular momentum. The triangle/contact division is not the kinetic electric/magnetic-force division. In the SOC-free Drude case the equality $\operatorname{Im} \sigma_0=(\omega/\Gamma)\operatorname{Re} \sigma_0$ makes triangle and contact each half the total. With Rashba structure that equality does not hold in general, although the \emph{sum} still reorganizes into the two equal force terms of Eq.~\eqref{eq:EB-halves-orb} within the stated analytical collision closure. These distinctions are essential when comparing a vector-potential calculation with an electric-force kinetic calculation.

\subsection{Final magnetization with the conductivity evaluated}
Substitute \eqref{orb:eq:ReSigmafinal} into the diagrammatic identity \eqref{orb:eq:Midentity}:
\begin{equation}
\boxed{\displaystyle
M_{\mathrm{orb},z}=\frac{e\sigma_D}{mc_0\Gamma\omega(1+X^2)}
\left[1+2\delta^2
-\frac{2\delta^2X^2B_R^2(2+3B_R^2-6X^2)}{\mathcal D}\right]\mathcal{H} .}\label{orb:eq:Mdimensionless}
\end{equation}
This is the orbital IFE with its real conductivity evaluated explicitly. The definitions are $\delta=\alpha_{\mathrm{so}}/v_F$, $X=\omega/\Gamma$, $B_R=\Delta_R/\Gamma$, and $\sigma_D=e^2\nu_0v_F^2/\Gamma$.

An equivalent expression in dimensional frequencies is obtained from
\begin{equation}
\mathcal D_{\rm phys}
=\Gamma^2(\Delta_R^2-4\omega^2)^2
+4\omega^2(\Gamma^2+\Delta_R^2-\omega^2)^2
=\Gamma^6\mathcal D.\label{orb:eq:Dphysical}
\end{equation}
Then
\begin{equation}
\boxed{\displaystyle
M_{\mathrm{orb},z}=\frac{e^3\nu_0v_F^2}{mc_0\omega(\Gamma^2+\omega^2)}
\left[1+2\delta^2
-\frac{2\delta^2\omega^2\Delta_R^2
(2\Gamma^2+3\Delta_R^2-6\omega^2)}{\mathcal D_{\rm phys}}\right]\mathcal{H} .}\label{orb:eq:Mphysical}
\end{equation}
All quantities inside the brackets are real and dimensionless. The sign of the magnetization includes the signed charge $e$; reversing optical helicity reverses $M_{\mathrm{orb},z}$.

\subsection{How the Lorentz force is represented in vector-potential diagrams}
Minimal coupling contains the orbital magnetic force already. The mechanical momenta obey $[\pi_i,\pi_j]=\mathrm{i}  e\varepsilon_{ijk}B_k/c_0$. Their Heisenberg equation, with a spatially uniform Rashba coefficient, is
\begin{equation}
 \dot\pi_i=e E_i+\frac{e}{2c_0}\varepsilon_{ijk}\{v_j,B_k\},
 \label{eq:operator-Lorentz}
\end{equation}
apart from the impurity force. The full Rashba velocity appears in this equation. Thus a calculation organized in $\mathbf A$ does not require a separate independent magnetic optical vertex to retain $\mathbf B=\boldsymbol\nabla\times\mathbf A$. It requires the correct momentum dependence and physical-current contacts. Differentiating the gauge-covariant Dyson equation reorganizes their sum into the density-force and Lorentz-force terms of Eq.~\eqref{eq:orb-ward-quadratic}.

The kinetic force decomposition gives two equal contributions $e\omega\operatorname{Re} \sigma/(2m\Gamma)$ to $\kappa_a$. The diagram decomposition instead gives
\begin{equation}
 \kappa_a^{\triangle}=\frac{e}{m}
 \left[\frac\omega\Gamma\operatorname{Re} \sigma-\frac12\operatorname{Im} \sigma\right],\qquad
 \kappa_a^{\rm ct}=\frac{e}{2m}\operatorname{Im} \sigma.
 \label{eq:force-versus-diagram}
\end{equation}
They are two organizations of the same complete response. For $\alpha_{\mathrm{so}}=0$, the Drude identity $\operatorname{Im} \sigma_0=(\omega/\Gamma)\operatorname{Re} \sigma_0$ makes the triangle and contact each half the total. With Rashba optical structure it does not generally hold, so the magnetic kinetic half cannot be assigned to the current contact alone. The survival of both kinetic forces is established by Eq.~\eqref{eq:EB-halves-orb}, while their sum is independently established by the differentiated diagrams.
\section{Frequency limits, angular-momentum balance, and matched comparisons}
\label{sec:scope}
\subsection{The two results and what controls their signs}
The calculations yield
\begin{equation}
 \boxed{M_{\mathrm{spin},z}=g\mu_B e^2\nu_0\mathcal C(\omega)\mathcal{H} ,
 \qquad M_{\mathrm{orb},z}=\frac{e}{mc_0\Gamma\omega}\operatorname{Re} \sigma(\omega)\mathcal{H} .}
 \label{eq:two-answers}
\end{equation}
The first probes a spin insertion and coherent precession; the second probes the current of a local orbital moment. The real conductivity measures absorption and is nonnegative for the passive exact system. The spin coefficient is a signed polarization response, not an absorption rate, and can change sign. Thus a sign reversal of $M_{\mathrm{spin},z}$ need not imply a reversal of $M_{\mathrm{orb},z}$. For a passive model, $\omega>0$, and $e<0$, the orbital sign is fixed by the helicity. This sign follows from the conductivity identity in the present parabolic-band approximation; it is not a universal restriction on multiband orbital IFE, whose material-dependent signs are established in first-principles calculations~\cite{Berritta2016,Adamantopoulos2025}. A negative truncated $\operatorname{Re} \sigma$ would indicate a failure of the approximation, not a new passive absorption channel.

For a compact description of the spin dynamics, eliminate the ladder functions:
\begin{align}
 s(\omega)&=\frac{2(z-\Gamma)(z^2+\Delta_R^2)}{P(\omega)},\nonumber\\
 P(\omega)&=2z^2(z-\Gamma)+\Delta_R^2(2z-\Gamma)
 =\Gamma(\Delta_R^2-4\omega^2)
 -2\mathrm{i} \omega(\Gamma^2+\Delta_R^2-\omega^2),\nonumber\\
 \mathcal C(\omega)&=\frac1{p_F^2}\operatorname{Im} 
 \frac{(z-\Gamma)(\Delta_R^2-2\Gamma z)}{zP(\omega)}.
 \label{eq:limits-compact}
\end{align}
The poles organize the discussion into three physical regimes. No expansion below replaces the full coefficient outside its stated hierarchy. All frequencies remain small compared with $\varepsilon_F$.

\subsection{Frequent momentum scattering: spin relaxation and the Edelstein regime}
For $\Delta_R\ll\Gamma$, spin turns through a small angle between collisions. The ladder gives the D'yakonov--Perel rates
\begin{equation}
 \gamma_\parallel=\frac{\Delta_R^2}{2\Gamma},\qquad
 \gamma_z=\frac{\Delta_R^2}{\Gamma}=2\gamma_\parallel,
 \qquad s(\omega)\simeq\frac{-\mathrm{i} \omega}{\gamma_\parallel-\mathrm{i} \omega}.
 \label{eq:dp-rates}
\end{equation}
These are eigenmode rates, not the one-particle width $\Gamma/2$. Near the slow optical relaxation feature,
\begin{equation}
 \mathcal C(\omega)\simeq\frac{1}{p_F^2\Gamma}
 \frac{\gamma_\parallel\omega}{\gamma_\parallel^2+\omega^2},
 \qquad \omega=O(\gamma_\parallel),\quad\Delta_R\ll\Gamma.
 \label{eq:dp-peak}
\end{equation}
Its maximum is $1/(2p_F^2\Gamma)$ at $\omega\simeq\gamma_\parallel$. This is a spin relaxation peak, not an interbranch absorption resonance. The approximation is not uniform over all $\omega\ll\Gamma$.

The same frequent-scattering material can be illuminated at $\omega\gg\Gamma$. In the hierarchy $\Delta_R\ll\Gamma\ll\omega$,
\begin{equation}
 \mathcal C\simeq-\frac{2\Gamma^2}{p_F^2\omega^3},\qquad
 S_z\simeq\frac{2e^2\nu_0\Gamma^2}{p_F^2\omega^3}\mathcal{H} .
 \label{eq:edelstein-hierarchy}
\end{equation}
The stationary weak-SOC limit is nonuniform. The output spin ladder diverges as the SOC spin relaxation rate vanishes, while its source also vanishes. At exactly $\alpha_{\mathrm{so}}=0$, scalar disorder and an electric field cannot create spin from an unpolarized state: $S_z=0$. One must retain a finite observation frequency or an explicitly specified additional spin relaxation channel before taking that limit. The nonzero $\Gamma^2$ term in Eq.~\eqref{eq:edelstein-hierarchy} is not spin generation in an SOC-free metal.

\subsection{Resolved splitting: coherent spin response and the interbranch resonance}
For $\Gamma\ll\Delta_R$, an electron can precess substantially between collisions. Away from the resonance and at fixed nonzero optical frequency,
\begin{equation}
 \mathcal C_{\rm cl}(\omega)=
 \frac{\Delta_R^2}{2p_F^2\omega(\Delta_R^2-\omega^2)},\qquad
 \Gamma\ll\omega,\ |\omega-\Delta_R|.
 \label{eq:spin-clean}
\end{equation}
It describes a dispersive spin response to virtual interbranch mixing: the coefficient changes sign across the splitting. Below the splitting but across the momentum-relaxation crossover, a useful single expression is
\begin{equation}
 \mathcal C\simeq\frac{\omega}{p_F^2}
 \left[\frac1{\Gamma^2+\omega^2}-\frac2{\Gamma^2+4\omega^2}\right],
 \qquad \Gamma,\omega\ll\Delta_R.
 \label{eq:resolved-crossover}
\end{equation}
It crosses zero at $\omega\simeq\Gamma/\sqrt2$ and tends to $1/(2p_F^2\omega)$ when $\Gamma\ll\omega\ll\Delta_R$. This intermediate $1/\omega$ law does not extend to a static electric drive.

Writing $\delta_\omega=\omega-\Delta_R=O(\Gamma)$ gives
\begin{equation}
 P\simeq\Delta_R^2(-3\Gamma+4\mathrm{i} \delta_\omega),\qquad
 \mathcal C\simeq-\frac{\delta_\omega}
 {4p_F^2[\delta_\omega^2+(3\Gamma/4)^2]}
 +O\!\left(\frac1{p_F^2\Delta_R}\right).
 \label{eq:spin-resonance}
\end{equation}
The dressed spin-precession pole has width $3\Gamma/4$, whereas the spin IFE line shape is dispersive. It must not be identified with the positive absorptive conductivity peak. The two actual Rashba contours allow transitions over an interval of width $4m\alpha_{\mathrm{so}}^2$. Representing them by one splitting is consistent near the resonance only when
\begin{equation}
 4m\alpha_{\mathrm{so}}^2\ll\Gamma\ll\Delta_R.
 \label{eq:resonance-validity}
\end{equation}
If disorder is narrower than this interval, the full momentum dependence of the splitting must be restored before assigning a resonance height or linewidth.

\subsection{Fast optical drive and the SOC-free orbital benchmark}
For $\max(\Gamma,\Delta_R)\ll\omega\ll\varepsilon_F$,
\begin{equation}
 \mathcal C=-\frac{\Delta_R^2+4\Gamma^2}{2p_F^2\omega^3}
 +O(\omega^{-5}),\qquad
 M_{\mathrm{orb},z}^{(0)}\longrightarrow
 \frac{n_0e^3}{m^2c_0\omega^3}\mathcal{H} .
 \label{eq:fast-tail}
\end{equation}
The clean spin tail is $\mathcal C\simeq-2\alpha_{\mathrm{so}}^2/\omega^3$, while the orbital response retains its spin-independent part at $\alpha_{\mathrm{so}}=0$. They are different mechanisms despite the same power of optical frequency in this window.

At fixed nonzero $\Gamma$ and $\Delta_R$, the exact small-optical-frequency expansion instead is
\begin{equation}
 \mathcal C=\frac{\omega}{p_F^2}
 \left(\frac2{\Delta_R^2}-\frac1{\Gamma^2}\right)+O(\omega^3).
 \label{eq:spin-static-check}
\end{equation}

For the same finite-SOC stationary limit, $s(0)=0$ and
\begin{equation}
 \sigma(0)=\frac{ne^2}{m\Gamma},\qquad
 n=n_0(1+2\delta^2),\qquad
 M_{\mathrm{orb},z}=
 \frac{ne^3}{m^2c_0\Gamma^2\omega}\mathcal{H}+O(\omega).
 \label{eq:orbital-static-limit}
\end{equation}
The spin response is generically linear in optical frequency at fixed nonzero SOC and disorder, while orbital accumulation has the $1/\omega$ law at fixed Fourier field amplitude. If $\Delta_R^2=2\Gamma^2$, the linear spin coefficient vanishes and the next odd power is required. The strict spin expansion requires frequencies below the slowest spin-relaxation scale; $\omega\ll\Gamma$ alone is insufficient when $\Delta_R\ll\Gamma$.

This is a useful check, rather than a fourth independent regime. The orbital expression grows as $1/\omega$ at fixed Fourier field amplitude; its derivation assumes a weak periodic drive and cannot be extrapolated to an arbitrarily slow field of fixed amplitude. The displacement and density modulations must remain perturbative. The limits of vanishing optical frequency, disorder, and SOC therefore need to be taken with their physical preparation specified.

\subsection{Optical helicity, absorbed power, and stored orbital angular momentum}
With our field amplitudes the absorbed power per unit area is
\begin{equation}
 \mathcal P_{\rm abs}=2\operatorname{Re} \sigma\,|\mathbf E|^2,
 \qquad \eta_{\rm pol}=\mathcal{H} /|\mathbf E|^2.
 \label{eq:power}
\end{equation}
$\eta_{\rm pol}$ is the signed degree of circular polarization: $\pm1$ for circular light and zero for linear light. It is optical spin angular momentum of the light, not an electron spin density. For the dipole response $\mathbf d_+=\mathrm{i} \sigma\mathbf E/\omega$, the cycle-averaged torque is
\begin{align}
 \mathcal T_z^{\rm opt}
 &=\overline{(\mathbf d\times\mathbf E_{\rm phys})_z}
 =\frac{\mathrm{i} \sigma}{\omega}(\mathbf E\times\mathbf E^*)_z
 -\frac{\mathrm{i} \sigma^*}{\omega}(\mathbf E^*\times\mathbf E)_z\nonumber\\
 &=\frac{2\operatorname{Re} \sigma}{\omega}\mathcal{H} 
 =\eta_{\rm pol}\frac{\mathcal P_{\rm abs}}{\omega}.
 \label{eq:optical-torque}
\end{align}
Torque is angular momentum transferred per unit time and area. The factor $1/\omega$ is the angular-momentum-to-energy ratio of a circular optical cycle. It explains why dissipation enters the accumulation interpretation through $\operatorname{Re} \sigma$.

In the SOC-free Drude model, let $\boldsymbol\xi$ denote optical displacement from an orbit center. The velocity and displacement amplitudes are
\begin{equation}
 \mathbf v_+=\frac{e\mathbf E}{mz},\qquad
 \boldsymbol\xi_+=\frac{\mathbf v_+}{-\mathrm{i} \omega},\qquad
 L_z=n_0m\,\overline{(\boldsymbol\xi\times\mathbf v)_z}
 =\frac{2n_0e^2}{m\omega(\Gamma^2+\omega^2)}\mathcal{H} .
 \label{eq:stored-orbital-L}
\end{equation}
Taking the cross product of $m\dot{\mathbf v}=e\mathbf E_{\rm phys}-m\Gamma\mathbf v$ with $\boldsymbol\xi$ gives
\begin{equation}
 \dot L_z=\mathcal T_z^{\rm opt}-\Gamma L_z,
 \qquad L_z^{\rm steady}=\tau_L\mathcal T_z^{\rm opt},\quad \tau_L=\Gamma^{-1},
 \qquad M_{\mathrm{orb},z}^{(0)}=\frac{e}{2mc_0}L_z.
 \label{eq:torque-lifetime}
\end{equation}
The gyromagnetic factor, torque, and lifetime now reproduce Eq.~\eqref{eq:orb-normal-final} with its full coefficient. This is a physical interpretation derived after the response calculation. At fixed nonzero $\omega$, taking $\Gamma\to0$ sends the absorbed torque to zero while the accumulation time diverges; their product has a finite limit. At exactly zero damping, the steady balance alone cannot select the orbital angular momentum. The periodic state must be specified by its preparation or switching prescription.

With Rashba SOC the orbital magnetic moment is measured by the full current, including $\alpha_{\mathrm{so}}\hat{\boldsymbol{\mathsf R}}$. Its factor $1/\Gamma$ follows from the gain/loss current moment in Eq.~\eqref{eq:impurity-current-moment} and the static diagrammatic resolvent. The conductivity result can consequently be written $M_{{\rm orb},z}=e\mathcal T_z^{\rm opt}/(2mc_0\Gamma)$. The mechanical angular momentum $\langle\mathbf r\times\mathbf{p} \rangle$ and the spin density are separately defined observables; the direct mechanical torque balance above concerns the SOC-free parabolic orbit.

\subsection{Comparisons needed for the main-text limits}
\label{sec:matched-comparisons}
In the hierarchy $\Delta_R\ll\Gamma\ll\omega$, Eq.~\eqref{eq:edelstein-hierarchy} has the $\Gamma^2\omega^{-3}$ scaling of Edelstein's dissipative spin-orientation theory~\cite{Edelstein:1998,Edelstein:2000}. That treatment retains the spin-output impurity vertex and neglects optical-vertex dressing in its high-frequency approximation. Here the optical factor $s(\omega)$ is retained before taking this limit. A coefficient comparison requires the same spin observable, field amplitude, and additional spin-relaxation channels: $\sigma_z$ and $\sigma_z/2$ differ by a factor of two, while a real-field phasor $\boldsymbol{\mathcal E}=2\mathbf E$ changes a quadratic coefficient by a factor of four. Agreement of the hierarchy is not an asserted equality of differently normalized coefficients.

The clean response in Eq.~\eqref{eq:spin-clean} also distinguishes coherent spin generation from a universally dissipative requirement. Static scalar disorder is time-reversal invariant; circular optical helicity is odd under time reversal. Disorder sets the impurity and spin-relaxation modes, whereas the drive supplies the time-reversal-odd perturbation. The SOC-free orbital response and the finite-SOC clean spin response therefore test distinct limits.

The perturbative spin calculations of Refs.~\cite{Taguchi:2011,Taguchi:2012} are compared only within their low-frequency and SOC expansions. The first uses light- and impurity-induced SOC; the second uses band Rashba SOC. Neither an angular-momentum-transfer interpretation nor a spin-density result evaluates the remaining orbital curl current. The stationary SOC limit described above also requires matching the output-frequency and spin-relaxation prescription before comparing a perturbative SOC coefficient.

For the ordinary-metal benchmark, the real-field phasor convention gives
\begin{equation}
 M_{\mathrm{orb},z}^{(0),\mathrm{clean}}=
 \frac{n_0e^3}{4m^2c_0\omega^3}
 i(\boldsymbol{\mathcal E}\times\boldsymbol{\mathcal E}^{*})_z,
 \qquad \boldsymbol{\mathcal E}=2\mathbf E.
 \label{eq:hertel-match}
\end{equation}
This recovers the local free-electron circulation benchmark~\cite{Hertel:2006}. In SI units, with volume density $n_{3D}$ in place of sheet density,
\begin{equation}
 \mathbf M_{\mathrm{SI}}^{(0)}=
 \frac{n_{3D}e^3}{4m^2\omega(\omega^2+\Gamma^2)}
 i(\boldsymbol{\mathcal E}\times\boldsymbol{\mathcal E}^{*}).
 \label{eq:SI-orbital}
\end{equation}
Comparison with the spin-independent quasiclassical response~\cite{Balatsky2024} is at $T=0$ and requires matching the signed magnetic moment, field convention, mass, and physical damping rate. The local coefficient is distinct from a net finite-beam moment with additional macroscopic drift currents~\cite{Hertel2015}; that drift is not the magnetic-force half or an individual current-contact diagram.

Reference~\cite{Maxim:2025} evaluates an SOC-dependent orbital curl-current coefficient, not the spin density. Its formal kinetic equation contains a collision convolution, but the evaluated response explicitly neglects driven collision feedback. The present calculation retains that feedback and both momentum and energy derivatives, as shown by the density conservation identity, the anomalous-current gain moment, and the differentiated vertices. Comparing frequency tails additionally requires distinguishing the full moment from an SOC subtraction and fixing the density convention. The explicit expansions below perform that distinction without identifying different current kernels merely because they share a power law.

\subsection{Full orbital moment and density-matched SOC correction}
\label{sec:orbital-subtractions}
The full orbital moment and an SOC-dependent subtraction have different asymptotic laws. Write $\sigma_D=e^2n_0/(m\Gamma)$ and $\delta=\alpha_{\mathrm{so}}/v_F$. Expanding Eq.~\eqref{orb:eq:image} at $\max(\Gamma,\Delta_R)\ll\omega\ll\varepsilon_F$ gives
\begin{align*}
 \frac{\operatorname{Re} \sigma}{\sigma_D}
 &=\frac{\Gamma^2}{\omega^2}(1+2\delta^2)
 +\frac{3\delta^2\Gamma^2\Delta_R^2-\Gamma^4(1+2\delta^2)}{\omega^4}
 +O(\omega^{-6}),\\
 M_{{\rm orb},z}
 &=\frac{n_0e^3\mathcal{H} }{m^2c_0\omega^3}
 \left[1+2\delta^2+
 \frac{3\delta^2\Delta_R^2-\Gamma^2(1+2\delta^2)}{\omega^2}
 +O(\omega^{-4})\right].
\end{align*}
These expansions concern the stated metallic window, not frequencies above the band scale. They also retain the same radial and spectral approximation as the preceding conductivity calculation.

At fixed chemical potential, define $\delta M_{\rm orb}^{(\mu)}=M_{\rm orb}(\alpha_{\mathrm{so}};\mu)-M_{\rm orb}(0;\mu)$. The two-branch density is $n(\mu,\alpha_{\mathrm{so}})=n_0(1+2\delta^2)$, so this subtraction includes a density change. The result is
\begin{equation*}
 \delta M_{{\rm orb},z}^{(\mu)}
 =\frac{n_0e^3\mathcal{H} }{m^2c_0}
 \left[\frac{2\delta^2}{\omega^3}
 +\frac{\delta^2(3\Delta_R^2-2\Gamma^2)}{\omega^5}
 +O(\omega^{-7})\right].
\end{equation*}
By contrast, subtracting a Drude reference at the \emph{same physical density},
\begin{equation*}
 M_{{\rm Drude},z}(n)
 =\frac{ne^3\mathcal{H} }{m^2c_0\omega(\Gamma^2+\omega^2)},
\end{equation*}
leaves
\begin{equation*}
 M_{{\rm orb},z}-M_{{\rm Drude},z}\bigl(n(\mu,\alpha_{\mathrm{so}})\bigr)
 =\frac{3n_0e^3\delta^2\Delta_R^2}{m^2c_0\omega^5}\mathcal{H} 
 +O(\omega^{-7}).
\end{equation*}
This density-matched residual can have an $\omega^{-5}$ tail even though the \emph{full} orbital magnetization has an $\omega^{-3}$ tail. Its displayed coefficient contains $\delta^2\Delta_R^2$; formally this is fourth order in $\alpha_{\mathrm{so}}$ when both factors are expanded. That statement describes the retained kernel, not an additional all-orders radial-expansion result. At fixed density, the chemical potential must be adjusted consistently before defining the SOC-free reference.

Consequently, the frequency power alone does not establish agreement between a full moment and a subtracted response. For a fixed-chemical-potential comparison the leading SOC correction above has an $\omega^{-3}$ density term. For a density-matched comparison the frequency power can agree with the earlier $\omega^{-5}$ coefficient while its SOC dependence and current kernel need not agree. The density-matched residual is not identified with the coefficient of Ref.~\cite{Maxim:2025}. A complete quantitative comparison requires the same subtraction, field amplitude, signed magnetic moment, and density prescription. These equations state precisely what is recovered and what remains different.

\newcounter{savedSupplementEquation}
\setcounter{savedSupplementEquation}{\value{equation}}
\appendix
\section{Collision and spectral bookkeeping at quadratic order}
\setcounter{equation}{\value{savedSupplementEquation}}
\label{app:nonlinear}
This appendix records the second-order products underlying the Born collision block and the physical distribution used in the main calculation. The single-splitting metallic and equilibrium-spectral approximation is the one specified in Section~\ref{sec:model}; its bookkeeping must be applied to both collisions and observables.

\subsection{An explicit second-order product expansion}
Write $G^{R,A}=G_0^{R,A}+G^{R,A,(1)}+G^{R,A,(2)}+\cdots$, $h=h^{(0)}+h^{(1)}+h^{(2)}+\cdots$, and similarly for $\Sigma$. There are no factorials. Let $\Phi_s[A,B]$ be the linear-field correction to the covariant product and $\Psi_{+-}[A,B]$ its mixed second-order correction, if retained at the chosen gradient order. At the mixed quadratic output,
\begin{align}
 (A\circ B)^{(2)}_{+-}={}&A_0\circ_0B^{(2)}+A^{(2)}\circ_0B_0
 +\sum_s A_s^{(1)}\circ_0B_{-s}^{(1)}\nonumber\\
 &+\sum_s\big[\Phi_s[A_0,B_{-s}^{(1)}]
 +\Phi_s[A_{-s}^{(1)},B_0]\big]+\Psi_{+-}[A_0,B_0].
 \label{eq:quadratic-product-full}
\end{align}
The half-momentum and half-frequency shifts in each $\circ_0$ are fixed by Eq.~\eqref{eq:shifts}. The local product in a particular projected calculation is an approximation to this expression, not its definition.

For example, the quadratic Keldysh self-energy contains
\begin{align}
 \Sigma^{K,(2)}=\gamma_{\rm imp}\int\!\frac{\mathrm{d}^2p}{(2\pi)^2} \Big\{&
 G_0^R\circ_0 h^{(2)}-h^{(2)}\circ_0G_0^A
 +G^{R,(2)}\circ_0h^{(0)}-h^{(0)}\circ_0G^{A,(2)}\nonumber\\
 &+\sum_s\big[G_s^{R,(1)}\circ_0h_{-s}^{(1)}
 -h_s^{(1)}\circ_0G_{-s}^{A,(1)}\big]\nonumber\\
 &+\sum_s\big[\Phi_s[G_0^R,h_{-s}^{(1)}]
 +\Phi_s[G_{-s}^{R,(1)},h^{(0)}]
 -\Phi_s[h_{-s}^{(1)},G_0^A]
 -\Phi_s[h^{(0)},G_{-s}^{A,(1)}]\big]\nonumber\\
 &+\Psi_{+-}[G_0^R,h^{(0)}]-\Psi_{+-}[h^{(0)},G_0^A]\Big\}.
 \label{eq:full-SigmaK2}
\end{align}
The corresponding loss part in $I^{(2)}$ is obtained from the same expansion with $G\to\Sigma$ and the prefactor $-\mathrm{i} $, as prescribed by Eq.~\eqref{eq:collision-exact}. This includes $\Sigma_s^{R,(1)}\circ_0h_{-s}^{(1)}$ and its advanced partner. Terms involving $G^{R,A,(2)}h^{(0)}$ cancel their corresponding loss self-energies by SCBA, just as at first order. The mixed spectral/distribution terms do not cancel by that argument alone, because $h^{(1)}$ depends on momentum and is a spin matrix. Their treatment must be stated at the selected radial/gradient order.

The same issue appears in reconstruction of the observable:
\begin{equation}
 G^{<,(2)}=\tfrac12\big[(G^R\circ h-h\circ G^A)^{(2)}-G^{R,(2)}+G^{A,(2)}\big].
 \label{eq:lesser2-full}
\end{equation}
Equation~\eqref{eq:quadratic-product-full} displays the additional terms beyond $\tfrac12(G_0^Rh^{(2)}-h^{(2)}G_0^A)$. The projected spin observable of Eq.~\eqref{eq:projected-lesser} and the static spectral gain of Eq.~\eqref{eq:projected-collision-orb} are the explicitly declared reductions used in this supplement. The mixed first-order spectral/distribution terms in Eqs.~\eqref{eq:full-SigmaK2} and \eqref{eq:lesser2-full} remain explicit in this record: their absence from the projected block is part of the stated spectral reduction, rather than a consequence of the first-order cancellation.

\subsection{Checking the collision moment used in the orbital balance}
In an unprojected calculation define
\begin{equation}
 \mathcal R_i^{(2)}=F_{{\rm imp},i}^{(2)}
 +\frac{m\Gamma}{e}j_i^{(2)}.
 \label{eq:orb-remainder}
\end{equation}
The momentum Ward identity then gives to first order in $Q$
\begin{equation}
 \Gamma j_i^{(2),H}
 =\mathrm{i} \varepsilon^{(2)}_{i\ell}Q_\ell\frac{e}{m\omega}\operatorname{Re} \sigma\,\mathcal{H} 
 -\frac{\mathrm{i}  e}{m}Q_jT_{ij}^{(2),H}(0)
 +\frac{e}{m}\mathcal R_i^{(2),H}.
 \label{eq:orb-remainder-Ward}
\end{equation}
For the collision block evaluated in Section~\ref{sec:orb-kinetics}, Eq.~\eqref{eq:impurity-current-moment} sets $\mathcal R_i^{(2)}=0$. The stationary Rashba spin identity sets the helicity stress at $Q=0$ to zero. The two optical forces then determine the current. If one evaluates the full spectral products instead of this projected block, the same momentum moment defines the residual $\mathcal R_i^{(2)}$; its helicity coefficient linear in $Q$ must be checked rather than removed by setting $\Omega=0$. This is the precise consistency check on the collision closure.

\setcounter{savedSupplementEquation}{\value{equation}}
\section{SOC-free phase-space calculation with explicit angular harmonics}
\setcounter{equation}{\value{savedSupplementEquation}}
\label{app:scalar}
Section~\ref{sec:normal-total} evaluated the SOC-free current moments with the energy-resolved Wigner density throughout. This appendix records additional harmonic and radial checks of the same phase-space calculation. It uses physical energy-resolved density, not the Keldysh kernel $h$. At $\alpha_{\mathrm{so}}=0$, write $\hat w=w\mathbf{1} $ and $\mathbf v=\mathbf{p} /m$. The equilibrium density is
\begin{equation}
 w^{(0)}_{\mathbf{p} \epsilon}=f_0(\epsilon)\delta(\epsilon-\xi_p),\qquad
 \mathscr D_i=\partial_{p_i}+v_i\partial_\epsilon,
 \quad\mathscr D_iw^{(0)}=v_i f_0'(\epsilon)\delta(\epsilon-\xi_p).
 \label{eq:scalar-equilibrium}
\end{equation}
The derivatives of the spectral delta function cancel: $\partial_{p_i}\delta(\epsilon-\xi_p)=-v_i\delta'$ and $v_i\partial_\epsilon\delta=v_i\delta'$. This is why the energy derivative is necessary in a phase-space density equation.

For scalar isotropic elastic scattering, $C[w]=-\Gamma(w-\langle w\rangle_\phi)$ at each radius and energy. The density harmonic is conserved, while all nonzero angular harmonics decay at $\Gamma$. The first-order equation is
\begin{equation}
 [d_s+\mathrm{i} \mathbf q_s\cdot\mathbf v+\Gamma(1-P_0)]w_s^{(1)}
 =-e\mathbf E_s\cdot\mathbf v\,f_0'\delta(\epsilon-\xi_p),
 \qquad d_s=-\mathrm{i} \omega_s,\quad z_s=\Gamma+d_s.
 \label{eq:scalar-linear-full}
\end{equation}
Here $P_0$ is the angular average. Expand $w_s^{(1)}=w_{s,0}+w_{s,1}+w_{s,2}+\cdots$ into angular harmonics, keeping the first gradient. The leading drift harmonic and the induced isotropic harmonic are
\begin{align}
 w_{s,1}&=-\frac{e}{z_s}(\mathbf E_s\cdot\mathbf v)f_0'\delta,\nonumber\\
 d_s w_{s,0}&=-\mathrm{i} \langle(\mathbf q_s\cdot\mathbf v)w_{s,1}\rangle
 =\frac{\mathrm{i}  ev^2}{2z_s}(\mathbf q_s\cdot\mathbf E_s)f_0'\delta,\nonumber\\
 w_{s,0}&=-\frac{ev^2}{2z_s\omega_s}
 (\mathbf q_s\cdot\mathbf E_s)f_0'\delta.
 \label{eq:scalar-l01}
\end{align}
The $1/\omega_s$ comes from conserved charge, not from a collision rate assigned to the density. The traceless second harmonic is
\begin{equation}
 w_{s,2}=\frac{\mathrm{i}  e}{z_s^2}
 \left[(\mathbf q_s\cdot\mathbf v)(\mathbf E_s\cdot\mathbf v)
 -\frac{v^2}{2}\mathbf q_s\cdot\mathbf E_s\right]f_0'\delta.
 \label{eq:scalar-l2}
\end{equation}
The scalar source has zero angular mean after subtraction. If a density propagator is needed beyond first order in $q_s$, feedback of $w_{s,0}$ through $w_{s,1}$ gives
\begin{equation}
 \left(d_s+\frac{v^2q_s^2}{2z_s}\right)w_{s,0}
 =\frac{\mathrm{i}  ev^2}{2z_s}(\mathbf q_s\cdot\mathbf E_s)f_0'\delta+O(q_s^3).
 \label{eq:scalar-density-diffusion}
\end{equation}
This is the conserving density ladder. Its diffusion coefficient is frequency dependent; taking $d_s+\Gamma$ in place of $d_s$ would give an incorrect electric contribution.

Define energy-resolved number, velocity, and stress moments with the spin trace included:
\begin{equation}
 N=2\int\!\frac{\mathrm{d}^2p}{(2\pi)^2}  w,\qquad \mathcal P_i=2\int\!\frac{\mathrm{d}^2p}{(2\pi)^2}  v_iw,
 \qquad \Pi_{ij}=2\int\!\frac{\mathrm{d}^2p}{(2\pi)^2}  v_iv_jw.
 \label{eq:scalar-moments}
\end{equation}
The linear drift moment is
\begin{equation}
 e\mathcal P_{s,i}^{(1)}=\sigma_\epsilon(s\omega)E_{s,i},\qquad
 \sigma_\epsilon(s\omega)=-\frac{e^2\nu_0v^2}{z_s}f_0'(\epsilon),\quad
 v^2=\frac{2(\epsilon+\varepsilon_F)}m,
 \quad \sigma(s\omega)=\int\mathrm{d} \epsilon\,\sigma_\epsilon(s\omega).
 \label{eq:scalar-spectral-conductivity}
\end{equation}
Equations~\eqref{eq:scalar-l01} and \eqref{eq:scalar-moments} give $e\omega_sN_s^{(1)}=\mathbf q_s\cdot\mathbf j_s^{(1)}$ at each energy, reproducing the charge Ward identity. At $T=0$, the integral in Eq.~\eqref{eq:scalar-spectral-conductivity} gives $e^2\nu_0v_F^2/z_s=n_0e^2/(mz_s)$.

\subsection{Explicit electric integration by parts}
The quadratic electric source is $-eE_{s,j}\mathscr D_jw_{-s}^{(1)}$. Its first velocity moment is
\begin{align}
 2\int\!\frac{\mathrm{d}^2p}{(2\pi)^2}  v_i[-eE_{s,j}\mathscr D_jw_{-s}^{(1)}]
 &=\frac{e}{m}E_{s,i}N_{-s}^{(1)}
 -eE_{s,j}\partial_\epsilon\Pi_{-s,ij}^{(1)}.
 \label{eq:scalar-electric-moment}
\end{align}
The first term follows from $\partial_{p_j}v_i=\delta_{ij}/m$ and vanishing momentum boundary terms. The second is the complete energy derivative of the stress moment; no derivative of a spectral delta function was discarded. Its integral over $\epsilon$ is a boundary term and vanishes for the localized perturbation in the degenerate metallic approximation. The $l=0$ harmonic therefore supplies the surviving electric density force. The $l=2$ stress is retained until this cancellation; it does not independently generate the final helicity current.

\subsection{Explicit magnetic integration by parts}
For the magnetic source $-(e/c_0)(\mathbf v\times\mathbf B_s)_j\partial_{p_j}w_{-s}^{(1)}$,
\begin{align}
 2\int\!\frac{\mathrm{d}^2p}{(2\pi)^2}  v_i\left[-\frac{e}{c_0}(\mathbf v\times\mathbf B_s)_j
 \partial_{p_j}w_{-s}^{(1)}\right]
 &=\frac{e}{mc_0}(\boldsymbol{\mathcal P}_{-s}^{(1)}\times\mathbf B_s)_i.
 \label{eq:scalar-magnetic-moment}
\end{align}
Here $\partial_{p_j}(\mathbf v\times\mathbf B_s)_j=0$ and the derivative acting on $v_i$ gives $\delta_{ij}/m$. Because $B_s$ is already first order in $Q$, only the $l=1$ drift at zeroth gradient is needed. This explains why the magnetic source comes from a different first-order harmonic than the electric density source.

The first velocity moment of the static helicity response relaxes at $\Gamma$. Its uniform stress is symmetric, so its helicity-odd part vanishes. Integrating Eqs.~\eqref{eq:scalar-electric-moment} and \eqref{eq:scalar-magnetic-moment} over energy, multiplying by $e$, and using Eq.~\eqref{eq:Faraday-explicit} gives Eqs.~\eqref{eq:jE-orb}--\eqref{eq:EB-halves-orb} with $\sigma=\sigma_0$. This completes the scalar calculation without identifying a triangle with one particular force term.

The radial integrals useful for checking each harmonic are, with $K=\epsilon+\varepsilon_F>0$,
\begin{equation}
 \int\!\frac{\mathrm{d}^2p}{(2\pi)^2}  v^2\delta(K-p^2/2m)=\frac K\pi,\qquad
 \int\!\frac{\mathrm{d}^2p}{(2\pi)^2}  v^4\delta(K-p^2/2m)=\frac{2K^2}{\pi m},\qquad
 \langle v_iv_jv_kv_\ell\rangle=\frac{v^4}{8}
 (\delta_{ij}\delta_{k\ell}+\delta_{ik}\delta_{j\ell}+\delta_{i\ell}\delta_{jk}).
 \label{eq:scalar-radial-checks}
\end{equation}
$K$ is kinetic energy measured from the bottom, whereas every Fermi function in this supplement uses $\epsilon$ measured from the chemical potential. Substituting $K=\epsilon$ would change the density normalization and remove the Fermi-surface contribution erroneously.

\setcounter{savedSupplementEquation}{\value{equation}}
\section{Checks that make the calculation reproducible}
\setcounter{equation}{\value{savedSupplementEquation}}
\label{app:checks}
The following checks test the derivation rather than merely fit its final frequency dependence.
\begin{enumerate}
\item Fix $G^<$, $e$, the spin vertex $\sigma_z/2$, and the positive-frequency field amplitude before evaluating a trace. An extra spin degeneracy or phasor factor changes a nonlinear response by a finite numerical factor.
\item Derive $\Sigma^{R,A}=\mp\mathrm{i} \Gamma/2$ and the $RA$ width $\Gamma$ from the same disorder correlator. Evaluate both branch radii and weights before performing a velocity moment.
\item Check the scalar eigenvalue $w$ and the spin eigenvalues $u,v$ of the ladder independently. Conservation fixes the unforced scalar dc mode through a density boundary condition; it does not permit division by $1-w(0)$.
\item Compare the kinetic source factor $s$ with the dressed optical velocity of Eq.~\eqref{eq:dressed-optical}. Calculate the four conductivity moments before attempting to identify a triangle with a bubble.
\item For spin, retain all three quadratic source terms, the first radial numerator moment, its contour arc, and the energy derivative of the output spin ladder. Use $\epsilon\delta'(\epsilon)=-\delta(\epsilon)$ before evaluating energy at zero.
\item For orbital response, differentiate shifted propagators, explicit mean-momentum vertices, and finite-transfer ladders. Verify the two Levi-Civita contractions and the factor $4$ in Eq.~\eqref{orb:eq:CtoK}. The zero-gradient value of a vertex does not determine its transfer derivative.
\item Keep the two optical orderings separate through the signed Fermi-window integration. Terms proportional to large powers of the radial numerator may cancel only after this exchange. Conjugate matrices in their reversed order.
\item Add the measurement-current contacts. For a Drude conductivity only, $\operatorname{Im} \sigma_0=(\omega/\Gamma)\operatorname{Re} \sigma_0$, so triangle and contact coefficients happen to be equal. For the Rashba conductivity this relation is not generally true; it cannot identify the contact with the magnetic kinetic half.
\item Verify $\sigma(0)=ne^2/(m\Gamma)$, Eq.~\eqref{eq:orb-normal-final}, helicity reversal, evenness under $\alpha_{\mathrm{so}}\to-\alpha_{\mathrm{so}}$, $\sigma(-\omega)=\sigma(\omega)^*$, and the disappearance of spin generation at exactly $\alpha_{\mathrm{so}}=0$. A singular steady-state limit is not a counterexample to the latter symmetry.
\item Check the collision moment by evaluating gain and loss separately before using Eq.~\eqref{eq:impurity-current-moment}. If full spectral products are retained, also evaluate their residual in Eq.~\eqref{eq:orb-remainder} at first order in output momentum.
\end{enumerate}

\setcounter{savedSupplementEquation}{\value{equation}}
\section{Full-momentum numerical resonance and comparison with the analytical reduction}
\setcounter{equation}{\value{savedSupplementEquation}}
\label{app:numerics}

The main-text resonance figure is obtained by evaluating the full-momentum nonlinear response across both physical Rashba Fermi contours. This section gives the numerical equations and explains why evaluating the final single-splitting metallic formulas is not a quantitatively adequate substitute at those parameters. The numerical curves are calculated independently of Eqs.~\eqref{eq:final-spin}, \eqref{orb:eq:image}, and \eqref{eq:orb-kinetic-final}; none of those coefficients is imposed on the numerical magnetization.

\subsection{Equilibrium propagators, regulator, and optical response}
We retain the full Rashba Hamiltonian $H_0(\mathbf{p} )=(p^2/2m-\varepsilon_F)\mathbf{1} +\alpha_{\mathrm{so}}(p_y\sigma_x-p_x\sigma_y)$ and the momentum-dependent splitting $2\alpha_{\mathrm{so}}p$. The equilibrium retarded propagator and impurity self-energy satisfy
\begin{align}
 G_0^R(\epsilon,\mathbf{p} )&=[\epsilon-H_0(\mathbf{p} )-\Sigma_{\rm imp}^R(\epsilon)+\mathrm{i} \eta]^{-1},\\
 \Sigma_{\rm imp}^R(\epsilon)&=\gamma_{\rm imp}\int\!\frac{\mathrm{d}^2p}{(2\pi)^2}  G_0^R(\epsilon,\mathbf{p} ),
 \qquad \gamma_{\rm imp}=\frac{\Gamma}{2\pi\nu_0}.
\end{align}
The scalar disorder is treated in noncrossing SCBA, with the real part and energy dependence of $\Sigma_{\rm imp}^R$ retained. The integral has the explicitly specified band cutoff $0\le p^2/(2m)\le4\varepsilon_F$, and the bare chemical potential is held fixed. A wide-band thermal reservoir has $\Sigma_{\rm bath}^R=-\mathrm{i} \eta$ and $\Sigma_{\rm bath}^<=2\mathrm{i} \eta f_0(\epsilon)$. It defines a stationary distribution with population relaxation rate $2\eta$. The production curves use $\eta/\Gamma=0.0025$; the reservoir remains finite.

We expand the minimally coupled Dyson--Keldysh system to second order in the positive- and negative-frequency fields. Both spectral and distribution responses of the impurity self-energy are retained. At each harmonic, each unknown self-energy matrix is determined by a linear equation of the form
\begin{equation}
 C=\gamma_{\rm imp}\int\!\frac{\mathrm{d}^2p}{(2\pi)^2} [S(\mathbf{p} )+L(\mathbf{p} )C R(\mathbf{p} )].
\end{equation}
The four Pauli components of $C$ give a $4\times4$ complex linear system. This procedure includes scattering-in feedback; it does not replace it by a scalar loss rate. In particular, the physical quadratic lesser density is reconstructed from the complete products of the retarded, advanced, and lesser responses, rather than identified with the distribution kernel $h^{(2)}$ alone.

For $\mathbf E_+=E_0(\mathbf e_x+\mathrm{i} \ell\mathbf e_y)/\sqrt2$, the optical vector potential is $\mathbf A_+=c_0\mathbf E_+/(\mathrm{i} \omega)$ and $\mathbf A_- =\mathbf A_+^*$. The Hamiltonian vertices are $-e\mathbf A\cdot\mathbf v/c_0$ and $e^2\mathbf A^2/(2mc_0^2)$. The quadratic Hamiltonian contact and the field-dependent physical measurement current are retained separately. Defining $\rho^{(n)}_\mathbf{p} =\int\mathrm{d} \epsilon[-\mathrm{i}  G^{<,(n)}]/(2\pi)$, the observables are
\begin{align}
 M_{{\rm spin},z}&=-\frac{g\mu_B}{2}\int\!\frac{\mathrm{d}^2p}{(2\pi)^2} \operatorname{Tr} [\sigma_z\rho^{(2)}_\mathbf{p} ],\\
 j_y^{(2)}&=e\int\!\frac{\mathrm{d}^2p}{(2\pi)^2} \operatorname{Tr} [v_y\rho^{(2)}_\mathbf{p} ]
 -\frac{e^2}{mc_0}\sum_{a=\pm} A_{y,a}\int\!\frac{\mathrm{d}^2p}{(2\pi)^2} \operatorname{Tr} \rho_{\mathbf{p} ,-a}^{(1)},\\
 M_{{\rm orb},z}&=\left.\frac{\mathrm{i} }{c_0}\partial_{Q_x}j_y^{(2),H}\right|_{Q=0}.
\end{align}
Here the response-kernel routing is $q_+=q_-=Q/2$, $\omega_+=\omega$, and $\omega_-=-\omega$, as in Eq.~\eqref{eq:orb-routing}. These are independent kernel arguments, not a claim about the wavevectors of conjugate plane waves. Outer electron momenta are $\mathbf{p} \pm\mathbf Q/2$, and optical mean momenta are shifted by $\pm\mathbf Q/4$. The derivative in $Q_x$ is propagated analytically through every matrix product and self-consistent feedback equation using dual matrices. There is no finite-$Q$ subtraction or fitted extrapolation. The linear conductivity is calculated from its own first-order current response, independently of the orbital moment.

At $T=0$, the real-energy interval $[-\omega,\omega]$ is split at zero. Below $-\omega$, all three retained optical harmonics are occupied. The corresponding lesser response equals its spectral discontinuity and is integrated on a vertical complex-energy contour starting at $-\omega$. The substitution $y=\Gamma t/(1-t)$ maps its infinite range to $0\le t<1$. This contour deformation retains the occupied sea and avoids cancellation errors in a coarse real-axis quadrature. Momentum integration uses composite Gauss--Legendre panels around all shifted Rashba poles. All propagators and feedback kernels at a given integration energy use the same quadrature.

\subsection{Normalization and numerical accuracy}
The numerical units are $\hbar=\varepsilon_F=m=c_0=|e|=1$, with signed charge $e=-1$. For comparison of the two moments, the main figure uses $g=2$ and $m=m_e$, so $\mu_B=1/2$. Both panels use $M_0=\mu_B e^2|E_0|^2/(2\pi\varepsilon_F^2)$; no separate amplitude multiplier is fitted. For another band mass or spin $g$ factor, the relative magnetic weights must be restored accordingly. The calculations are zero-field response coefficients. Changing $\ell$ reverses both signed moments while leaving conductivity unchanged; the plotted helicity is $\ell=-1$.

There are 81 directly computed frequencies per disorder value, $\omega/\varepsilon_F=0.18,0.182,\ldots,0.34$, for $\varepsilon_F\tau=50,100,150$. A $C^2$ not-a-knot cubic interpolant passes through every calculated point and is displayed at 1601 frequencies. No smoothing filter, susceptibility fit, baseline subtraction, frequency shift, or selected replacement of data is used. Independent midpoint calculations test the interpolation: the largest spin and orbital deviations are $0.00492M_0$ and $0.0367M_0$, respectively, both at $\varepsilon_F\tau=150$. They are below $1.3\times10^{-4}$ of the spin panel height and $4.6\times10^{-5}$ of the orbital panel height. These are executed spot checks, not uniform error bounds at every frequency.

Independent uniform radial quadrature and refined pole-aware quadrature agree within $6\times10^{-6}$ relative for the tested spin and orbital coefficients at $\omega/\varepsilon_F=0.2575,0.265,0.2725$ and $\varepsilon_F\tau=150$. Fresh runs reproduce helicity reversal at $0.265$. Halving $\eta$ there changes spin by $0.036\%$ and orbital magnetization by $0.555\%$. Additional archived checks test angular resolution, the SOC-free orbital benchmark, the cutoff, and the impurity feedback residual. All raw scans, executed checks, and the C++ solver are included in the source package.

\subsection{Why the numerical curves are used in the Letter}
The two Rashba Fermi radii are $p_\lambda=\sqrt{p_F^2+m^2\alpha_{\mathrm{so}}^2}-\lambda m\alpha_{\mathrm{so}}$. Their clean vertical transition energies fill the interval $2\alpha_{\mathrm{so}}p_+\le\omega\le2\alpha_{\mathrm{so}}p_-$, of width $4m\alpha_{\mathrm{so}}^2$. With $\beta=\alpha_{\mathrm{so}}p_F/\varepsilon_F=0.13$, the interval is $0.24365\lesssim\omega/\varepsilon_F\lesssim0.27745$. It is shaded in the figures as a guide; finite-disorder self-energy effects can shift the features.

The single-splitting metallic equations replace this interval by a single splitting. Quantitative use of that replacement near a resolved resonance requires Eq.~\eqref{eq:resonance-validity}, particularly $r=4m\alpha_{\mathrm{so}}^2/\Gamma\ll1$. For the main-text parameters, $r=1.69,3.38,5.07$, respectively. Thus evaluating the final equations more densely cannot repair their unresolved radial transition spread. Numerical momentum integration is needed to determine the peak shapes at these parameters.

\begin{figure}[t]
\centering
\includegraphics[width=.95\textwidth]{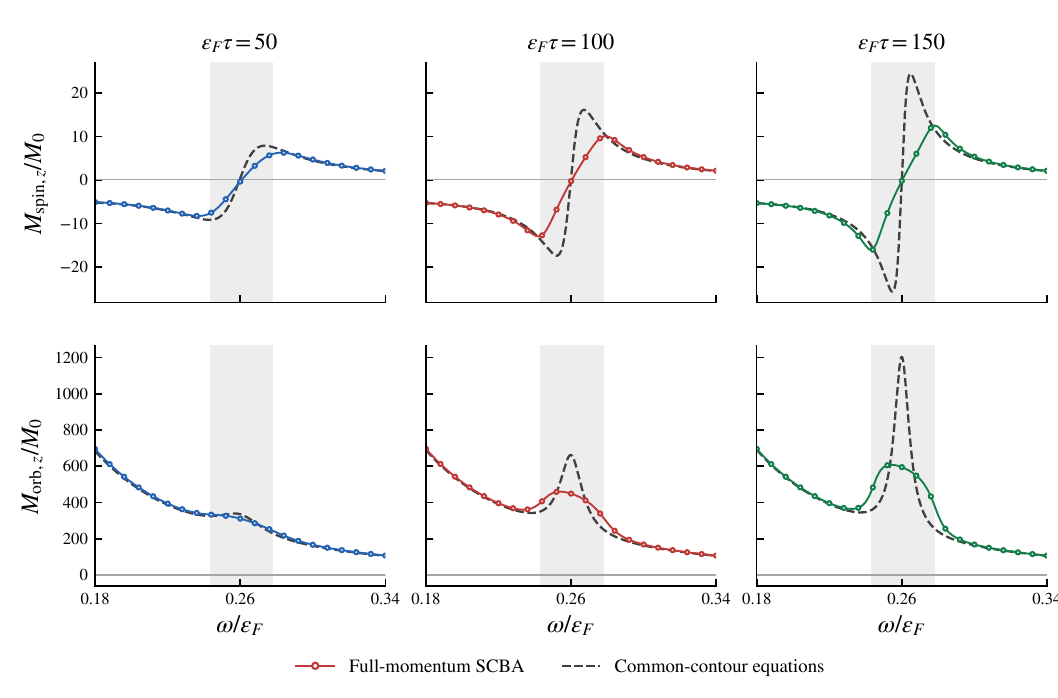}
\caption{Comparison at identical field convention, helicity, SOC, disorder, and magnetic normalization. Solid colored curves and open symbols show the full-momentum numerical response; dashed curves evaluate the single-splitting metallic equations directly. Top: spin. Bottom: orbital. Gray shading is the clean interbranch transition interval. No amplitude or frequency fit is applied. The analytical reduction sharpens features when the finite transition spread exceeds the disorder width. The physical contrast between dispersive spin reversal and fixed-sign orbital dominance survives in the full-momentum calculation.}
\label{fig:num-analytic}
\end{figure}

\begin{center}
\begin{tabular}{ccccc}
\toprule
$\varepsilon_F\tau$ & $r$ & spin difference & orbital difference & conductivity diagnostic\\
\midrule
50 & 1.69 & $40.1\%$ & $3.55\%$ & $0.456\%$\\
100 & 3.38 & $68.8\%$ & $30.8\%$ & $0.906\%$\\
150 & 5.07 & $83.5\%$ & $50.6\%$ & $1.73\%$\\
\bottomrule
\end{tabular}
\end{center}
Here $M_{\mathrm{an}}$ denotes the single-splitting analytical response. The spin and orbital differences in this table mean $\max_\omega|M_{\rm num}-M_{\mathrm{an}}|/\max_\omega|M_{\mathrm{an}}|$ on the shared numerical grid. They are discrepancies between two approximations, not numerical quadrature errors. The last column is $\max_\omega|M_{{\rm orb},{\rm num}}/[e\operatorname{Re} \sigma_{\rm num}\mathcal{H} /(mc_0\Gamma\omega)]-1|$. Its percent-level size is a numerical consistency diagnostic, not proof that the projected conductivity identity is exact for this finite-band, finite-reservoir model.

A freshly executed overlap check uses $\beta=0.02$, $\varepsilon_F\tau=100$, and $\omega/\varepsilon_F=0.03,0.04,0.05$. Here $r=0.08$. The largest spin difference is $2.44\%$ of the largest sampled analytical spin magnitude, and the orbital pointwise relative difference is below $0.95\%$. Relative spin errors near its zero can be larger; the stated spin metric is not a pointwise error bound. This check illustrates the more controlled radial regime without claiming exact agreement of the different spectral closures.

The numerical result retains the finite reservoir and cutoff. A unique reservoir-free stationary limit, a cutoff-independent fixed-density continuum limit, and an exact unprojected electric/magnetic decomposition have not been established. In particular, Eq.~\eqref{eq:EB-halves-orb} remains an analytical result of the declared collision closure; the full-momentum curves are not a separate numerical proof of that equality. The remainder in Eq.~\eqref{eq:orb-remainder} still specifies the additional unprojected consistency check. What the curves establish directly is the simultaneous frequency evolution of physical spin density and orbital curl current: spin reversal, orbital enhancement, and a total moment whose sign does not change over the displayed interval.

\bibliographystyle{apsrev4-2}
\bibliography{biblio}